\documentclass[letterpaper,11pt]{article}
\pdfoutput=1 

\usepackage{jheppub} 

\usepackage{float}
\usepackage{setspace}
\usepackage{subfigure}
\usepackage{hyperref}
\usepackage{dsfont}
\usepackage{amsmath}
\usepackage{amsmath, amssymb}
\usepackage{physics}
\usepackage{braket}
\usepackage{tikz}
\usepackage{xcolor}                    
\usetikzlibrary{arrows.meta}           
\usetikzlibrary{patterns,patterns.meta}
\usetikzlibrary{arrows.meta}   
\definecolor{cutred}{RGB}{220, 45, 45} 
\definecolor{bodyblue}{RGB}{0, 70, 180} 
\usepackage{xcolor,tikz}
\usetikzlibrary{patterns,patterns.meta}
\usepackage{physics}
\usepackage{amsmath}
\usepackage{tikz}
\usepackage{mathdots}
\usepackage{yhmath}
\usepackage{cancel}
\usepackage{color}
\usepackage{siunitx}
\usepackage{array}
\usepackage{multirow}
\usepackage{amssymb}
\usepackage{gensymb}
\usepackage{tabularx}
\usepackage{extarrows}
\usepackage{booktabs}
\usetikzlibrary{fadings}
\usetikzlibrary{patterns}
\usetikzlibrary{shadows.blur}
\usetikzlibrary{shapes}
\usepackage{subcaption}
\usepackage{graphicx}
\usepackage{subcaption}   
\usepackage{hyperref}     

\def\ba{\begin{align}}\def\ea{\end{align}}
\def\beq{\begin{eqnarray}}\def\eeq{\end{eqnarray}}
\def\be{\begin{equation}}\def\ee{\end{equation}}
\def\ben{\begin{equation}}
\def\een{\end{equation}}
\def\bea{\begin{eqnarray}}
\def\eea{\end{eqnarray}}

\title{Generalized Replica Manifolds I: Surgery and Averaging}
\author{Mohamed Hany Radwan}

\affiliation{Department of Physics and Astronomy, University of Kentucky, Lexington, KY 40506, U.S.A.}
\emailAdd{m.radwan@uky.edu}

\abstract{We develop a simple framework for implementing a type of path integral ``surgery'' through a correlated averaging over codimension-one defects/extended operators. This technique is then used to construct replica manifolds by effectively cutting and gluing the path integral without explicitly modifying the underlying manifold. We argue that restricted forms of this averaging may be used to calculate Rényi entanglement entropy corresponding to a wide range of choices of subsystem partitioning. When the entanglement entropy being calculated in this way does not simply correspond to entanglement between subregions, we call the resulting objects from this surgery ``generalized replica manifolds''.  We show how this framework extends to gauge theories and, in particular, how in non‑Abelian gauge theories it establishes a connection between replica calculations of a gauge-invariant notion of entanglement between color degrees of freedom and a quiver gauge theory structure. Finally, we discuss how this framework looks in the context of large-$N$ theories and holography, with a bird’s-eye view of potential future directions. This paper focuses on averaging over operators that form a representation of the Heisenberg group; a subsequent paper will focus on how we can extend this framework to more general operator averaging.}

\begin{document}
\maketitle

\section{Introduction}
\label{sec:intro}

Over the past several decades, quantum entanglement measures have emerged as an incredibly useful tool across many different areas of physics, from quantum information and quantum computing \cite{Horodecki:2009zz, Pezze:2016nxl, Steane:1997kb}, to condensed matter \cite{Amico:2007ag, Laflorencie:2015eck}, holography \cite{Nishioka:2009un, Headrick:2019eth, Nishioka:2018khk, Rangamani:2016dms, VanRaamsdonk:2016exw}, and black holes \cite{Witten:2024upt, Harlow:2014yka, Wall:2018ydq}. Indeed, if we focus our attention on holography and the idea of emergent gravity in general, it is difficult to overstate the progress that has been made since the discovery of the Ryu–Takayanagi formula, which established a concrete connection between entanglement in the boundary theory and bulk geometry \cite{Ryu:2006bv}. Later developments in this direction further enforced the deep connection between quantum information and gravity, with ideas like islands/replica wormholes \cite{Almheiri:2019psf, Penington:2019npb, Almheiri:2019qdq}, entanglement wedge reconstruction \cite{Dong:2016eik, Jafferis:2015del, Cotler:2017erl, Faulkner:2017vdd}, holographic tensor networks \cite{Swingle:2009bg, Hayden:2016cfa}, and the link between bulk locality and quantum error correction \cite{Almheiri:2014lwa, Pastawski:2015qua, Harlow:2016vwg}, being just some of the resulting highlights.

Given how fruitful investigating this topic turned out to be, one may wonder what kind of lessons we can learn from studying entanglement structures corresponding to more general ways of partitioning the degrees of freedom in the boundary theory, beyond spatial subregions. In particular, we may want to start by focusing on cases where the degrees of freedom are partitioned in a way that does not reference spatial subregions at all. We will refer to this kind of entanglement as entanglement between internal degrees of freedom.

There are many reasons to believe that studying this kind of entanglement in holography is worthwhile. One obvious reason is that this may be the only way to understand the link between entanglement and geometry in gauge-gravity duality examples where the gauge theory does not possess a base space, like D$0$-brane holography \cite{Anous:2019rqb, Das:2020jhy, Das:2020xoa, Frenkel:2023aft, Fliss:2025kzi}. In fact, exploring this direction can be important even in examples where the gauge theory does have a base space and the standard Ryu–Takayanagi formula can be applied. For example, in arguably the most famous concrete realization of gauge-gravity duality, the gravitational dual to the $\mathcal{N}=4$ supersymmetric Yang-Mills theory carries an internal space piece (the $S^5$ in $AdS_5 \times S^5$) which has been argued to be linked to internal degrees of freedom entanglement \cite{Mollabashi:2014qfa, Karch:2014pma, Das:2022njy, Bohra:2024wrq}. A perhaps even stronger motivation to pursue this direction arises from the fact that it is believed that this kind of entanglement is key in understanding sub-AdS locality and holography in the flat space limit \cite{Balasubramanian:2014sra, Caminiti:2024ctd, Susskind:1998vk}. 

Despite the many reasons making it an appealing research topic, the rich variety of entanglement between internal degrees of freedom, along with the roles they might play in bulk space emergence, remain largely unexplored. In part, this is due to the fact that progress in this direction faces one big obstacle: it is not at all clear how, using the AdS/CFT dictionary, one can derive the correct bulk dual to this kind of entanglement entropy in a similar way to how the Ryu–Takayanagi formula was derived in \cite{Lewkowycz:2013nqa}.

To better explain the source of this challenge, let us review some basic facts about entanglement entropy calculations. Entanglement entropy can be typically understood as the entropy of the reduced density matrix that one gets by performing a partial trace on a pure state, keeping only the information corresponding to the degrees of freedom of some subsystem. Regardless of how one obtains a density matrix $\hat{\rho}$, we can define many different measures of its entropy. Arguably, the most studied entropy measure is the von Neumann entropy given by  
\begin{equation}
    S_v=-\Tr{\hat{\rho}\ln\hat{\rho}}
\end{equation}
Another useful measure of entropy is the Rényi entropy defined by 
\begin{equation}
    S_n=\frac{1}{1-n}\ln\Tr{\hat{\rho}^n}
\end{equation}
In practice, it is often easier to calculate the von Neumann entropy by evaluating the Rényi entropy first and then taking the limit $n \rightarrow 1$. In this case, we have the simple relation
\begin{equation}
    \lim_{n\rightarrow 1} S_n=S_v
\end{equation}
This is typically referred to as the replica trick and has proven to be very useful in getting answers for entropy calculations that would have otherwise been very difficult.

\begin{figure}[!t]
  \centering

  \begin{subfigure}[]{\linewidth 0pt}
    \centering

    \begingroup
    \tikzset{every picture/.style={line width=0.75pt}} 

    \begin{tikzpicture}[x=0.75pt,y=0.75pt,yscale=-1,xscale=1.2]

    \draw [color={rgb, 255:red, 0; green, 0; blue, 255 }  ,draw opacity=1 ]  [dash pattern={on 2.25pt off 2.25pt on 2.25pt off 2.25pt}]  (256.09,127.5) -- (244.94,127.5) -- (231.21,127.5) ;
    \draw  [dash pattern={on 2.25pt off 2.25pt on 2.25pt off 2.25pt}]  (206.33,116.74) -- (217.48,116.74) -- (231.21,116.71) ;
    \draw [color={rgb, 255:red, 0; green, 0; blue, 255 }  ,draw opacity=1 ] [dash pattern={on 2.25pt off 2.25pt on 2.25pt off 2.25pt}]  (256.09,116.71) -- (244.94,116.71) -- (231.21,116.71) ;
    \draw  [fill={rgb, 255:red, 0; green, 0; blue, 0 }  ,fill opacity=1 ] (230.51,116.71) .. controls (230.51,116.33) and (230.83,116.02) .. (231.21,116.02) .. controls (231.59,116.02) and (231.9,116.33) .. (231.9,116.71) .. controls (231.9,117.1) and (231.59,117.41) .. (231.21,117.41) .. controls (230.83,117.41) and (230.51,117.1) .. (230.51,116.71) -- cycle ;
    \draw    (206.79,57.55) -- (206.79,117.02) ;
    \draw    (255.97,57.38) -- (256.09,117.19) ;
    \draw   [dash pattern={on 2.25pt off 2.25pt on 2.25pt off 2.25pt}]  (206.33,127.47) -- (217.48,127.47) -- (231.21,127.5) ;
    \draw  [fill={rgb, 255:red, 0; green, 0; blue, 0 }  ,fill opacity=1 ] (230.51,127.5) .. controls (230.51,127.88) and (230.83,128.19) .. (231.21,128.19) .. controls (231.59,128.19) and (231.9,127.88) .. (231.9,127.5) .. controls (231.9,127.11) and (231.59,126.8) .. (231.21,126.8) .. controls (230.83,126.8) and (230.51,127.11) .. (230.51,127.5) -- cycle ;
    \draw    (206.79,186.66) -- (206.79,127.19) ;
    \draw    (255.97,186.84) -- (256.09,127.02) ;
    \draw    (206.33,186.39) -- (256.31,186.33) ;
    \draw [color={rgb, 255:red, 0; green, 0; blue, 255 }  ,draw opacity=1 ] [dash pattern={on 2.25pt off 2.25pt on 2.25pt off 2.25pt}]  (316.08,127.5) -- (304.93,127.5) -- (291.2,127.5) ;
    \draw  [dash pattern={on 2.25pt off 2.25pt on 2.25pt off 2.25pt}]  (266.31,116.74) -- (277.46,116.74) -- (291.2,116.71) ;
    \draw [color={rgb, 255:red, 0; green, 0; blue, 255 }  ,draw opacity=1 ] [dash pattern={on 2.25pt off 2.25pt on 2.25pt off 2.25pt}]  (316.08,116.71) -- (304.93,116.71) -- (291.2,116.71) ;
    \draw  [fill={rgb, 255:red, 0; green, 0; blue, 0 }  ,fill opacity=1 ] (290.5,116.71) .. controls (290.5,116.33) and (290.81,116.02) .. (291.2,116.02) .. controls (291.58,116.02) and (291.89,116.33) .. (291.89,116.71) .. controls (291.89,117.1) and (291.58,117.41) .. (291.2,117.41) .. controls (290.81,117.41) and (290.5,117.1) .. (290.5,116.71) -- cycle ;
    \draw    (266.78,57.55) -- (266.78,117.02) ;
    \draw    (315.96,57.38) -- (316.08,117.19) ;
    \draw    [dash pattern={on 2.25pt off 2.25pt on 2.25pt off 2.25pt}] (266.31,127.47) -- (277.46,127.47) -- (291.2,127.5) ;
    \draw  [fill={rgb, 255:red, 0; green, 0; blue, 0 }  ,fill opacity=1 ] (290.5,127.5) .. controls (290.5,127.88) and (290.81,128.19) .. (291.2,128.19) .. controls (291.58,128.19) and (291.89,127.88) .. (291.89,127.5) .. controls (291.89,127.11) and (291.58,126.8) .. (291.2,126.8) .. controls (290.81,126.8) and (290.5,127.11) .. (290.5,127.5) -- cycle ;
    \draw    (266.78,186.66) -- (266.78,127.19) ;
    \draw    (315.96,186.84) -- (316.08,127.02) ;
    \draw    (266.31,186.39) -- (316.3,186.33) ;
    \draw [color={rgb, 255:red, 0; green, 0; blue, 255 }  ,draw opacity=1 ]    (243.15,131.8) .. controls (245.77,150.15) and (269.46,141.1) .. (299.44,119.44) ;
    \draw [shift={(301.76,117.75)}, rotate = 143.64] [fill={rgb, 255:red, 0; green, 0; blue, 255 }  ,fill opacity=1 ][line width=0.08]  [draw opacity=0] (3.57,-1.72) -- (0,0) -- (3.57,1.72) -- cycle    ;
    \draw [shift={(242.92,128.59)}, rotate = 89.22] [fill={rgb, 255:red, 0; green, 0; blue, 255 }  ,fill opacity=1 ][line width=0.08]  [draw opacity=0] (3.57,-1.72) -- (0,0) -- (3.57,1.72) -- cycle    ;
    \draw [color={rgb, 255:red, 0; green, 0; blue, 255 }  ,draw opacity=1 ]   (308.32,123.02) .. controls (307.59,107.99) and (301.75,95.47) .. (251.79,114.37) ;
    \draw [shift={(249.48,115.25)}, rotate = 338.83] [fill={rgb, 255:red, 0; green, 0; blue, 255 }  ,fill opacity=1 ][line width=0.08]  [draw opacity=0] (3.57,-1.72) -- (0,0) -- (3.57,1.72) -- cycle    ;
    \draw [shift={(308.42,126.23)}, rotate = 269.23] [fill={rgb, 255:red, 0; green, 0; blue, 255 }  ,fill opacity=1 ][line width=0.08]  [draw opacity=0] (3.57,-1.72) -- (0,0) -- (3.57,1.72) -- cycle    ;
    \draw [line width=1.5]    (326.1,122.11) -- (392.1,122.11) ;
    \draw [shift={(396.1,122.11)}, rotate = 180] [fill={rgb, 255:red, 0; green, 0; blue, 0 }  ][line width=0.08]  [draw opacity=0] (6.97,-3.35) -- (0,0) -- (6.97,3.35) -- cycle    ;
\draw   (407.6,61.39) -- (493.06,61.39) -- (493.06,182.82) -- (407.6,182.82) -- cycle ;
\draw  [fill={rgb, 255:red, 0; green, 0; blue, 0 }  ,fill opacity=1 ] (448.18,122.11) .. controls (448.18,121) and (449.14,120.1) .. (450.33,120.1) .. controls (451.52,120.1) and (452.48,121) .. (452.48,122.11) .. controls (452.48,123.21) and (451.52,124.11) .. (450.33,124.11) .. controls (449.14,124.11) and (448.18,123.21) .. (448.18,122.11) -- cycle ;
    \draw    (218.62,120.48) -- (218.69,123.73) ;
    \draw [shift={(218.75,126.73)}, rotate = 268.82] [fill={rgb, 255:red, 0; green, 0; blue, 0 }  ][line width=0.08]  [draw opacity=0] (3.57,-1.72) -- (0,0) -- (3.57,1.72) -- cycle    ;
    \draw [shift={(218.56,117.48)}, rotate = 88.82] [fill={rgb, 255:red, 0; green, 0; blue, 0 }  ][line width=0.08]  [draw opacity=0] (3.57,-1.72) -- (0,0) -- (3.57,1.72) -- cycle    ;
    \draw    (279.07,120.48) -- (279.13,123.73) ;
    \draw [shift={(279.2,126.73)}, rotate = 268.82] [fill={rgb, 255:red, 0; green, 0; blue, 0 }  ][line width=0.08]  [draw opacity=0] (3.57,-1.72) -- (0,0) -- (3.57,1.72) -- cycle    ;
    \draw [shift={(279.01,117.48)}, rotate = 88.82] [fill={rgb, 255:red, 0; green, 0; blue, 0 }  ][line width=0.08]  [draw opacity=0] (3.57,-1.72) -- (0,0) -- (3.57,1.72) -- cycle    ;

    \draw    (206.33,57.82) -- (256.31,57.89) ;
    \draw    (266.31,57.82) -- (316.3,57.89) ;

    \draw (465,137) node  [font=\scriptsize] [align=center] {\begin{minipage}[lt]{80pt}\setlength\topsep{0pt}
    {\scriptsize Conical Defect}
    \end{minipage}};
    \draw (130,118.51) node [anchor=north west][inner sep=0.75pt]    {$\Tr{\rho _{x >0}^{2} }=$};

    \end{tikzpicture}
    \endgroup

    \label{fig:top}
  \end{subfigure}

  \vspace{0.9em}

  \begin{subfigure}[]{\linewidth 0pt}
    \centering

    \begingroup
     
    \tikzset{
    pattern size/.store in=\mcSize, 
    pattern size = 5pt,
    pattern thickness/.store in=\mcThickness, 
    pattern thickness = 0.3pt,
    pattern radius/.store in=\mcRadius, 
    pattern radius = 1pt}
    \makeatletter
    \pgfutil@ifundefined{pgf@pattern@name@_0bla0s63e}{
    \pgfdeclarepatternformonly[\mcThickness,\mcSize]{_0bla0s63e}
    {\pgfqpoint{0pt}{0pt}}
    {\pgfpoint{\mcSize+\mcThickness}{\mcSize+\mcThickness}}
    {\pgfpoint{\mcSize}{\mcSize}}
    {
    \pgfsetcolor{\tikz@pattern@color}
    \pgfsetlinewidth{\mcThickness}
    \pgfpathmoveto{\pgfqpoint{0pt}{0pt}}
    \pgfpathlineto{\pgfpoint{\mcSize+\mcThickness}{\mcSize+\mcThickness}}
    \pgfusepath{stroke}
    }}
    \makeatother

     
    \tikzset{
    pattern size/.store in=\mcSize, 
    pattern size = 5pt,
    pattern thickness/.store in=\mcThickness, 
    pattern thickness = 0.3pt,
    pattern radius/.store in=\mcRadius, 
    pattern radius = 1pt}
    \makeatletter
    \pgfutil@ifundefined{pgf@pattern@name@_sqsvzp7om}{
    \pgfdeclarepatternformonly[\mcThickness,\mcSize]{_sqsvzp7om}
    {\pgfqpoint{0pt}{0pt}}
    {\pgfpoint{\mcSize+\mcThickness}{\mcSize+\mcThickness}}
    {\pgfpoint{\mcSize}{\mcSize}}
    {
    \pgfsetcolor{\tikz@pattern@color}
    \pgfsetlinewidth{\mcThickness}
    \pgfpathmoveto{\pgfqpoint{0pt}{0pt}}
    \pgfpathlineto{\pgfpoint{\mcSize+\mcThickness}{\mcSize+\mcThickness}}
    \pgfusepath{stroke}
    }}
    \makeatother

     
    \tikzset{
    pattern size/.store in=\mcSize, 
    pattern size = 5pt,
    pattern thickness/.store in=\mcThickness, 
    pattern thickness = 0.3pt,
    pattern radius/.store in=\mcRadius, 
    pattern radius = 1pt}
    \makeatletter
    \pgfutil@ifundefined{pgf@pattern@name@_c576sb0ws}{
    \pgfdeclarepatternformonly[\mcThickness,\mcSize]{_c576sb0ws}
    {\pgfqpoint{0pt}{0pt}}
    {\pgfpoint{\mcSize+\mcThickness}{\mcSize+\mcThickness}}
    {\pgfpoint{\mcSize}{\mcSize}}
    {
    \pgfsetcolor{\tikz@pattern@color}
    \pgfsetlinewidth{\mcThickness}
    \pgfpathmoveto{\pgfqpoint{0pt}{0pt}}
    \pgfpathlineto{\pgfpoint{\mcSize+\mcThickness}{\mcSize+\mcThickness}}
    \pgfusepath{stroke}
    }}
    \makeatother

     
    \tikzset{
    pattern size/.store in=\mcSize, 
    pattern size = 5pt,
    pattern thickness/.store in=\mcThickness, 
    pattern thickness = 0.3pt,
    pattern radius/.store in=\mcRadius, 
    pattern radius = 1pt}
    \makeatletter
    \pgfutil@ifundefined{pgf@pattern@name@_7vlo9xmie}{
    \pgfdeclarepatternformonly[\mcThickness,\mcSize]{_7vlo9xmie}
    {\pgfqpoint{0pt}{0pt}}
    {\pgfpoint{\mcSize+\mcThickness}{\mcSize+\mcThickness}}
    {\pgfpoint{\mcSize}{\mcSize}}
    {
    \pgfsetcolor{\tikz@pattern@color}
    \pgfsetlinewidth{\mcThickness}
    \pgfpathmoveto{\pgfqpoint{0pt}{0pt}}
    \pgfpathlineto{\pgfpoint{\mcSize+\mcThickness}{\mcSize+\mcThickness}}
    \pgfusepath{stroke}
    }}
    \makeatother
    \tikzset{every picture/.style={line width=0.75pt}} 

    \begin{tikzpicture}[x=0.75pt,y=0.75pt,yscale=-1,xscale=1.2]

    \draw  [draw opacity=0][pattern=_0bla0s63e,pattern size=1.6500000000000001pt,pattern thickness=0.75pt,pattern radius=0pt, pattern color={rgb, 255:red, 128; green, 128; blue, 128}] (297.82,149.15) -- (302.42,149.15) -- (302.42,208.36) -- (297.82,208.36) -- cycle ;
    \draw  [draw opacity=0][pattern=_sqsvzp7om,pattern size=1.6500000000000001pt,pattern thickness=0.75pt,pattern radius=0pt, pattern color={rgb, 255:red, 128; green, 128; blue, 128}] (238,149.25) -- (242.61,149.25) -- (242.61,208.46) -- (238,208.46) -- cycle ;
    \draw  [draw opacity=0][pattern=_c576sb0ws,pattern size=1.6500000000000001pt,pattern thickness=0.75pt,pattern radius=0pt, pattern color={rgb, 255:red, 128; green, 128; blue, 128}] (298.07,78.59) -- (302.67,78.59) -- (302.67,137.8) -- (298.07,137.8) -- cycle ;
    \draw  [draw opacity=0][pattern=_7vlo9xmie,pattern size=1.6500000000000001pt,pattern thickness=0.75pt,pattern radius=0pt, pattern color={rgb, 255:red, 128; green, 128; blue, 128}] (238.12,78.59) -- (242.73,78.59) -- (242.73,137.8) -- (238.12,137.8) -- cycle ;

    \draw  [dash pattern={on 2.25pt off 2.25pt on 2.25pt off 2.25pt}]  (215.49,137.43) -- (226.1,137.53) -- (237.68,137.52) ;
    \draw    (215.59,78.59) -- (215.59,137.88) ;
    \draw    (215.21,78.88) -- (238.35,78.89) ;
    \draw    (237.82,78.94) -- (237.68,137.86) ;

    \draw  [dash pattern={on 2.25pt off 2.25pt on 2.25pt off 2.25pt}]  (215.49,149.62) -- (226.1,149.52) -- (237.68,149.53) ;
    \draw    (215.59,208.46) -- (215.59,149.17) ;
    \draw    (215.21,208.17) -- (238.35,208.16) ;
    \draw    (237.82,208.11) -- (237.68,149.19) ;

    \draw [color={rgb, 255:red, 0; green, 0; blue, 255 }  ,draw opacity=1 ] [dash pattern={on 2.25pt off 2.25pt on 2.25pt off 2.25pt}]  (242.57,137.43) -- (253.18,137.53) -- (264.75,137.52) ;
    \draw    (242.66,78.59) -- (242.66,137.88) ;
    \draw    (242.28,78.88) -- (265.43,78.89) ;
    \draw    (264.9,78.94) -- (264.75,137.86) ;

    \draw [color={rgb, 255:red, 0; green, 0; blue, 255 }  ,draw opacity=1 ] [dash pattern={on 2.25pt off 2.25pt on 2.25pt off 2.25pt}]  (242.57,149.62) -- (253.18,149.52) -- (264.75,149.53) ;
    \draw    (242.66,208.46) -- (242.66,149.17) ;
    \draw    (242.28,208.17) -- (265.43,208.16) ;
    \draw    (264.9,208.11) -- (264.75,149.19) ;

    \draw  [dash pattern={on 2.25pt off 2.25pt on 2.25pt off 2.25pt}]  (275.48,137.43) -- (286.09,137.53) -- (297.66,137.52) ;
    \draw    (275.57,78.59) -- (275.57,137.88) ;
    \draw    (275.19,78.88) -- (298.34,78.89) ;
    \draw    (297.81,78.94) -- (297.66,137.86) ;

    \draw  [dash pattern={on 2.25pt off 2.25pt on 2.25pt off 2.25pt}]  (275.48,149.52) -- (286.09,149.42) -- (297.66,149.43) ;
    \draw    (275.57,208.36) -- (275.57,149.07) ;
    \draw    (275.19,208.07) -- (298.34,208.06) ;
    \draw    (297.81,208.01) -- (297.66,149.09) ;

    \draw [color={rgb, 255:red, 0; green, 0; blue, 255 }  ,draw opacity=1 ] [dash pattern={on 2.25pt off 2.25pt on 2.25pt off 2.25pt}]  (302.56,137.43) -- (313.16,137.53) -- (324.74,137.52) ;
    \draw    (302.65,78.59) -- (302.65,137.88) ;
    \draw    (302.27,78.88) -- (325.42,78.89) ;
    \draw    (324.89,78.94) -- (324.74,137.86) ;

    \draw [color={rgb, 255:red, 0; green, 0; blue, 255 }  ,draw opacity=1 ] [dash pattern={on 2.25pt off 2.25pt on 2.25pt off 2.25pt}]  (302.56,149.52) -- (313.16,149.42) -- (324.74,149.43) ;
    \draw    (302.65,208.36) -- (302.65,149.07) ;
    \draw    (302.27,208.07) -- (325.42,208.06) ;
    \draw    (324.89,208.01) -- (324.74,149.09) ;

    \draw [color={rgb, 255:red, 0; green, 0; blue, 255 }  ,draw opacity=1 ]   (317.59,144.46) .. controls (316.95,129.34) and (311.71,116.06) .. (258.26,134.28) ;
    \draw [shift={(255.79,135.14)}, rotate = 340.74] [fill={rgb, 255:red, 0; green, 0; blue, 255 }  ,fill opacity=1 ][line width=0.08]  [draw opacity=0] (3.57,-1.72) -- (0,0) -- (3.57,1.72) -- cycle    ;
    \draw [shift={(317.67,147.68)}, rotate = 269.23] [fill={rgb, 255:red, 0; green, 0; blue, 255 }  ,fill opacity=1 ][line width=0.08]  [draw opacity=0] (3.57,-1.72) -- (0,0) -- (3.57,1.72) -- cycle    ;
    \draw [color={rgb, 255:red, 0; green, 0; blue, 255 }  ,draw opacity=1 ]   (255.19,153.7) .. controls (257.73,172.07) and (280.71,163.22) .. (310.66,141.56) ;
    \draw [shift={(312.98,139.87)}, rotate = 143.64] [fill={rgb, 255:red, 0; green, 0; blue, 255 }  ,fill opacity=1 ][line width=0.08]  [draw opacity=0] (3.57,-1.72) -- (0,0) -- (3.57,1.72) -- cycle    ;
    \draw [shift={(254.96,150.48)}, rotate = 89.22] [fill={rgb, 255:red, 0; green, 0; blue, 255 }  ,fill opacity=1 ][line width=0.08]  [draw opacity=0] (3.57,-1.72) -- (0,0) -- (3.57,1.72) -- cycle    ;
    \draw    (286.73,141.81) -- (286.8,145.07) ;
    \draw [shift={(286.86,148.07)}, rotate = 268.82] [fill={rgb, 255:red, 0; green, 0; blue, 0 }  ][line width=0.08]  [draw opacity=0] (3.57,-1.72) -- (0,0) -- (3.57,1.72) -- cycle    ;
    \draw [shift={(286.67,138.81)}, rotate = 88.82] [fill={rgb, 255:red, 0; green, 0; blue, 0 }  ][line width=0.08]  [draw opacity=0] (3.57,-1.72) -- (0,0) -- (3.57,1.72) -- cycle    ;
    \draw    (226.75,142.06) -- (226.81,145.32) ;
    \draw [shift={(226.88,148.32)}, rotate = 268.82] [fill={rgb, 255:red, 0; green, 0; blue, 0 }  ][line width=0.08]  [draw opacity=0] (3.57,-1.72) -- (0,0) -- (3.57,1.72) -- cycle    ;
    \draw [shift={(226.68,139.06)}, rotate = 88.82] [fill={rgb, 255:red, 0; green, 0; blue, 0 }  ][line width=0.08]  [draw opacity=0] (3.57,-1.72) -- (0,0) -- (3.57,1.72) -- cycle    ;

    \draw [line width=1.5]    (335.1,143.52) -- (401.1,143.52) ;
    \draw [shift={(405.1,143.52)}, rotate = 180] [fill={rgb, 255:red, 0; green, 0; blue, 0 }  ][line width=0.08]  [draw opacity=0] (6.97,-3.35) -- (0,0) -- (6.97,3.35) -- cycle    ;

    \draw (477,137.12) node  [font=\Huge] [align=right] {\begin{minipage}[lt]{67pt}\setlength\topsep{0pt}
    \textbf{{\fontfamily{lmtt}\selectfont {\fontsize{70}{90}\selectfont   ?}}}
    \end{minipage}};
    \draw (138,139.57) node [anchor=north west][inner sep=0.75pt]    {$~\,~\Tr{\rho _{\psi }^{2}} =$};
    \draw (310,176.4) node [anchor=north west][inner sep=0.75pt]  [font=\footnotesize]  {$\psi $};
    \draw (250,176.4) node [anchor=north west][inner sep=0.75pt]  [font=\footnotesize]  {$\psi $};
    \draw (282,176.4) node [anchor=north west][inner sep=0.75pt]  [font=\footnotesize]  {$\phi $};
    \draw (222,176.4) node [anchor=north west][inner sep=0.75pt]  [font=\footnotesize]  {$\phi $};
    \draw (282,106.36) node [anchor=north west][inner sep=0.75pt]  [font=\footnotesize]  {$\phi $};
    \draw (222,106.36) node [anchor=north west][inner sep=0.75pt]  [font=\footnotesize]  {$\phi $};
    \draw (310,106.36) node [anchor=north west][inner sep=0.75pt]  [font=\footnotesize]  {$\psi $};
    \draw (250,106.36) node [anchor=north west][inner sep=0.75pt]  [font=\footnotesize]  {$\psi $};

    \end{tikzpicture}
    \endgroup
    \label{fig:bottom}
  \end{subfigure}

                \caption{(a) A representation of a Euclidean path integral relevant for an evaluation of the half-space entanglement entropy in the ground state of a quantum field theory in $1+1$-dimensional Minkowski space. The vertical direction can be understood as Euclidean time evolution, while the double-sided arrows represent identifications of the color-coded cuts they are pointing to. The path integral can be interpreted as introducing a conical singularity in the Euclidean manifold at the entangling point separating the two halves of space. This in turn allows us to understand the boundary conditions we need to impose in the corresponding bulk calculation. (b) A representation of the analogous path integral for the entanglement entropy between two interacting fields. The shading pattern here represents interactions between the two fields. Our goal will be to try to replace the question mark with something more suitable for an application of the AdS/CFT dictionary.}
  \label{fig:stacked}
\end{figure}
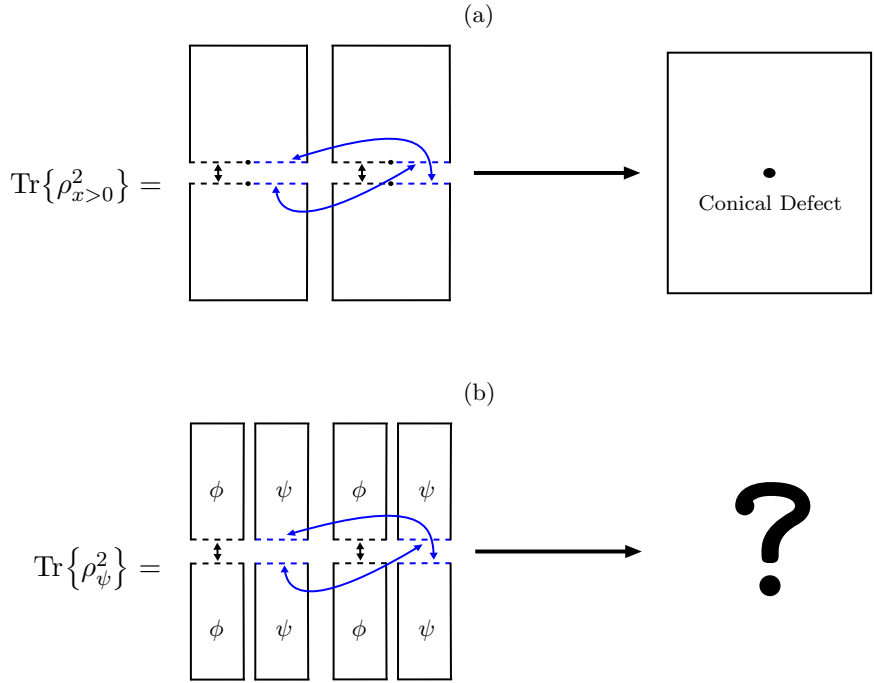

In situations where the entanglement entropy being calculated corresponds to entanglement between spatial subregions, the replica trick can be given a nice geometric interpretation \cite{Calabrese:2009qy, Calabrese:2004eu, Casini:2009sr}. In this case, the Rényi entropy can be obtained by performing a path integral over a special ``replica'' manifold built out of $n$ copies of the original system that have been geometrically glued together in a particular way. This provides a very useful approach to tackling these calculations, as these manufactured path integrals typically correspond to manifolds with some conical singularities along the entangling surface separating the two subregions. An example of such a path integral is shown in fig. \ref{fig:stacked}(a).\footnote{This approach, while in many ways successful, does not lead to a well-defined notion of entanglement between spatial subregions in relativistic field theories. The more rigorous way to tackle this subject is by thinking in terms of von Neumann algebras \cite{Witten:2018zxz, Sorce:2023fdx, Casini:2022rlv} }

This replica manifold picture played a very important role in the derivation of the Ryu–Takayanagi formula in \cite{Lewkowycz:2013nqa}. Crucially, the purely geometric nature of the replica manifold path integral makes it possible to identify the appropriate bulk boundary conditions needed to carefully translate the calculation from the boundary to the bulk by using the standard ingredients of the AdS/CFT dictionary.      

Now compare this to a situation where, starting with two interacting scalar fields, we want to calculate the entanglement entropy between them.\footnote{For discussions of this kind of entanglement, see \cite{Mollabashi:2014qfa, Taylor:2015kda, Nakai:2017qos, MohammadiMozaffar:2015clv, Huffel:2017ewr, MohammadiMozaffar:2024uiy}.} That is, given the ground state for example, we trace over the degrees of freedom associated to one of the fields everywhere while leaving the other one intact. An example of a path integral relevant to such a calculation is shown in fig \ref{fig:stacked}(b). It should be immediately clear that we can no longer have a purely geometric interpretation corresponding to putting the theory on a replica manifold with some conical singularities. This is because the correct path integral has to treat the two fields differently, and so a simple manifold manipulation, which treats all fields in the same way, cannot give the correct answer. 

Clearly, this failure to interpret the calculation as nothing more than placing the theory on a replica manifold will persist for any partitioning of the degrees of freedom that is not solely based on spatial subregions. A question now arises: is there any way to perform a similar analysis to the one in \cite{Lewkowycz:2013nqa} but for internal degrees of freedom? One of the main goals of this paper is to investigate a framework on the field theory side that can potentially help us in using the AdS/CFT dictionary to derive the bulk duals of internal degrees of freedom entanglement in the boundary theory.

To get a hint of how we may potentially make progress, let us look at what the exact statement of what the AdS/CFT correspondence tells us. In particular, using the GKPW dictionary \cite{Witten:1998qj, Gubser:1998bc, Aharony:1999ti}, we have 
\begin{equation} \label{GKPW}
\left\langle e^{\int d^d x \phi_0(\vec{x}) \mathcal{O}(\vec{x})}\right\rangle_{CFT}=
Z_{\text{string/gravity}}\Big[\left.\phi\right|_{\partial \text {AdS }}=\phi_{0}\Big]
\end{equation}
which tells us how to relate sources we couple to certain operators in the CFT to boundary conditions we impose on the string theory (or supergravity) partition function.\footnote{Given the form we used in \eqref{GKPW}, it typically refers to sourcing single trace operators in the CFT. However, much is also known about how the AdS/CFT dictionary can accommodate more general CFT deformations \cite{Witten:2001ua, Berkooz:2002ug, Gubser:2002vv, Papadimitriou:2007sj}.} Perhaps then, if we can find a way to represent path integrals like the one in fig \ref{fig:stacked}(b) in terms of some sources (or some averaging over these sources), then we may have some hope of understanding it in terms of a bulk calculation. 

Having identified our goal, our strategy in pursuing it will be rather simple. Focusing our attention on the field theory side, we will start by asking if one can engineer some averaging over sources, similar to $\phi_0$ in \eqref{GKPW}, to effectively cut and glue the path integral manifold. This kind of manipulation is typically referred to as performing a ``surgery''. In its simplest form, the kind of surgery we are talking about corresponds to simple single particle quantum mechanics identities like 
\ben
\frac{Z(n\beta)}{Z(\beta)^n}=\left(\prod_{j=0}^{n-1} \int_{-\infty}^{\infty} \frac{dK_j\, dJ_j}{2\pi}\right)
\prod_{j=0}^{n-1}
\left\langle e^{i J_j X}\, e^{i K_j P}\, e^{-i J_{j+1} X}\right\rangle_{\beta},
\een
where $Z(\beta)$ is the thermal partition at a temperature $1/\beta$ and the right-hand side is a correlated average over some finite temperature expectation values.    

After exploring how these surgery operations can be implemented using this kind of engineered averaging, we will explore how we can use it to manufacture replica manifolds for Rényi entropy calculations. Since we have a lot more control as to which operators we couple the sources to, we will find that we can perform a sort of ``selective'' surgery that only targets the degrees of freedom we are interested in. This strategy will allow us to construct path integrals like the one shown in \ref{fig:stacked}(b) out of averaging over sources in the original theory. As we mentioned earlier, the path integrals we construct in this way are not really interpretable as simply living on a replica manifold. That being said, we will find that maintaining some of the geometric intuition one gets in the standard replica manifold case can be very useful, and so we will refer to these constructions as ``generalized replica manifolds''.   

Our focus in this paper will be on averaging over operator insertions that form a representation of the Heisenberg group. An obvious example of such an averaging would correspond to averaging over sources coupled to a fundamental scalar field operator and its conjugate momentum.\footnote{More precisely, if we put the scalar field on a lattice, we will get one representation of the Heisenberg group for each lattice point.} While these are not the kind of sources one typically discusses in the context of the AdS/CFT correspondence, they are still a good starting point. In a subsequent paper, we will discuss how to extend these ideas beyond the Heisenberg group, using sources coupled to more general operators.  

A couple of important points. First, we note that an alternative way to think of these sources is as operator insertions or some kind of defects placed inside the path integral. As such, throughout the text, we will refer to them as any one of these things interchangeably, picking whichever term better fits the context of the discussion at hand. Second, while the author's main motivation in pursuing these ideas had its origins in holography, the framework we will present could be in principle applied to any local non-gravitational theory. In fact, apart from a brief discussion towards the end of section \ref{sec:lookahead}, we will not be discussing holography at all. Instead, we will focus our efforts on understanding how we can construct these generalized replica manifolds on the field theory side.

This paper is organized as follows. In section \ref{sec:simple}, we start by considering the rather elementary example of two coupled harmonic oscillators. In this case, the ground state wave function is Gaussian and can be straightforwardly treated in multiple different ways \cite{Weedbrook:2012cjy, Srednicki:1993im}. Nevertheless, the simplicity of this setup will offer us a convenient playground to develop the main ideas behind our path integral surgery technique and how it can be used in entanglement entropy calculations. As this is meant to serve as a prototype for the rest of the paper, we will carry out the calculation in (perhaps too much) detail all the way through to the end. Naturally, this will involve reproducing a lot of well known facts along the way that some readers may want to skim over.  

In section \ref{sec:MIS}, we sketch how, working at some fixed finite cutoff, the lessons learned from section \ref{sec:simple} can be extended to local field theories’ path integrals. One of the main ideas in this section is that we can understand these surgery operations in terms of introducing codimension-one auxiliary fields that effectively cut and glue the manifold on which we are performing the path integral. We will use real scalar fields as our first example and then extend the discussion to gauge theories. 

In section \ref{sec:constraints}, we argue that one can develop a picture where a connection is made between certain choices of density matrix partial tracing and imposing constraints on the auxiliary fields introduced in section \ref{sec:MIS}. While this section will not play an essential role in the rest of the paper, we argue that this picture can be useful in future calculations.

Section \ref{sec:gaugetheory} addresses a potential application of these ideas to better understand the concept of entanglement between color degrees of freedom in non-Abelian gauge theories. Within the context of matrix quantum mechanics, the aim is to try to address how one can perform a partial trace over degrees of freedom corresponding to matrix elements inside some block in a manifestly gauge-invariant way. We will see how using auxiliary bifundamental matter, we can define notions of partial tracing degrees of freedom associated to ``a'' block instead of a specific block. We will then show how the replica calculation in this case is related to a quiver gauge theory structure.

Finally, in section \ref{sec:lookahead} we discuss how applying these ideas may be useful in the context of large-$N$ models and holography. This is meant to offer a bird’s eye view of future directions and how this all ties to the motivation outlined in the introduction. 

\subsection{Relation to previous work}

While to the author's best knowledge, the approach and framework presented here is novel, some of the formulas we will derive have appeared in different forms before in the literature \cite{Shiba:2014uia, Chakraborty:2018wkx, Chakraborty:2020gjs, Moitra:2020cty, 
Moitra:2023gjm, Sarkar:2025uoe, Haldar:2020ymg, Klich:2024iug}. As such, it is worthwhile to point out exactly what the similarities and differences are between this paper and these previous works.

 Let us give a quick overview of what was done in those works. In \cite{Shiba:2014uia}, an operator formalism was developed for Rényi entanglement entropy calculations where the calculation can be expressed as the expectation value of an operator that the author called the ``glueing operator''. The derivation is based on a careful analysis of what $\Tr{\hat{\rho}_\Omega^n}$, for some reduced density matrix $\hat{\rho}_\Omega$, looks like when written in a basis of canonically conjugate variables. By comparing the resulting expression to the trace of the tensor product of $n$ copies of the unreduced density matrix $\Tr{\hat{\rho}^{(n)}}$, the form of this glueing operator $E_\Omega$ is identified by demanding that it satisfies the following relation
\begin{equation}
\Tr{\hat{\rho}_\Omega^n}=\Tr{\hat{\rho}^{(n)}E_\Omega}
\end{equation}
The resulting expressions from this approach are essentially identical to some of the formulas we will derive for Rényi entropy calculations.
 
Alternatively, the work done in \cite{Chakraborty:2018wkx, Chakraborty:2020gjs, Moitra:2020cty, 
Moitra:2023gjm, Sarkar:2025uoe, Haldar:2020ymg} is mostly based on exploiting Wigner function techniques. These techniques are used to represent Rényi entanglement entropy calculations in terms of some integral/averaging over some Euclidean time ``kick'' sources that act on the subsystem of interest only at specific Euclidean time instants. These works also discuss how this formalism can be applied to fermions, which will not be discussed in this paper. The formulas in these papers can be related to some of the expressions we will derive here. In fact, we will occasionally use the same ``Euclidean time kick sources'' terminology. 
 
Finally, in this note \cite{Klich:2024iug}, it was pointed out that the SWAP operation, widely used in Rényi entanglement entropy calculations, can be represented in terms of an average over displacement operators. This offers one way to interpret some of the formulas we will derive.  
 
Compared to these works, the main novelties in this paper will mostly come down to three main ideas: developing a path integral surgery technique, showing how this can be used for a general class of entropy calculations, and illustrating an application of these concepts to how we can think about notions of color entanglement in non-Abelian gauge theory in section \ref{sec:gaugetheory}. In fact, to the author's best knowledge, this path integral surgery technique itself has not been discussed in the literature before, which may be interesting in its own right independently of Rényi entropy calculations. The framework we will develop here will also serve as the foundation of the discussion in a follow-up paper where results will have very little overlap with these previous works.

\section{A simple example}
\label{sec:simple}

\begin{figure}[!t] 
	\begin{center}

 
\tikzset{
pattern size/.store in=\mcSize, 
pattern size = 5pt,
pattern thickness/.store in=\mcThickness, 
pattern thickness = 0.3pt,
pattern radius/.store in=\mcRadius, 
pattern radius = 1pt}
\makeatletter
\pgfutil@ifundefined{pgf@pattern@name@_u1hd2979x}{
\pgfdeclarepatternformonly[\mcThickness,\mcSize]{_u1hd2979x}
{\pgfqpoint{0pt}{0pt}}
{\pgfpoint{\mcSize+\mcThickness}{\mcSize+\mcThickness}}
{\pgfpoint{\mcSize}{\mcSize}}
{
\pgfsetcolor{\tikz@pattern@color}
\pgfsetlinewidth{\mcThickness}
\pgfpathmoveto{\pgfqpoint{0pt}{0pt}}
\pgfpathlineto{\pgfpoint{\mcSize+\mcThickness}{\mcSize+\mcThickness}}
\pgfusepath{stroke}
}}
\makeatother
\tikzset{every picture/.style={line width=0.75pt}} 

\begin{tikzpicture}[x=0.75pt,y=0.75pt,yscale=-1.2,xscale=1.2]

\draw  [pattern=_u1hd2979x,pattern size=1.5pt,pattern thickness=0.75pt,pattern radius=0pt, pattern color={rgb, 255:red, 193; green, 193; blue, 193}] (300.43,130.34) .. controls (300.43,91.55) and (331.7,60.11) .. (370.29,60.11) .. controls (408.87,60.11) and (440.14,91.55) .. (440.14,130.34) .. controls (440.14,169.13) and (408.87,200.57) .. (370.29,200.57) .. controls (331.7,200.57) and (300.43,169.13) .. (300.43,130.34) -- cycle ;
\draw  [draw opacity=0][fill={rgb, 255:red, 255; green, 255; blue, 255 }  ,fill opacity=1 ] (306.3,130.44) .. controls (306.3,94.89) and (334.84,66.08) .. (370.04,66.08) .. controls (405.25,66.08) and (433.79,94.89) .. (433.79,130.44) .. controls (433.79,165.98) and (405.25,194.8) .. (370.04,194.8) .. controls (334.84,194.8) and (306.3,165.98) .. (306.3,130.44) -- cycle ;
\draw    (303.14,97.77) .. controls (307.75,88.71) and (314.24,78.9) .. (326.12,70.72) ;
\draw [shift={(328.46,69.18)}, rotate = 147.7] [fill={rgb, 255:red, 0; green, 0; blue, 0 }  ][line width=0.08]  [draw opacity=0] (5.36,-2.57) -- (0,0) -- (5.36,2.57) -- (3.56,0) -- cycle    ;
\draw  [draw opacity=0][fill={rgb, 255:red, 255; green, 255; blue, 255 }  ,fill opacity=1 ] (301.16,136.36) .. controls (301,134.39) and (300.91,132.4) .. (300.91,130.39) .. controls (300.91,128.35) and (301,126.33) .. (301.17,124.34) -- (306.67,124.82) .. controls (306.51,126.65) and (306.43,128.51) .. (306.43,130.39) .. controls (306.43,132.24) and (306.51,134.07) .. (306.66,135.89) -- cycle ;
\draw  [draw opacity=0][fill={rgb, 255:red, 255; green, 255; blue, 255 }  ,fill opacity=1 ] (376.47,199.86) .. controls (374.36,200.05) and (372.21,200.15) .. (370.04,200.15) .. controls (368.11,200.15) and (366.19,200.07) .. (364.3,199.92) -- (364.75,194.45) .. controls (366.49,194.59) and (368.26,194.66) .. (370.04,194.66) .. controls (372.04,194.66) and (374.02,194.57) .. (375.97,194.4) -- cycle ;
\draw  [draw opacity=0][fill={rgb, 255:red, 255; green, 255; blue, 255 }  ,fill opacity=1 ] (439.39,136.4) .. controls (439.56,134.4) and (439.65,132.38) .. (439.65,130.34) .. controls (439.65,128.33) and (439.56,126.34) .. (439.4,124.37) -- (433.13,124.9) .. controls (433.28,126.69) and (433.36,128.51) .. (433.36,130.34) .. controls (433.36,132.2) and (433.28,134.04) .. (433.12,135.86) -- cycle ;
\draw  [draw opacity=0][fill={rgb, 255:red, 255; green, 255; blue, 255 }  ,fill opacity=1 ] (364.13,60.84) .. controls (366.26,60.65) and (368.42,60.55) .. (370.6,60.55) .. controls (372.52,60.55) and (374.43,60.63) .. (376.32,60.78) -- (375.86,66.37) .. controls (374.13,66.23) and (372.37,66.16) .. (370.6,66.16) .. controls (368.59,66.16) and (366.61,66.25) .. (364.65,66.42) -- cycle ;
\draw  [draw opacity=0] (306.44,124.75) .. controls (309.1,93.69) and (333.75,68.94) .. (364.68,66.25) -- (370.29,130.34) -- cycle ; \draw   (306.44,124.75) .. controls (309.1,93.69) and (333.75,68.94) .. (364.68,66.25) ;  
\draw  [draw opacity=0] (364.68,194.43) .. controls (333.79,191.73) and (309.18,166.98) .. (306.53,135.92) -- (370.29,130.34) -- cycle ; \draw   (364.68,194.43) .. controls (333.79,191.73) and (309.18,166.98) .. (306.53,135.92) ;  
\draw  [draw opacity=0] (434.04,135.92) .. controls (431.39,166.98) and (406.78,191.73) .. (375.89,194.43) -- (370.29,130.34) -- cycle ; \draw   (434.04,135.92) .. controls (431.39,166.98) and (406.78,191.73) .. (375.89,194.43) ;  
\draw  [draw opacity=0] (375.89,66.25) .. controls (406.78,68.94) and (431.39,93.7) .. (434.04,124.76) -- (370.29,130.34) -- cycle ; \draw   (375.89,66.25) .. controls (406.78,68.94) and (431.39,93.7) .. (434.04,124.76) ;  
\draw [color={rgb, 255:red, 0; green, 0; blue, 255 }  ,draw opacity=1 ]   (306.45,124.72) .. controls (327.98,124.71) and (364.46,87.35) .. (364.71,66.24) ;
\draw [color={rgb, 255:red, 0; green, 0; blue, 255 }  ,draw opacity=1 ]   (375.87,66.28) .. controls (376.15,87.39) and (413.66,124.74) .. (434.04,124.76) ;
\draw [color={rgb, 255:red, 0; green, 0; blue, 255 }  ,draw opacity=1 ]   (375.86,194.43) .. controls (375.8,173.85) and (413.48,136.33) .. (434.04,135.95) ;
\draw [color={rgb, 255:red, 0; green, 0; blue, 255 }  ,draw opacity=1 ]   (364.71,194.43) .. controls (364.24,173.85) and (327.45,135.97) .. (306.53,135.95) ;

\draw (297.84,71.59) node [anchor=north west][inner sep=0.75pt]  [xscale=1,yscale=1]  {$4\beta $};
\draw (325,164) node [anchor=north west][inner sep=0.75pt]  [xscale=1,yscale=1]  {$x_{1}$};
\draw (306,181) node [anchor=north west][inner sep=0.75pt]  [xscale=1,yscale=1]  {$x_{2}$};

\end{tikzpicture}
	\end{center}
	\caption{A representation of the Euclidean path integral that evaluates $Z_4$. The $x_2$ oscillator lives on the outer Euclidean time circle which has length $4\beta$, while $x_1$ lives on the inner arcs each with length $\beta$. The blue lines represent that for each arc the endpoints are identified. The shading pattern between the inner arcs and the outer circle represents the fact that the two oscillators are interacting.}
	\label{repfigsph}
\end{figure}
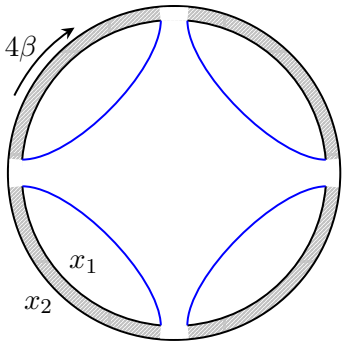

Consider a system of two coupled oscillators described by the following Hamiltonian
\begin{equation} \label{hamiltonian}
H=\frac{p_1^2+p_2^2}{2 m}+\frac{1}{2} m \alpha_1^2 x_1^2+\frac{1}{2} m \alpha_2^2 x_2^2+\frac{1}{2} k\left(x_1-x_2\right)^2
\end{equation}
As we mentioned in the introduction, the entanglement entropy of this system in the ground state is well known and straightforward to calculate directly. However, for our purposes, we would like to set up the calculation in terms of something that is analogous to a path integral over some replica manifold which calculates the $n$-th Rényi entropy for us, analytically continue in $n$, and then take the $n \rightarrow 1$ limit to get the von Neumann entropy. How would such a path integral look like?  
It is actually easier to visualize this in the case where the two oscillators are at some finite temperature $1/\beta$. If we are interested in ground state results, we can always take the limit $\beta \rightarrow \infty$. 
Since the entire system is described by a thermal state with temperature $1/\beta$, the (unnormalized) density matrix is simple, given by 
\begin{equation}
\rho={\mathrm e}^{-\beta H}
\end{equation}
We can then obtain a reduced density matrix $\rho_{x_2}$ by performing a partial trace over the $x_1$ oscillator
\begin{equation}
\rho_{x_2} \equiv \Tr_{x_1}\{ \rho\}=\int_{-\infty}^{\infty}dx_1\bra{x_1}\rho\ket{x_1}
\end{equation}
 The $n$-th Rényi entropy of this reduced state can then be written as 
\begin{align}\label{Rényi1}
S_n=\frac{1}{1-n} \Big(\ln Z_n -n\ln Z_1\Big),
\end{align}
with
\begin{align}
Z_n \equiv \Tr{\rho_{x_2}^n}=\int_{-\infty}^{\infty}dx_2\bra{x_2}\rho_{x_2}^n\ket{x_2}.
\end{align}
$Z_1$ here is nothing but the partition function of the entire system at the given temperature and is simply given by a path integral of the entire system over a closed euclidean time cycle of size $\beta$. The key quantity we need to represent as a path integral is $Z_n$. This seems like a rather strange object to consider as a path integral over some manifold. We need to somehow find a way to evaluate the path integral with the $x_1$ oscillator on closed Euclidean time cycles of size $\beta$ while the $x_2$ oscillator lives on a Euclidean time cycle of size $n \beta$. A schematic representation of such a path integral for $Z_4$ is shown in fig. \ref{repfigsph}.

As we argued earlier, a path integral representation of this object cannot be obtained by simply manipulating the manifold, as changes to the manifold do not differentiate between different ``internal'' degrees of freedom. A path integral representation of $Z_n$ should treat the two oscillators in different ways. 

One way to proceed would be to try to enforce the necessary boundary conditions on the oscillators by hand in a similar way to how one would deal with enforcing constraints. This is essentially the approach discussed in \cite{Mollabashi:2014qfa}. We will take a different approach using path integral insertions. 

\subsection{Cutting and gluing the partition function}
To illustrate the main idea, we will turn to an even simpler system as a first step: a single harmonic oscillator.  For a single harmonic oscillator at a temperature $1/\beta$, the partition function is given by the path integral\footnote{The notation $\int_{\beta}$ here is a shorthand for the integral being defined over a closed Euclidean time circle of size $\beta$.}
\ben
Z(\beta)=\int\mathcal{D} x ~e^{-\int_{\beta} L_E ~d\tau },
\een
where $L_E$ denotes the Euclidean Lagrangian of the system. The Rényi entropy here is given by
\begin{equation}\label{thermalRényi}
S_{n,\beta}=\frac{1}{1-n} \Big(\ln Z(n\beta) -n\ln Z(\beta)\Big).
\end{equation}
For such a simple system, we can immediately write down the path integral representing $Z(n\beta)$ for any $n$ by simply manipulating the length of the Euclidean time circle. Our goal, however, is to figure out how we can go from $Z(n\beta)$ to $Z(\beta)^n$ (or vice versa) without explicitly changing the manifold the path integral is being evaluated on. In this section, we will mostly focus on showing how to go from $Z(n\beta)$ to $Z(\beta)^n$.
\subsubsection{$Z(n\beta)$ to $Z(\beta)^n$}

\begin{figure}[t!]
\centering
\tikzset{every picture/.style={line width=0.75pt}} 
\begin{tikzpicture}[x=0.75pt,y=0.75pt,yscale=-0.85,xscale=0.85]

\draw  [draw opacity=0] (291.12,144.34) .. controls (293.78,113.28) and (318.42,88.52) .. (349.35,85.83) -- (354.96,149.92) -- cycle ; \draw   (291.12,144.34) .. controls (293.78,113.28) and (318.42,88.52) .. (349.35,85.83) ;  
\draw  [draw opacity=0] (418.71,155.5) .. controls (416.06,186.56) and (391.46,211.32) .. (360.57,214.01) -- (354.96,149.92) -- cycle ; \draw   (418.71,155.5) .. controls (416.06,186.56) and (391.46,211.32) .. (360.57,214.01) ;  
\draw  [draw opacity=0] (349.35,214.01) .. controls (318.47,211.32) and (293.86,186.56) .. (291.21,155.5) -- (354.96,149.92) -- cycle ; \draw   (349.35,214.01) .. controls (318.47,211.32) and (293.86,186.56) .. (291.21,155.5) ;  
\draw  [draw opacity=0] (360.57,85.83) .. controls (391.46,88.53) and (416.06,113.28) .. (418.71,144.34) -- (354.96,149.92) -- cycle ; \draw   (360.57,85.83) .. controls (391.46,88.53) and (416.06,113.28) .. (418.71,144.34) ;  

\draw  [draw opacity=0] (521.4,144.34) .. controls (524.07,113.28) and (548.71,88.52) .. (579.64,85.83) -- (585.25,149.92) -- cycle ; \draw   (521.4,144.34) .. controls (524.07,113.28) and (548.71,88.52) .. (579.64,85.83) ;  
\draw  [draw opacity=0] (649,155.5) .. controls (646.35,186.56) and (621.74,211.32) .. (590.85,214.01) -- (585.25,149.92) -- cycle ; \draw   (649,155.5) .. controls (646.35,186.56) and (621.74,211.32) .. (590.85,214.01) ;  
\draw  [draw opacity=0] (579.64,214.01) .. controls (548.75,211.32) and (524.15,186.56) .. (521.49,155.5) -- (585.25,149.92) -- cycle ; \draw   (579.64,214.01) .. controls (548.75,211.32) and (524.15,186.56) .. (521.49,155.5) ;  
\draw  [draw opacity=0] (590.85,85.83) .. controls (621.74,88.53) and (646.35,113.28) .. (649,144.34) -- (585.25,149.92) -- cycle ; \draw   (590.85,85.83) .. controls (621.74,88.53) and (646.35,113.28) .. (649,144.34) ;  
\draw [color={rgb, 255:red, 0; green, 0; blue, 255 }  ,draw opacity=1 ]   (521.41,144.31) .. controls (542.94,144.29) and (579.42,106.94) .. (579.67,85.83) ;
\draw [color={rgb, 255:red, 0; green, 0; blue, 255 }  ,draw opacity=1 ]   (579.67,214.02) .. controls (579.2,193.44) and (542.41,155.55) .. (521.5,155.53) ;
\draw [color={rgb, 255:red, 0; green, 0; blue, 255 }  ,draw opacity=1 ]   (590.83,85.86) .. controls (591.12,106.97) and (628.62,144.32) .. (649,144.34) ;
\draw [color={rgb, 255:red, 0; green, 0; blue, 255 }  ,draw opacity=1 ]   (590.82,214.02) .. controls (590.76,193.44) and (628.44,155.91) .. (649,155.53) ;

\draw   (60.16,149.92) .. controls (60.16,114.41) and (88.94,85.63) .. (124.45,85.63) .. controls (159.96,85.63) and (188.75,114.41) .. (188.75,149.92) .. controls (188.75,185.43) and (159.96,214.22) .. (124.45,214.22) .. controls (88.94,214.22) and (60.16,185.43) .. (60.16,149.92) -- cycle ;

\draw [line width=1.5]    (210.09,149.92) -- (265.54,149.92) ;
\draw [shift={(269.54,149.92)}, rotate = 180] [fill={rgb, 255:red, 0; green, 0; blue, 0 }  ][line width=0.08]  [draw opacity=0] (8.75,-4.2) -- (0,0) -- (8.75,4.2) -- (5.81,0) -- cycle    ;
\draw [line width=1.5]    (440.38,149.92) -- (495.84,149.92) ;
\draw [shift={(499.84,149.92)}, rotate = 180] [fill={rgb, 255:red, 0; green, 0; blue, 0 }  ][line width=0.08]  [draw opacity=0] (8.75,-4.2) -- (0,0) -- (8.75,4.2) -- (5.81,0) -- cycle    ;
\draw    (63.31,119) .. controls (72.48,102.45) and (77.53,96.91) .. (95.36,87.67) ;
\draw [shift={(97.97,86.33)}, rotate = 153.21] [fill={rgb, 255:red, 0; green, 0; blue, 0 }  ][line width=0.08]  [draw opacity=0] (5.36,-2.57) -- (0,0) -- (5.36,2.57) -- (3.56,0) -- cycle    ;

\draw (100,142.82) node [anchor=north west][inner sep=0.75pt]  [xscale=1,yscale=1]  {$Z( 4\beta )$};
\draw (564,141.82) node [anchor=north west][inner sep=0.75pt]  [xscale=1,yscale=1]  {$Z( \beta )^{4}$};
\draw (60,84) node [anchor=north west][inner sep=0.75pt]  [font=\footnotesize,xscale=1,yscale=1]  {$4\beta $};

\end{tikzpicture}
\caption{Schematic representation of the path integral surgery we wish to perform. By cutting the Euclidean time circle into four pieces and then gluing each piece to itself we go from a path integral evaluating $Z(4\beta)$ to one evaluating $Z(\beta)^4$. }
\label{surgery2}
\end{figure}
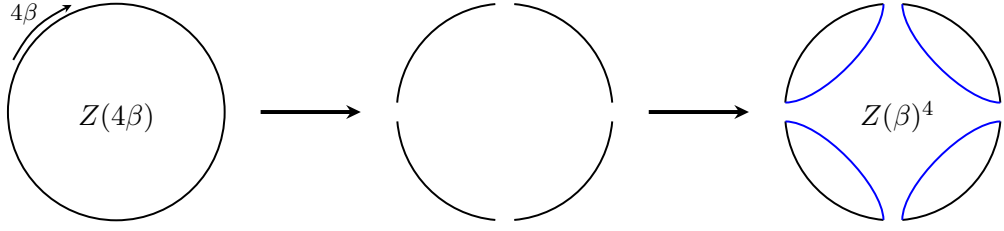

 Our strategy will be very simple: starting from a path integral representing $Z(n\beta)$, we will try to find a way to ``cut'' the path integral at a set of Euclidean time points given by $\tau_j=j \beta$ where $j$ is an integer that takes values between $0$ and $n-1$. The next step is to ``glue'' the cut pieces by identifying the points $\tau_j+\epsilon$ and $\tau_{j+1}-\epsilon$ with limit $\epsilon \rightarrow 0$ being taken at the end of the calculation.\footnote{For a different perspective on how to approach quantum mechanical implementations of cutting, gluing and topology change, see \cite{Shapere:2012wn}.} A schematic representation of this procedure is shown in fig. \ref{surgery2}. Our goal now is to understand how we can implement these operations using insertions in the path integral, without manipulating the manifold directly, in the hope that this will eventually allow us to perform something of a “selective surgery” that can pick out specific internal degrees of freedom.

Let us first address how the gluing procedure can be implemented as it is the more straightforward of the two. Gluing two point on the Euclidean time circle can be implemented in a pretty straightforward way by simply introducing Dirac delta function insertions into the path integral as follows\footnote{The notation on the left-hand side here means that this is the $Z(n\beta)$ partition function with every Euclidean time point $\tau_j+\epsilon$ identified with $\tau_{j+1}-\epsilon$. On the right-hand side, inside the Dirac delta functions, the subscript on $x$ means this is $x(\tau)$ evaluated at this Euclidean time so that something like $x_a$ just means $x$ at Euclidean time $\tau=a$.}
\ben
\label{constrain1}
Z\left(n\beta\right)|_{\tau_j+\epsilon \sim \tau_{j+1}-\epsilon} = \int \mathcal{D} x ~e^{-\int_0^{n\beta} L_E d\tau }\prod_{j=0}^{n-1} \delta\left(x_{\tau_j+\epsilon}-x_{\tau_{j+1}-\epsilon }\right).
\een
At this point, this is just some constrained path integral, but it is not quite $Z(\beta)^n$. We will address that eventually, but let us first massage this expression into something a bit more useful. We can represent the Dirac delta functions in the following way
\begin{align*}
\prod_{j=0}^{n-1} \delta(x_{\tau_j+\epsilon}-x_{\tau_{j+1}-\epsilon})&=\left( \prod_{j=0}^{n-1} \int_{-\infty}^{\infty} \frac{d J_j}{2\pi}~ \right)  e^{i \sum_{j=0}^{n-1} J_j\big[x_{\tau_j+\epsilon}-x_{\tau_{j+1}-\epsilon}\big]}  \\
&=\left( \prod_{j=0}^{n-1} \int_{-\infty}^{\infty} \frac{d J_j}{2\pi}~ \right) ~e^{i \int_{n\beta} d\tau\sum_{j=0}^{n-1} J_j\Big[\delta\big(\tau-(\tau_j+\epsilon)\big)-\delta\big(\tau-(\tau_{j+1}-\epsilon)\big)\Big]x}. 
\end{align*}
The constrained path integral partition function can then be written as 
\begin{equation*}
 \left( \prod_{j=0}^{n-1} \int_{-\infty}^{\infty} \frac{d J_j}{2\pi}~ \right) \int \mathcal{D} x ~e^{-\int_{n\beta} d\tau \left\{L_E-i  \sum_{j=0}^{n-1}J_j\Big[\delta\big(\tau-(\tau_j+\epsilon)\big)-\delta\big(\tau-(\tau_{j+1}-\epsilon)\big)\Big]x\right\}}. 
\end{equation*}
This can be then interpreted as just the original path integral with some Euclidean time ``kicks'' sources that we should integrate over.\footnote{The idea of integrating over Euclidean time ``kick'' sources to calculate Rényi entropy has been used extensively in \cite{Chakraborty:2018wkx,Moitra:2020cty,Chakraborty:2020gjs,Moitra:2023gjm,Sarkar:2025uoe, Haldar:2020ymg}. In these works, this was derived using Wigner function representations of the Rényi entropy.} Provided that we can perform the path integral first, we can just calculate the partition function in the presence of these external sources and integrate over them at the end of the calculation.  

Let us now address how to ``cut'' the path integral. The issue here is that the action contains derivative terms that couple the points at $\tau_j+\epsilon$ to $\tau_j-\epsilon$. But in a calculation of $Z(\beta)^n$, these two points should exist in two different copies of the manifold in the path integral. Another way to see this is to point out that, for example, $Z(2\beta)$ can be written as
\begin{equation*}
Z(2\beta)=\Tr{\rho^2}=\int dx_1 dx_2 ~\rho_{x_1x_2}~\rho_{x_2x_1},
\end{equation*}
where again $\rho$ is the (unnormalized) density matrix obtained by a Euclidean time evolution from $0$ to $\beta$ and $\rho_{x_1x_2}$ denotes its position-space matrix elements. Introducing the Dirac delta function constraints of \eqref{constrain1} into this object will not produce $Z(\beta)^2$. Instead, the object we need is a ``cut'' path integral that should look like
\begin{equation*}
\int dx_1 dx_2 dx_3 dx_4 ~\rho_{x_1x_2}~\rho_{x_3x_4}.
\end{equation*}
As it stands, this object, which we need as something of an intermediate step, does not look like the trace of a string of matrices which stops us from interpreting it as some modified path integral on a manifold with Euclidean time $2 \beta$. However, we can easily rewrite it as 
\begin{equation*}
\int dx_1 dx_2 dx_3 dx_4 ~\rho_{x_1x_2}\Omega_{x_2x_3} \rho_{x_3x_4}\Omega_{x_4x_1}, 
\end{equation*}
with 
\begin{equation*}
\Omega_{xx\prime} =1.
\end{equation*}
The $1$ in this equation is not to be confused with the identity operator. This is rather a ``matrix of ones'', where all entries written in the position basis are equal to one.  It appears that our cutting procedure would amount to inserting such matrices at all points with $\tau_j=j\beta$. How can this be achieved? Well, this matrix is nothing but a projection (up to a constant) onto the zero momentum state of the harmonic oscillator. Indeed, in the momentum basis we have 
\ben
\Omega_{pp\prime} =2 \pi \delta(p) \delta(p-p\prime).
\een
Now we just need to find a way to insert operators of this form at every point where we want to disconnect the path integral. This is best implemented within the phase space path integral 
\begin{equation}
Z(n\beta)=\int \mathcal{D} x \mathcal{D} p ~e^{ \int_{n\beta} d \tau \big\{ip \dot{x}-H\big\}}.
\end{equation}
Similarly to how we implemented the position constraints, we can implement a zero momentum constraint at the points $\tau_j=j\beta$ 
\begin{align*}
 \int \mathcal{D}x\mathcal{D} p~ e^{\int_{n\beta} d \tau\big\{ ip \dot{x}-H\big\}}\left(\prod_{j=0}^{n-1} 2\pi \delta\left(p_{\tau_j}\right)\right)=& \left( \prod_{j=0}^{n-1} \int_{-\infty}^{\infty} {d K_j} \right)    \\
 &\times \int \mathcal{D} x \mathcal{D} p~  e^{\int_{n\beta} d \tau \left\{ ip \left[\dot{x}+\left(\sum_{j=0}^{n-1}K_j\delta(\tau-\tau_j)\right)\right]-H\right\}}.
\end{align*}
Now putting everything together we get 
\begin{align}\label{result1}
    Z(\beta)^n=\lim_{\epsilon \rightarrow 0}\left( \prod_{j=0}^{n-1} \int_{-\infty}^{\infty} \frac{d K_j dJ_j}{2\pi}\right)\int &\mathcal{D} x \mathcal{D} p ~\exp \Bigg\{ \int_{n\beta} d \tau \, \textstyle ip \Big[\dot{x}+\left(\sum_{j=0}^{n-1}K_j\delta(\tau-\tau_j)\right)\Big] \! -\!H\nonumber\\&~\,~~~~~~~~\,+\textstyle i\!\sum_{j=0}^{n-1} \! J_j\Big[\delta\big(\tau\!-\!(\tau_j \!+ \! \epsilon)\!\big)-\delta\big(\tau \! - \!(\tau_{j+1} \!- \! \epsilon)\!\big)\Big]x \Bigg\}.
\end{align}

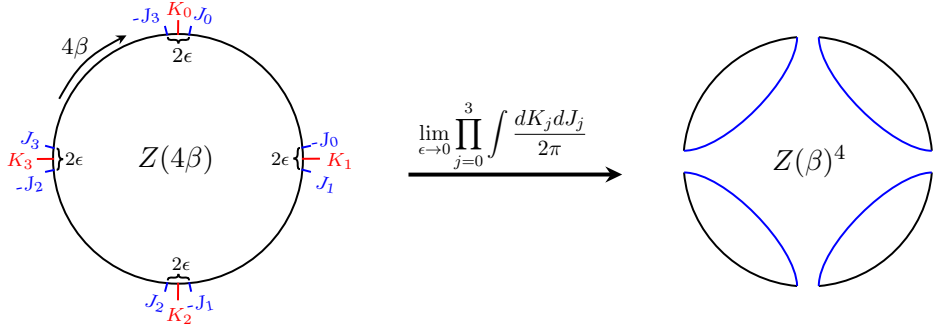
\begin{figure}[!t]
\centering

\tikzset{every picture/.style={line width=0.75pt}} 

\begin{tikzpicture}[x=0.75pt,y=0.75pt,yscale=-0.98,xscale=0.98]

\draw  [draw opacity=0] (433.8,158.17) .. controls (436.47,127.11) and (461.11,102.35) .. (492.04,99.66) -- (497.65,163.75) -- cycle ; \draw   (433.8,158.17) .. controls (436.47,127.11) and (461.11,102.35) .. (492.04,99.66) ;  
\draw  [draw opacity=0] (561.4,169.33) .. controls (558.75,200.39) and (534.14,225.15) .. (503.25,227.84) -- (497.65,163.75) -- cycle ; \draw   (561.4,169.33) .. controls (558.75,200.39) and (534.14,225.15) .. (503.25,227.84) ;  
\draw  [draw opacity=0] (492.04,227.84) .. controls (461.15,225.15) and (436.55,200.39) .. (433.89,169.33) -- (497.65,163.75) -- cycle ; \draw   (492.04,227.84) .. controls (461.15,225.15) and (436.55,200.39) .. (433.89,169.33) ;  
\draw  [draw opacity=0] (503.25,99.66) .. controls (534.14,102.36) and (558.75,127.11) .. (561.4,158.17) -- (497.65,163.75) -- cycle ; \draw   (503.25,99.66) .. controls (534.14,102.36) and (558.75,127.11) .. (561.4,158.17) ;  
\draw [color={rgb, 255:red, 0; green, 0; blue, 255 }  ,draw opacity=1 ]   (433.81,158.14) .. controls (455.34,158.12) and (491.82,120.76) .. (492.07,99.66) ;
\draw [color={rgb, 255:red, 0; green, 0; blue, 255 }  ,draw opacity=1 ]   (492.07,227.85) .. controls (491.6,207.27) and (454.81,169.38) .. (433.9,169.36) ;
\draw [color={rgb, 255:red, 0; green, 0; blue, 255 }  ,draw opacity=1 ]   (503.23,99.69) .. controls (503.52,120.8) and (541.02,158.15) .. (561.4,158.17) ;
\draw [color={rgb, 255:red, 0; green, 0; blue, 255 }  ,draw opacity=1 ]   (503.22,227.85) .. controls (503.16,207.27) and (540.84,169.74) .. (561.4,169.36) ;

\draw   (109.16,162.32) .. controls (109.16,126.81) and (137.94,98.03) .. (173.45,98.03) .. controls (208.96,98.03) and (237.75,126.81) .. (237.75,162.32) .. controls (237.75,197.83) and (208.96,226.62) .. (173.45,226.62) .. controls (137.94,226.62) and (109.16,197.83) .. (109.16,162.32) -- cycle ;

\draw    (112.31,131.4) .. controls (121.48,114.85) and (126.53,109.31) .. (144.36,100.07) ;
\draw [shift={(146.97,98.73)}, rotate = 153.21] [fill={rgb, 255:red, 0; green, 0; blue, 0 }  ][line width=0.08]  [draw opacity=0] (5.36,-2.57) -- (0,0) -- (5.36,2.57) -- (3.56,0) -- cycle    ;
\draw [color={rgb, 255:red, 255; green, 0; blue, 0 }  ,draw opacity=1 ]   (173.45,235.19) -- (173.52,226.12) ;
\draw [color={rgb, 255:red, 0; green, 0; blue, 255 }  ,draw opacity=1 ]   (166.75,230.76) -- (168.04,225.9) ;
\draw [color={rgb, 255:red, 0; green, 0; blue, 255 }  ,draw opacity=1 ]   (180.25,230.76) -- (179.19,225.9) ;
\draw [color={rgb, 255:red, 255; green, 0; blue, 0 }  ,draw opacity=1 ]   (100.82,162.4) -- (109.69,162.34) ;
\draw [color={rgb, 255:red, 0; green, 0; blue, 255 }  ,draw opacity=1 ]   (105.05,155.57) -- (109.91,156.86) ;
\draw [color={rgb, 255:red, 0; green, 0; blue, 255 }  ,draw opacity=1 ]   (105.05,169.07) -- (109.91,168.01) ;
\draw [color={rgb, 255:red, 255; green, 0; blue, 0 }  ,draw opacity=1 ]   (173.56,90.87) -- (173.47,98.51) ;
\draw [color={rgb, 255:red, 0; green, 0; blue, 255 }  ,draw opacity=1 ]   (166.7,93.87) -- (167.99,98.73) ;
\draw [color={rgb, 255:red, 0; green, 0; blue, 255 }  ,draw opacity=1 ]   (180.2,93.87) -- (179.14,98.73) ;
\draw [color={rgb, 255:red, 255; green, 0; blue, 0 }  ,draw opacity=1 ]   (246.51,162.24) -- (237.22,162.3) ;
\draw [color={rgb, 255:red, 0; green, 0; blue, 255 }  ,draw opacity=1 ]   (241.86,169.07) -- (237,167.78) ;
\draw [color={rgb, 255:red, 0; green, 0; blue, 255 }  ,draw opacity=1 ]   (241.86,155.57) -- (237,156.63) ;

\draw [line width=1.5]    (292.07,170.44) -- (398.06,170.4) ;
\draw [shift={(402.06,170.39)}, rotate = 179.98] [fill={rgb, 255:red, 0; green, 0; blue, 0 }  ][line width=0.08]  [draw opacity=0] (8.75,-4.2) -- (0,0) -- (8.75,4.2) -- (5.81,0) -- cycle    ;
\draw    (168.26,99.55) .. controls (167.95,102.9) and (173.67,99.1) .. (173.67,102.9) ;
\draw    (179.07,99.55) .. controls (179.38,102.9) and (173.67,99.1) .. (173.67,102.9) ;

\draw    (236.22,156.75) .. controls (233.69,156.45) and (236.77,162.16) .. (232.88,162.16) ;
\draw    (236.22,167.56) .. controls (233.69,167.87) and (236.77,162.16) .. (232.88,162.16) ;

\draw    (110.99,167.9) .. controls (113.53,168.21) and (110.45,162.49) .. (114.34,162.49) ;
\draw    (110.99,157.09) .. controls (113.53,156.78) and (110.45,162.49) .. (114.34,162.49) ;

\draw    (168.26,224.93) .. controls (167.95,221.68) and (173.67,225.48) .. (173.67,221.58) ;
\draw    (179.07,224.93) .. controls (179.38,221.68) and (173.67,225.48) .. (173.67,221.58) ;

\draw (151,154) node [anchor=north west][inner sep=0.75pt]  [xscale=1,yscale=1]  {$Z( 4\beta )$};
\draw (477.65,154) node [anchor=north west][inner sep=0.75pt]  [xscale=1,yscale=1]  {$Z( \beta )^{4}$};
\draw (112.03,98.1) node [anchor=north west][inner sep=0.75pt]  [font=\footnotesize,xscale=1,yscale=1]  {$4\beta $};
\draw (248.56,158) node [anchor=north west][inner sep=0.75pt]  [font=\scriptsize,color={rgb, 255:red, 255; green, 0; blue, 0 }  ,opacity=1 ,xscale=0.9,yscale=0.9]  {$K_{1}$};
\draw (244,169) node [anchor=north west][inner sep=0.75pt]  [font=\scriptsize,color={rgb, 255:red, 0; green, 0; blue, 255 }  ,opacity=1 ,rotate=-10,xscale=0.9,yscale=0.9]  {$J_{1}$};
\draw (240,149) node [anchor=north west][inner sep=0.75pt]  [font=\scriptsize,color={rgb, 255:red, 0; green, 0; blue, 255 }  ,opacity=1 ,rotate=-350,xscale=0.9,yscale=0.9]  {$\text{-}J_{0}$};
\draw (166,238) node [anchor=north west][inner sep=0.75pt]  [font=\scriptsize,color={rgb, 255:red, 255; green, 0; blue, 0 }  ,opacity=1 ,xscale=0.9,yscale=0.9]  {$K_{2}$};
\draw (156,230) node [anchor=north west][inner sep=0.75pt]  [font=\scriptsize,color={rgb, 255:red, 0; green, 0; blue, 255 }  ,opacity=1 ,rotate=-10,xscale=0.9,yscale=0.9]  {$J_{2}$};
\draw (176,233) node [anchor=north west][inner sep=0.75pt]  [font=\scriptsize,color={rgb, 255:red, 0; green, 0; blue, 255 }  ,opacity=1 ,rotate=-350,xscale=0.9,yscale=0.9]  {$\text{-}J_{1}$};
\draw (84,158) node [anchor=north west][inner sep=0.75pt]  [font=\scriptsize,color={rgb, 255:red, 255; green, 0; blue, 0 }  ,opacity=1 ,xscale=0.9,yscale=0.9]  {$K_{3}$};
\draw (92,146) node [anchor=north west][inner sep=0.75pt]  [font=\scriptsize,color={rgb, 255:red, 0; green, 0; blue, 255 }  ,opacity=1 ,rotate=-10,xscale=0.9,yscale=0.9]  {$J_{3}$};
\draw (88,172) node [anchor=north west][inner sep=0.75pt]  [font=\scriptsize,color={rgb, 255:red, 0; green, 0; blue, 255 }  ,opacity=1 ,rotate=-350,xscale=0.9,yscale=0.9]  {$\text{-}J_{2}$};
\draw (166,80) node [anchor=north west][inner sep=0.75pt]  [font=\scriptsize,color={rgb, 255:red, 255; green, 0; blue, 0 }  ,opacity=1 ,xscale=0.9,yscale=0.9]  {$K_{0}$};
\draw (180.93,82) node [anchor=north west][inner sep=0.75pt]  [font=\scriptsize,color={rgb, 255:red, 0; green, 0; blue, 255 }  ,opacity=1 ,rotate=-10,xscale=0.9,yscale=0.9]  {$J_{0}$};
\draw (148,85) node [anchor=north west][inner sep=0.75pt]  [font=\scriptsize,color={rgb, 255:red, 0; green, 0; blue, 255 }  ,opacity=1 ,rotate=-350,xscale=0.9,yscale=0.9]  {$\text{-}J_{3}$};
\draw (296,131) node [anchor=north west][inner sep=0.75pt]  [xscale=0.75,yscale=0.75]  {$\displaystyle\lim _{\epsilon \rightarrow 0}\prod _{j=0}^{3}\int \frac{dK_{j} dJ_{j}}{2\pi }$};
\draw (168.96,104.03) node [anchor=north west][inner sep=0.75pt]  [font=\scriptsize,xscale=1,yscale=1]  {$2\epsilon $};
\draw (220,158) node [anchor=north west][inner sep=0.75pt]  [font=\scriptsize,xscale=0.9,yscale=0.9]  {$2\epsilon $};
\draw (168.96,212) node [anchor=north west][inner sep=0.75pt]  [font=\scriptsize,xscale=0.9,yscale=0.9]  {$2\epsilon $};
\draw (114,158) node [anchor=north west][inner sep=0.75pt]  [font=\scriptsize,xscale=0.9,yscale=0.9]  {$2\epsilon $};

\end{tikzpicture}
\caption{A source distribution along the Euclidean time circle that, after being integrated over, takes us from $Z(4\beta)$ to $Z(\beta)^4$. Sources labeled by $J_j$ are linearly coupled to the position operator while sources labeled by $K_j$ are linearly coupled to the momentum operator in the precise sense given in \eqref{result1}.}
\label{sourcedistribution}
\end{figure}

Let us recap what we have done so far. We found that in order to calculate $Z(\beta)^n$ from $Z(n\beta)$, we need to couple $Z(n\beta)$ to some external Euclidean time ``kick'' sources, some of which are coupled to position and others to momentum. These sources should then be integrated over in the precise way of \eqref{result1}.\footnote{A similar expression, derived in the context of operator formalism, appears in \cite{Shiba:2014uia}.} An illustration of this identity is shown in fig. \ref{sourcedistribution}.

We can make further progress by developing the configuration space path integral corresponding to \eqref{result1}. This can be done by the usual integration over momenta but it is easier to realize that the effect of the $K_j$ sources in the phase space path integral is merely to shift the dependence on the derivatives by $K_j\delta(\tau-j\beta)$. After all, the $\dot{x}$ term in the phase space path integral is itself an external source as far as the integrals over momenta are concerned. We can then write down the configuration space path integral version of \eqref{result1} right away
\begin{align}\label{result2}
    Z(\beta)^n=\lim_{\epsilon \rightarrow 0}\left( \prod_{j=0}^{n-1} \int_{-\infty}^{\infty} \frac{d K_j dJ_j}{2\pi}\right)\int &\mathcal{D} x  ~\exp \Bigg\{- \int_{n\beta} d \tau \textstyle L_E\Big(x, \dot{x}+\sum_{j=0}^{n-1}K_j\delta(\tau-\tau_j)\Big) \nonumber\\&~~~\textstyle -i\sum_{j=0}^{n-1}J_j\Big[\delta\big(\tau-(\tau_j+\epsilon)\big)-\delta\big(\tau-(\tau_{j+1}-\epsilon)\big)\Big]x  \Bigg\}.
\end{align}
Now that we have a nice expression, let us try to actually calculate it for the harmonic oscillator and see if we actually get $Z(\beta)^n$. We first point out that we have 
\begin{align*}
    \int_{n\beta} \left( \! \dot{x}+\sum_{j=0}^{n-1}K_j\delta(\tau-\tau_j) \! \right)^2  d \tau= \int_{n\beta}  \dot{x}^2-2x\sum_{j=0}^{n-1}K_j\delta'(\tau-\tau_j)+ \left(\sum_{j=0}^{n-1}K_j\delta(\tau-\tau_j) \! \right)^2  d \tau
\end{align*}
This allows us to just write \eqref{result2} for the harmonic oscillator simply as  
\begin{align}
        Z(\beta)^n=\lim_{\epsilon \rightarrow 0}\left( \prod_{j=0}^{n-1}  \int_{-\infty}^{\infty} \frac{d K_j dJ_j}{2\pi}\right)\int \mathcal{D} x  \,&\exp \Bigg\{ - \int_{n\beta} \textstyle  d \tau \,\frac{1}{2}m \dot{x}^2+\frac{1}{2}m \omega^2 x^2 \nonumber\\
        &~~~~~~~~~~~~~~~~~~~~+\textstyle \frac{1}{2}m \left(\sum_{j=0}^{n-1}K_j\delta(\tau-\tau_j)\right)^2-\mathcal{J}x \Bigg\},
\end{align}
with $\mathcal{J}$ being given by
\ben \label{finalsource}
\mathcal{J}=\sum_{j=0}^{n-1}\Bigg\{mK_j\delta'(\tau-\tau_j)+iJ_j\Big[\delta\big(\tau-(\tau_j+\epsilon)\big)-\delta\big(\tau-(\tau_{j+1}-\epsilon)\big)\Big]\Bigg\}.
\een
Ignoring the quadratic term in $K_j$ for a second, this just becomes a standard problem of an external source coupled to $x$. The standard functional analysis treatment of such a source then allows us to write 
\begin{align}\label{freeresult}
   \nonumber \frac{Z(\beta)^n}{Z(n\beta)}= \lim_{\epsilon \rightarrow 0}\left( \prod_{j=0}^{n-1} \int_{-\infty}^{\infty} \frac{d K_j dJ_j}{2\pi}\right)\exp \Bigg\{ \frac{1}{2} \int \textstyle d&\tau d\tau' \mathcal{J}(\tau) G_{n\beta}(\tau-\tau') \mathcal{J}(\tau') \\ 
    &\textstyle ~~~~~~~~~-m \left(\sum_{j=0}^{n-1}K_j\delta(\tau-\tau_j)\right)^2\Bigg\},
\end{align}
where $G_{n\beta}(\tau-\tau')$ is the Euclidean time thermal Green's function given by
\ben
G_{n\beta}(\tau-\tau')=\frac{\cosh \left(\omega \left(\frac{n  \beta}{2}-|\tau-\tau'|\right)\right)}{2  m \omega  \sinh \left(\frac{\omega  n  \beta}{2}\right)}.
\een
To avoid clutter in our notation, we will suppress the $\epsilon \rightarrow 0$ limit in the rest of this calculation, but we will keep it in mind whenever it may be relevant. 

We would like to verify that \eqref{freeresult} is indeed correct. The right hand side is just a bunch of Gaussian integrals similar to the ones that typically show up in entanglement entropy calculations of coupled oscillators. Let us see how we may evaluate them  efficiently.  First we consider what kind of terms we would get when the source in \eqref{freeresult} is expanded. We first have\footnote{Derivatives with a unprimed and primed $\tau$ subscript correspond to Green's function arguments with unprimed and primed index $j$ subscripts, respectively.}
\begin{equation}
 m^2K_jK_{j'} \partial_{\tau} \partial_{\tau'}G_{n\beta}(\tau_j-\tau_{j'})=K_jK_{j'}\left(-m^2\omega^2G_{n\beta}(\tau_j-\tau_{j'})+m\delta(\tau_j-\tau_{j'})\right).
\end{equation}
The delta function terms simply cancel the ones showing up in the exponent of \eqref{freeresult} and the only relevant contributions are the ones proportional to the Green's function. We also have the following terms
\begin{equation}
 imK_jJ_{j'} \partial_{\tau} G_{n\beta}(\tau_j-\tau_{j'})=-i\operatorname{sgn}(\tau_j-\tau_{j'})K_jJ_{j'}\frac{\sinh \left(\omega \left(\frac{n  \beta}{2}-|\tau_j-\tau_{j'}|\right)\right)}{2   \sinh \left(\frac{n \omega  \beta}{2}\right)},
\end{equation}
and finally we have terms of the form 
\begin{equation}
 -J_jJ_{j'} G_{n\beta}(\tau_j-\tau_{j'}).
\end{equation}
We can collect all of this into a $2n \times 2n$ matrix 
\begin{equation}
    \mathbf{\Omega} =
\begin{bmatrix}
~\mathbf{K} &~~ \mathbf{B}^\top \\
~\mathbf{B} & \mathbf{J}
\end{bmatrix},
\end{equation}
where the elements of each block are given by
\[
\mathbf{K_{jj'}} = m^2 \omega^2 G_{n \beta}\big((j-j')\beta\big),
\]

\[
\mathbf{J_{jj'}} =  2G_{n \beta}\big((j-j')\beta\big)-\left[G_{n \beta}\big((j-j'-1)\beta\big)+G_{n \beta}\big((j-j'+1)\beta\big)\right],
\]

\[
\mathbf{B_{jj'}} = im\left[ G'_{n \beta}\big((j-j')\beta\big)-G'_{n \beta}\big((j-j'+1)\beta\big)\right].
\]
This allows as to write \eqref{freeresult} as
\begin{equation}
    \frac{Z(\beta)^n}{Z(n\beta)}= \left( \prod_{j=0}^{n-1} \int_{-\infty}^{\infty} \frac{d K_j dJ_j}{2\pi}\right)\exp \Bigg\{- \frac{1}{2} \begin{bmatrix} K \\ J \end{bmatrix}^\top \begin{bmatrix}
~\mathbf{K} &~~ \mathbf{B}^\top \\
~\mathbf{B} & \mathbf{J}
\end{bmatrix} \begin{bmatrix}K \\ J \end{bmatrix}\Bigg\},
\end{equation}
with
\[
K^\top \equiv [K_1, K_2, \dots, K_n],~~~~~~~~~~~~~~~~~~~~J^\top \equiv [J_1, J_2, \dots, J_n].
\]
The good news here is that all four blocks of $\mathbf{\Omega}$ are circulant matrices. This means that they all commute and are all diagonalized by the same discrete Fourier transform basis. In other words, diagonalizing $\mathbf{K}$ and $\mathbf{J}$ using the discrete Fourier transform basis will automatically diagonalize $\mathbf{B}$ and $\mathbf{B}^\top$.
The diagonal elements of these matrices are given by 
\begin{align*}
    \mathbf{K}_k &= \frac{m \omega}{2} \cdot \frac{\sinh(\omega \beta)}{\cosh(\omega \beta) - \cos\left(\frac{2 \pi k}{n}\right)} , \\[1em]
    \mathbf{J}_k &= \frac{2}{m \omega} \cdot \frac{\sinh(\omega \beta) \sin^2\left(\frac{\pi k}{n}\right)}{1 + \cosh(\omega \beta) - 2 \cos^2\left(\frac{\pi k}{n}\right)} , \\[1em]
    \mathbf{B}_k &= \frac{i}{2} \cdot \frac{(\cosh(\omega \beta) - 1)\left(1 + e^{-i \frac{2 \pi k}{n}}\right)}{\cosh(\omega \beta) - \cos\left(\frac{2 \pi k}{n}\right)} , \\[1em]
    \mathbf{B}^\top_k &= \frac{i}{2} \cdot \frac{(\cosh(\omega \beta) - 1)\left(1 + e^{i \frac{2 \pi k}{n}}\right)}{\cosh(\omega \beta) - \cos\left(\frac{2 \pi k}{n}\right)},
\end{align*}
where the subscript $k$ denotes the $k$-th diagonal element and goes from $0$ to $n-1$.
Since all of the block entries commute, we have 
\begin{align}
\det\left(\mathbf{\Omega}\right)&=\det\left(\mathbf{K}\mathbf{J}-\mathbf{B}\mathbf{B}^\top\right) \nonumber\\
    &=\prod_{k=0}^{n-1}\frac{\cosh (\omega \beta)-1}{\cosh (\omega \beta)-\cos \left(\frac{2  \pi k}{n}\right)}\nonumber\\
    &=2^{n-1}\frac{\left(\cosh (\omega \beta)-1\right)^n}{\cosh (\omega n \beta)-1}.
\end{align}
The Gaussian integrals then give
\begin{equation}
    \sqrt{\frac{\left(2 \pi \right)^{2n}}{\det\left(\mathbf{\Omega}\right)}}=\left(2 \pi \right)^{n} \sqrt{\frac{2\left(\cosh (\omega n \beta)-1\right)}{\left( 2\left(\cosh (\omega \beta)-1\right) \right)^n}}.
\end{equation}
The $\left(2 \pi \right)^n$ factor exactly cancels with the one in the integration measure of \eqref{freeresult} and we get
\begin{equation} \label{deter}
    \frac{Z(\beta)^n}{Z(n\beta)}=  \sqrt{\frac{2\left(\cosh (\omega n \beta)-1\right)}{\left( 2\left(\cosh (\omega \beta)-1\right) \right)^n}}.
\end{equation}
This is the exact ratio we would get if we replaced the partition functions on the left hand side with their well known elementary expressions.

This seems like a lot of work to derive something elementary that we knew from the start. But the lesson here is that now we are starting to have a picture of how we can enforce boundary conditions on select degrees of freedom and design a path integral like the one in fig. \ref{repfigsph}. By thinking carefully about what it means for one of the oscillators to be living on a bunch of smaller Euclidean time cycles of length $\beta$ and how the path integral registers the difference between this case where the oscillator lives on the bigger $n\beta$ Euclidean time cycle, we can carefully design insertions to plug into the path integral to effectively go from $Z(n \beta)$ to $Z(\beta)^n$.

\subsubsection{$Z(\beta)^n$ to $Z(n\beta)$}

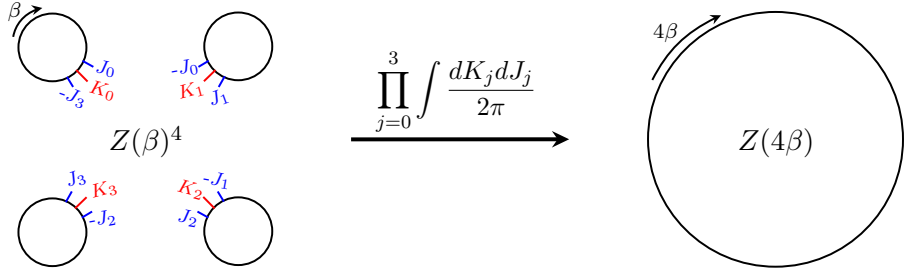
\begin{figure}
    \centering

\tikzset{every picture/.style={line width=0.75pt}} 

\begin{tikzpicture}[x=0.75pt,y=0.75pt,yscale=-1,xscale=1]

\draw   (123,138.33) .. controls (123,128.9) and (130.67,121.24) .. (140.14,121.24) .. controls (149.6,121.24) and (157.28,128.9) .. (157.28,138.33) .. controls (157.28,147.77) and (149.6,155.42) .. (140.14,155.42) .. controls (130.67,155.42) and (123,147.77) .. (123,138.33) -- cycle ;
\draw [color={rgb, 255:red, 250; green, 4; blue, 4 }  ,draw opacity=1 ]   (157.86,155.54) -- (152.14,149.73) ;
\draw [color={rgb, 255:red, 0; green, 0; blue, 255 }  ,draw opacity=1 ]   (160.78,148.14) -- (155.31,145.1) ;
\draw [color={rgb, 255:red, 0; green, 0; blue, 255 }  ,draw opacity=1 ]   (150.56,158.28) -- (147.46,153.16) ;

\draw   (123,231.62) .. controls (123,222.18) and (130.67,214.53) .. (140.14,214.53) .. controls (149.6,214.53) and (157.28,222.18) .. (157.28,231.62) .. controls (157.28,241.06) and (149.6,248.71) .. (140.14,248.71) .. controls (130.67,248.71) and (123,241.06) .. (123,231.62) -- cycle ;
\draw   (216.72,137.8) .. controls (216.72,128.36) and (224.4,120.71) .. (233.86,120.71) .. controls (243.33,120.71) and (251,128.36) .. (251,137.8) .. controls (251,147.24) and (243.33,154.89) .. (233.86,154.89) .. controls (224.4,154.89) and (216.72,147.24) .. (216.72,137.8) -- cycle ;
\draw   (216.72,231.09) .. controls (216.72,221.65) and (224.4,214) .. (233.86,214) .. controls (243.33,214) and (251,221.65) .. (251,231.09) .. controls (251,240.53) and (243.33,248.18) .. (233.86,248.18) .. controls (224.4,248.18) and (216.72,240.53) .. (216.72,231.09) -- cycle ;
\draw [color={rgb, 255:red, 250; green, 4; blue, 4 }  ,draw opacity=1 ]   (216.54,155.4) -- (222.39,149.73) ;
\draw [color={rgb, 255:red, 0; green, 0; blue, 255 }  ,draw opacity=1 ]   (224,158.3) -- (227.05,152.88) ;
\draw [color={rgb, 255:red, 0; green, 0; blue, 255 }  ,draw opacity=1 ]   (213.78,148.16) -- (218.93,145.08) ;

\draw [color={rgb, 255:red, 250; green, 4; blue, 4 }  ,draw opacity=1 ]   (216.2,213.79) -- (221.92,219.59) ;
\draw [color={rgb, 255:red, 0; green, 0; blue, 255 }  ,draw opacity=1 ]   (213.28,221.19) -- (218.75,224.22) ;
\draw [color={rgb, 255:red, 0; green, 0; blue, 255 }  ,draw opacity=1 ]   (223.5,211.04) -- (226.6,216.16) ;

\draw [color={rgb, 255:red, 250; green, 4; blue, 4 }  ,draw opacity=1 ]   (157.37,213.93) -- (151.53,219.6) ;
\draw [color={rgb, 255:red, 0; green, 0; blue, 255 }  ,draw opacity=1 ]   (149.92,211.02) -- (146.86,216.45) ;
\draw [color={rgb, 255:red, 0; green, 0; blue, 255 }  ,draw opacity=1 ]   (160.14,221.17) -- (154.98,224.25) ;

\draw    (120.35,133) .. controls (121.39,127) and (125.52,121.63) .. (132.19,119.22) ;
\draw [shift={(135.06,118.41)}, rotate = 168.07] [fill={rgb, 255:red, 0; green, 0; blue, 0 }  ][line width=0.08]  [draw opacity=0] (5.36,-2.57) -- (0,0) -- (5.36,2.57) -- (3.56,0) -- cycle    ;
\draw   (440.52,184.71) .. controls (440.52,149.2) and (469.31,120.42) .. (504.81,120.42) .. controls (540.32,120.42) and (569.11,149.2) .. (569.11,184.71) .. controls (569.11,220.22) and (540.32,249.01) .. (504.81,249.01) .. controls (469.31,249.01) and (440.52,220.22) .. (440.52,184.71) -- cycle ;

\draw    (442.68,154.76) .. controls (450.3,139.93) and (459.03,131.07) .. (474.8,123.31) ;
\draw [shift={(477.34,122.09)}, rotate = 155.14] [fill={rgb, 255:red, 0; green, 0; blue, 0 }  ][line width=0.08]  [draw opacity=0] (5.36,-2.57) -- (0,0) -- (5.36,2.57) -- (3.56,0) -- cycle    ;
\draw [line width=1.5]    (290.76,184.74) -- (396.75,184.69) ;
\draw [shift={(400.75,184.69)}, rotate = 179.98] [fill={rgb, 255:red, 0; green, 0; blue, 0 }  ][line width=0.08]  [draw opacity=0] (8.75,-4.2) -- (0,0) -- (8.75,4.2) -- (5.81,0) -- cycle    ;

\draw (302,138) node [anchor=north west][inner sep=0.75pt]  [xscale=0.9,yscale=0.9]  {$\displaystyle \prod _{j=0}^{3}\int \frac{dK_{j} dJ_{j}}{2\pi }$};
\draw (484.31,177.61) node [anchor=north west][inner sep=0.75pt]  [xscale=1,yscale=1]  {$Z( 4\beta )$};
\draw (116.22,113.89) node [anchor=north west][inner sep=0.75pt]  [font=\scriptsize,xscale=1,yscale=1]  {$\beta $};
\draw (201,157) node [anchor=north west][inner sep=0.75pt]  [font=\scriptsize,color={rgb, 255:red, 255; green, 0; blue, 0 }  ,opacity=1 ,rotate=-345,xscale=0.9,yscale=0.9]  {$K_{1}$};
\draw (216,161) node [anchor=north west][inner sep=0.75pt]  [font=\scriptsize,color={rgb, 255:red, 0; green, 0; blue, 255 }  ,opacity=1 ,rotate=-340,xscale=0.9,yscale=0.9]  {$J_{1}$};
\draw (198,145) node [anchor=north west][inner sep=0.75pt]  [font=\scriptsize,color={rgb, 255:red, 0; green, 0; blue, 255 }  ,opacity=1 ,rotate=-350,xscale=0.9,yscale=0.9]  {$\text{-}J_{0}$};
\draw (203.5,203) node [anchor=north west][inner sep=0.75pt]  [font=\scriptsize,color={rgb, 255:red, 255; green, 0; blue, 0 }  ,opacity=1 ,rotate=-15,xscale=0.9,yscale=0.9]  {$K_{2}$};
\draw (203,218) node [anchor=north west][inner sep=0.75pt]  [font=\scriptsize,color={rgb, 255:red, 0; green, 0; blue, 255 }  ,opacity=1 ,rotate=-10,xscale=0.9,yscale=0.9]  {$J_{2}$};
\draw (216,198.5) node [anchor=north west][inner sep=0.75pt]  [font=\scriptsize,color={rgb, 255:red, 0; green, 0; blue, 255 }  ,opacity=1 ,rotate=-10,xscale=0.9,yscale=0.9]  {\text{-}$J_{1}$};
\draw (157,207) node [anchor=north west][inner sep=0.75pt]  [font=\scriptsize,color={rgb, 255:red, 255; green, 0; blue, 0 }  ,opacity=1 ,rotate=-345,xscale=0.9,yscale=0.9]  {$K_{3}$};
\draw (156.5,221) node [anchor=north west][inner sep=0.75pt]  [font=\scriptsize,color={rgb, 255:red, 0; green, 0; blue, 255 }  ,opacity=1 ,rotate=-350,xscale=0.9,yscale=0.9]  {\text{-}$J_{2}$};
\draw (144,202) node [anchor=north west][inner sep=0.75pt]  [font=\scriptsize,color={rgb, 255:red, 0; green, 0; blue, 255 }  ,opacity=1 ,rotate=-340,xscale=0.9,yscale=0.9]  {$J_{3}$};
\draw (159,153) node [anchor=north west][inner sep=0.75pt]  [font=\scriptsize,color={rgb, 255:red, 255; green, 0; blue, 0 }  ,opacity=1 ,rotate=-15,xscale=0.9,yscale=0.9]  {$K_{0}$};
\draw (161.5,142.5) node [anchor=north west][inner sep=0.75pt]  [font=\scriptsize,color={rgb, 255:red, 0; green, 0; blue, 255 }  ,opacity=1 ,rotate=-10,xscale=0.9,yscale=0.9]  {$J_{0}$};
\draw (144.5,155.5) node [anchor=north west][inner sep=0.75pt]  [font=\scriptsize,color={rgb, 255:red, 0; green, 0; blue, 255 }  ,opacity=1 ,rotate=-20,xscale=0.9,yscale=0.9]  {$\text{-}J_{3}$};
\draw (167,176.61) node [anchor=north west][inner sep=0.75pt]  [xscale=1,yscale=1]  {$Z( \beta )^{4}$};
\draw (441.59,124.51) node [anchor=north west][inner sep=0.75pt]  [font=\scriptsize,xscale=1,yscale=1]  {$4\beta $};

\end{tikzpicture}
\caption{Illustration of how using the same techniques of integrating over external sources can take us from $Z(\beta)^n$ to $Z(4\beta)$. Both the parameter $\epsilon$, used to separate sources in Euclidean time, and the limit $\epsilon \rightarrow 0$ are being suppressed here for illustration clarity.}
\label{betantonbeta}
\end{figure}

Now let us briefly sketch how we could have gone in the opposite direction to what we just did. That is, by starting from $Z(\beta)^n$, we could have decided to cut open every Euclidean time circle and then glue them together in such a way to form $Z(n\beta)$ using the same techniques. From this perspective, the identity relating $Z(\beta)^n$ to $Z(n\beta)$ can be written as a form of averaging over a product of $n$ $3$-point thermal correlators as follows 
\ben \label{betatonbeta}
\frac{Z(n\beta)}{Z(\beta)^n}=\left(\prod_{j=0}^{n-1} \int_{-\infty}^{\infty} \frac{dK_j\, dJ_j}{2\pi}\right)
\prod_{j=0}^{n-1}
\left\langle e^{i J_j X}\, e^{i K_j P}\, e^{-i J_{j+1} X}\right\rangle_{\beta}.
\een
An illustration of this operation is shown in fig. \ref{betantonbeta}. It is straightforward to verify that \eqref{betatonbeta} is indeed true for the harmonic oscillator. Depending on the calculation we are interested in, performing this kind of operation can be more convenient than going from $Z(n\beta)$ to $Z(\beta)^n$. 

Inspecting the expectation values in \eqref{betatonbeta}, the reader may recognize them as essentially a form of Wigner functions, making contact with the ideas in \cite{Chakraborty:2018wkx, Moitra:2020cty, Chakraborty:2020gjs, Moitra:2023gjm, Sarkar:2025uoe, Haldar:2020ymg}. However, we will mostly refrain from thinking of them in this way, choosing to emphasize instead the path integral surgery picture. 

\subsection{Surgery in the energy basis and the spectral form factor}

Before applying these ideas to entropy calculations, we would like to briefly comment on the freedom in the choice of basis in performing these surgery operations. The point here is that we are not confined to the use of sources coupled to the position and momentum operators to establish the relationship between $Z(n \beta)$ and $Z(\beta)^n$. The main idea is simply to manipulate $Z(n\beta)$ to break off correlations between states right before and right after every cycle of $\beta$, and then attach the now ``loose'' ends to each other in such a way that form closed loops each of size $\beta$. Typically, this would involve some insertions into the phase space path integral that, in some basis, look like a combination of matrices of ones to break off the correlation and delta functions to establish the reattachment. This procedure is obviously independent of the choice of basis. So instead of our earlier choice of the position basis implementation, let us see what happens if we try to implement it in the energy basis. In this basis, we choose to represent the matrix of ones as follows 
\begin{equation}
   \sum_{l=0}^\infty \sum_{m=0}^\infty \ketbra{E_l}{E_m}= \sum_{l=0}^\infty \sum_{m=0}^\infty \frac{{a^\dagger}^l}{\sqrt{l!}} \ketbra{0}{0} \frac{{a}^m}{\sqrt{m!}},
\end{equation}
 while the delta function insertion is implemented by coupling the Hamiltonian operator to our usual Euclidean time kick sources just like we did with the position operator. The sources here however are nothing but a Lorentzian time evolution with time playing the role of the source. And so we just have an identity that looks something like\footnote{The form of the integral over sources here is slightly different than what we used before to account for the fact that they are coupled to an operator, in this case the Hamiltonian, with a discrete eigenvalue spectrum. Another way to say this is that the relevant insertion here being represented by this integral is a Kronecker delta instead of a Dirac delta function.} 
\begin{align*}
  & \lim_{T\rightarrow \infty}\left(\prod_{j=0}^{n-1}\frac{1 }{2T} \int_{-T}^{T} dt_j\right)\Tr{\prod_{j=0}^{n-1}\ketbra{0}{0} \left(\sum_{m=0}^\infty \frac{{a}^m}{\sqrt{m!}}\right)e^{iHt_j}e^{-\beta H}e^{-iHt_j}\left(\sum_{l=0}^\infty\frac{{a^\dagger}^l}{\sqrt{l!}}\right)} \\
   &=\Tr{\prod_{j=1}^{n}\left(\ketbra{0}{0} \sum_{m=0}^\infty \frac{{a}^m}{\sqrt{m!}}e^{-\beta H}\frac{{a^\dagger}^m}{\sqrt{m!}} \right)}=Z(\beta)^n.
\end{align*}
From this perspective this identity looks trivial. Nevertheless, this picture can be quite useful as we will now argue. So far our approach was based on starting from $Z(n\beta)$ and introducing some sources or insertions that effectively transform it into $Z(\beta)^n$. But, as discussed earlier, one can also go in the opposite direction, starting from $n$ copies of the system each with a partition function $Z(\beta)$, we can introduce insertions that cut and reconnect them in such a way to form $Z(n\beta)$.
\begin{align*}
   &\lim_{T \rightarrow \infty} \left( \prod_{k=0}^{n-1}\frac{1}{2T} \int_{-T}^{T} dt_j\right)\prod_{j=0}^{n-1}\Tr{e^{-\beta H}~ e^{it_jH}\left(\sum_{l=0}^\infty\sum_{m=0}^\infty\frac{{a^\dagger}^l}{\sqrt{l!}}\ketbra{0}{0}  \frac{{a}^m}{\sqrt{m!}}\right)e^{-it_{j+1}H}} \\&
   = \lim_{T \rightarrow \infty} \left( \prod_{k=0}^{n-1}\frac{1}{2T} \int_{-T}^{T} dt_j\right)\prod_{j=0}^{n-1}\Tr{ e^{\left(i(t_j-t_{j+1})-\beta\right)H}}=Z(n\beta).
\end{align*}
Using the energy basis implementation in this case is especially simple as no ``cutting'' is needed. We only need to insert sources that identify the state on some Euclidean time slices between the different copies in a way that we will make precise shortly. This is just a direct result of the density matrix being diagonal in the energy basis. Now, let us focus on the case of having just two copies. In this case, our sources which are just Lorentzian time evolutions and can be rewritten as
\begin{equation} \label{nodegenracy}
   \lim_{T\rightarrow \infty} \frac{1 }{2T} \int_{-T}^{T} dt ~Z(\beta-it) Z(\beta+it)=Z(2\beta).
\end{equation}
  The integrand on the left-hand side is nothing but the quantity known as the spectral form factor. This identity is well known to hold when the energy spectrum has no degeneracy. In our framework, what one gains by having no degeneracy is that the integral over the Lorentzian time sources completely identifies the state on some Euclidean time slice  in the first copy with the state on Euclidean time slice in the second copy giving us $Z(2 \beta)$. If the energy spectrum is degenerate, then we need more sources to break the degeneracy and form $Z(2 \beta)$ from the two $Z(\beta)$ copies. Even without doing this, this picture offers a nice interpretation for the time-averaged spectral form factor in terms of a second Rényi entropy anyway which would be given by 
\begin{align} \label{Rényi2}
S_2=- \ln \left(\lim_{T\rightarrow \infty}\frac{1 }{2T} \int_{-T}^{T} dt\frac{ ~Z(\beta-it) Z(\beta+it)}{Z(\beta)^2}\right).
\end{align}
This can just be thought of as the second Rényi entropy associated with a reduced density matrix seen by an observer probing the system at temperature $1/\beta$ but having access to nothing but energy measurements. In the absence of energy degeneracy, the thermal density matrix cannot be meaningfully reduced for such an observer, which means that \eqref{Rényi2} should just produce the standard thermal second Rényi entropy of the system. Clearly, this is going to be the case whenever \eqref{nodegenracy} holds.

\subsection{Entanglement entropy from partial surgery}
To conclude this section, we will go back and solve the problem we set out to solve in the beginning which is to calculate the entanglement entropy between the two coupled oscillators described by \eqref{hamiltonian}.

The details of the Hamiltonian here are not particularly relevant for us, the only important part is that it can be written in terms of two decoupled oscillators $X_1$ and $X_2$ with frequencies $\omega_1$ and $\omega_2$ respectively with the original coupled oscillators given by
\begin{align*}
x_1=X_1 \cos{\theta} +X_2 \sin{\theta} ,\\
x_2=X_2 \cos{\theta} -X_1 \sin{\theta} ,
\end{align*}
for some angle $\theta$. 

The partition function of this entire system is just the partition functions of the decoupled oscillators multiplied together $Z_1(\beta) Z_2(\beta)$, where $Z_1(\beta)$ and $Z_2(\beta)$ are the thermal partition functions of $X_1$ and $X_2$ respectively. 
If we wanted to calculate the thermal Rényi entropy of the system we would simply write 

\begin{figure} [!t]
 \centering

 
\tikzset{
pattern size/.store in=\mcSize, 
pattern size = 5pt,
pattern thickness/.store in=\mcThickness, 
pattern thickness = 0.3pt,
pattern radius/.store in=\mcRadius, 
pattern radius = 1pt}
\makeatletter
\pgfutil@ifundefined{pgf@pattern@name@_cndzivqrx}{
\pgfdeclarepatternformonly[\mcThickness,\mcSize]{_cndzivqrx}
{\pgfqpoint{0pt}{0pt}}
{\pgfpoint{\mcSize+\mcThickness}{\mcSize+\mcThickness}}
{\pgfpoint{\mcSize}{\mcSize}}
{
\pgfsetcolor{\tikz@pattern@color}
\pgfsetlinewidth{\mcThickness}
\pgfpathmoveto{\pgfqpoint{0pt}{0pt}}
\pgfpathlineto{\pgfpoint{\mcSize+\mcThickness}{\mcSize+\mcThickness}}
\pgfusepath{stroke}
}}
\makeatother

 
\tikzset{
pattern size/.store in=\mcSize, 
pattern size = 5pt,
pattern thickness/.store in=\mcThickness, 
pattern thickness = 0.3pt,
pattern radius/.store in=\mcRadius, 
pattern radius = 1pt}
\makeatletter
\pgfutil@ifundefined{pgf@pattern@name@_edw2lmiby}{
\pgfdeclarepatternformonly[\mcThickness,\mcSize]{_edw2lmiby}
{\pgfqpoint{0pt}{0pt}}
{\pgfpoint{\mcSize+\mcThickness}{\mcSize+\mcThickness}}
{\pgfpoint{\mcSize}{\mcSize}}
{
\pgfsetcolor{\tikz@pattern@color}
\pgfsetlinewidth{\mcThickness}
\pgfpathmoveto{\pgfqpoint{0pt}{0pt}}
\pgfpathlineto{\pgfpoint{\mcSize+\mcThickness}{\mcSize+\mcThickness}}
\pgfusepath{stroke}
}}
\makeatother
\tikzset{every picture/.style={line width=0.75pt}} 

\begin{tikzpicture}[x=0.75pt,y=0.75pt,yscale=-1,xscale=1]

\draw  [pattern=_cndzivqrx,pattern size=1.5pt,pattern thickness=0.75pt,pattern radius=0pt, pattern color={rgb, 255:red, 193; green, 193; blue, 193}] (439.43,144.76) .. controls (439.43,105.97) and (470.7,74.52) .. (509.29,74.52) .. controls (547.87,74.52) and (579.14,105.97) .. (579.14,144.76) .. controls (579.14,183.54) and (547.87,214.99) .. (509.29,214.99) .. controls (470.7,214.99) and (439.43,183.54) .. (439.43,144.76) -- cycle ;
\draw  [draw opacity=0][fill={rgb, 255:red, 255; green, 255; blue, 255 }  ,fill opacity=1 ] (445.3,144.85) .. controls (445.3,109.31) and (473.84,80.49) .. (509.04,80.49) .. controls (544.25,80.49) and (572.79,109.31) .. (572.79,144.85) .. controls (572.79,180.4) and (544.25,209.21) .. (509.04,209.21) .. controls (473.84,209.21) and (445.3,180.4) .. (445.3,144.85) -- cycle ;
\draw  [draw opacity=0][fill={rgb, 255:red, 255; green, 255; blue, 255 }  ,fill opacity=1 ] (440.16,150.78) .. controls (440,148.81) and (439.91,146.82) .. (439.91,144.81) .. controls (439.91,142.77) and (440,140.75) .. (440.17,138.75) -- (445.67,139.23) .. controls (445.51,141.07) and (445.43,142.93) .. (445.43,144.81) .. controls (445.43,146.66) and (445.51,148.49) .. (445.66,150.3) -- cycle ;
\draw  [draw opacity=0][fill={rgb, 255:red, 255; green, 255; blue, 255 }  ,fill opacity=1 ] (515.47,214.28) .. controls (513.36,214.47) and (511.21,214.57) .. (509.04,214.57) .. controls (507.11,214.57) and (505.19,214.49) .. (503.3,214.34) -- (503.75,208.87) .. controls (505.49,209.01) and (507.26,209.08) .. (509.04,209.08) .. controls (511.04,209.08) and (513.02,208.99) .. (514.97,208.81) -- cycle ;
\draw  [draw opacity=0][fill={rgb, 255:red, 255; green, 255; blue, 255 }  ,fill opacity=1 ] (578.39,150.81) .. controls (578.56,148.82) and (578.65,146.8) .. (578.65,144.76) .. controls (578.65,142.74) and (578.56,140.75) .. (578.4,138.78) -- (572.13,139.32) .. controls (572.28,141.11) and (572.36,142.92) .. (572.36,144.76) .. controls (572.36,146.62) and (572.28,148.46) .. (572.12,150.28) -- cycle ;
\draw  [draw opacity=0][fill={rgb, 255:red, 255; green, 255; blue, 255 }  ,fill opacity=1 ] (503.13,75.26) .. controls (505.26,75.07) and (507.42,74.97) .. (509.6,74.97) .. controls (511.52,74.97) and (513.43,75.04) .. (515.32,75.2) -- (514.86,80.78) .. controls (513.13,80.64) and (511.37,80.57) .. (509.6,80.57) .. controls (507.59,80.57) and (505.61,80.66) .. (503.65,80.84) -- cycle ;
\draw  [draw opacity=0] (445.44,139.17) .. controls (448.1,108.11) and (472.75,83.36) .. (503.68,80.66) -- (509.29,144.76) -- cycle ; \draw   (445.44,139.17) .. controls (448.1,108.11) and (472.75,83.36) .. (503.68,80.66) ;  
\draw  [draw opacity=0] (503.68,208.85) .. controls (472.79,206.15) and (448.18,181.4) .. (445.53,150.33) -- (509.29,144.76) -- cycle ; \draw   (503.68,208.85) .. controls (472.79,206.15) and (448.18,181.4) .. (445.53,150.33) ;  
\draw  [draw opacity=0] (573.04,150.33) .. controls (570.39,181.4) and (545.78,206.15) .. (514.89,208.85) -- (509.29,144.76) -- cycle ; \draw   (573.04,150.33) .. controls (570.39,181.4) and (545.78,206.15) .. (514.89,208.85) ;  
\draw  [draw opacity=0] (514.89,80.66) .. controls (545.78,83.36) and (570.39,108.12) .. (573.04,139.18) -- (509.29,144.76) -- cycle ; \draw   (514.89,80.66) .. controls (545.78,83.36) and (570.39,108.12) .. (573.04,139.18) ;  
\draw [color={rgb, 255:red, 0; green, 0; blue, 255 }  ,draw opacity=1 ]   (445.45,139.14) .. controls (466.98,139.13) and (503.46,101.77) .. (503.71,80.66) ;
\draw [color={rgb, 255:red, 0; green, 0; blue, 255 }  ,draw opacity=1 ]   (514.87,80.69) .. controls (515.15,101.81) and (552.66,139.16) .. (573.04,139.18) ;
\draw [color={rgb, 255:red, 0; green, 0; blue, 255 }  ,draw opacity=1 ]   (514.86,208.85) .. controls (514.8,188.27) and (552.48,150.74) .. (573.04,150.36) ;
\draw [color={rgb, 255:red, 0; green, 0; blue, 255 }  ,draw opacity=1 ]   (503.71,208.85) .. controls (503.24,188.27) and (466.45,150.39) .. (445.53,150.36) ;

\draw    (442.46,111.43) .. controls (447.08,102.37) and (453.56,92.56) .. (465.44,84.39) ;
\draw [shift={(467.78,82.84)}, rotate = 147.7] [fill={rgb, 255:red, 0; green, 0; blue, 0 }  ][line width=0.08]  [draw opacity=0] (5.36,-2.57) -- (0,0) -- (5.36,2.57) -- (3.56,0) -- cycle    ;

\draw  [pattern=_edw2lmiby,pattern size=1.5pt,pattern thickness=0.75pt,pattern radius=0pt, pattern color={rgb, 255:red, 193; green, 193; blue, 193}] (131.85,144.76) .. controls (131.85,105.97) and (163.12,74.52) .. (201.7,74.52) .. controls (240.28,74.52) and (271.56,105.97) .. (271.56,144.76) .. controls (271.56,183.54) and (240.28,214.99) .. (201.7,214.99) .. controls (163.12,214.99) and (131.85,183.54) .. (131.85,144.76) -- cycle ;
\draw  [draw opacity=0][fill={rgb, 255:red, 255; green, 255; blue, 255 }  ,fill opacity=1 ] (137.96,144.76) .. controls (137.96,109.21) and (166.5,80.4) .. (201.7,80.4) .. controls (236.91,80.4) and (265.45,109.21) .. (265.45,144.76) .. controls (265.45,180.3) and (236.91,209.11) .. (201.7,209.11) .. controls (166.5,209.11) and (137.96,180.3) .. (137.96,144.76) -- cycle ;
\draw  [draw opacity=0] (137.62,144.51) .. controls (137.75,109.09) and (166.39,80.42) .. (201.7,80.42) .. controls (201.74,80.42) and (201.78,80.42) .. (201.83,80.42) -- (201.7,144.76) -- cycle ; \draw   (137.62,144.51) .. controls (137.75,109.09) and (166.39,80.42) .. (201.7,80.42) .. controls (201.74,80.42) and (201.78,80.42) .. (201.83,80.42) ;  
\draw  [draw opacity=0] (201.68,209.09) .. controls (166.35,209.08) and (137.71,180.28) .. (137.71,144.76) .. controls (137.71,144.53) and (137.71,144.31) .. (137.71,144.09) -- (201.7,144.76) -- cycle ; \draw   (201.68,209.09) .. controls (166.35,209.08) and (137.71,180.28) .. (137.71,144.76) .. controls (137.71,144.53) and (137.71,144.31) .. (137.71,144.09) ;  
\draw  [draw opacity=0] (265.69,144.23) .. controls (265.69,144.41) and (265.69,144.58) .. (265.69,144.76) .. controls (265.69,180.29) and (237.04,209.09) .. (201.7,209.09) .. controls (201.65,209.09) and (201.59,209.09) .. (201.53,209.09) -- (201.7,144.76) -- cycle ; \draw   (265.69,144.23) .. controls (265.69,144.41) and (265.69,144.58) .. (265.69,144.76) .. controls (265.69,180.29) and (237.04,209.09) .. (201.7,209.09) .. controls (201.65,209.09) and (201.59,209.09) .. (201.53,209.09) ;  
\draw  [draw opacity=0] (201.68,80.42) .. controls (201.69,80.42) and (201.69,80.42) .. (201.7,80.42) .. controls (237.01,80.42) and (265.64,109.17) .. (265.69,144.66) -- (201.7,144.76) -- cycle ; \draw   (201.68,80.42) .. controls (201.69,80.42) and (201.69,80.42) .. (201.7,80.42) .. controls (237.01,80.42) and (265.64,109.17) .. (265.69,144.66) ;  
\draw  [draw opacity=0][fill={rgb, 255:red, 0; green, 0; blue, 255 }  ,fill opacity=1 ] (136.29,138.08) .. controls (136.29,137.25) and (136.99,136.57) .. (137.86,136.57) .. controls (138.73,136.57) and (139.43,137.25) .. (139.43,138.08) .. controls (139.43,138.92) and (138.73,139.6) .. (137.86,139.6) .. controls (136.99,139.6) and (136.29,138.92) .. (136.29,138.08) -- cycle ;
\draw  [draw opacity=0][fill={rgb, 255:red, 0; green, 0; blue, 255 }  ,fill opacity=1 ] (136.37,150.76) .. controls (136.37,149.93) and (137.08,149.25) .. (137.95,149.25) .. controls (138.82,149.25) and (139.52,149.93) .. (139.52,150.76) .. controls (139.52,151.6) and (138.82,152.28) .. (137.95,152.28) .. controls (137.08,152.28) and (136.37,151.6) .. (136.37,150.76) -- cycle ;
\draw  [draw opacity=0][fill={rgb, 255:red, 255; green, 0; blue, 0 }  ,fill opacity=1 ] (136.14,144.37) .. controls (136.14,143.53) and (136.85,142.85) .. (137.72,142.85) .. controls (138.59,142.85) and (139.29,143.53) .. (139.29,144.37) .. controls (139.29,145.21) and (138.59,145.88) .. (137.72,145.88) .. controls (136.85,145.88) and (136.14,145.21) .. (136.14,144.37) -- cycle ;

\draw  [draw opacity=0][fill={rgb, 255:red, 0; green, 0; blue, 255 }  ,fill opacity=1 ] (267.09,138.08) .. controls (267.09,137.25) and (266.39,136.57) .. (265.52,136.57) .. controls (264.65,136.57) and (263.95,137.25) .. (263.95,138.08) .. controls (263.95,138.92) and (264.65,139.6) .. (265.52,139.6) .. controls (266.39,139.6) and (267.09,138.92) .. (267.09,138.08) -- cycle ;
\draw  [draw opacity=0][fill={rgb, 255:red, 0; green, 0; blue, 255 }  ,fill opacity=1 ] (267.01,150.76) .. controls (267.01,149.93) and (266.3,149.25) .. (265.43,149.25) .. controls (264.56,149.25) and (263.86,149.93) .. (263.86,150.76) .. controls (263.86,151.6) and (264.56,152.28) .. (265.43,152.28) .. controls (266.3,152.28) and (267.01,151.6) .. (267.01,150.76) -- cycle ;
\draw  [draw opacity=0][fill={rgb, 255:red, 255; green, 0; blue, 0 }  ,fill opacity=1 ] (267.24,144.37) .. controls (267.24,143.53) and (266.53,142.85) .. (265.66,142.85) .. controls (264.79,142.85) and (264.09,143.53) .. (264.09,144.37) .. controls (264.09,145.21) and (264.79,145.88) .. (265.66,145.88) .. controls (266.53,145.88) and (267.24,145.21) .. (267.24,144.37) -- cycle ;

\draw  [draw opacity=0][fill={rgb, 255:red, 0; green, 0; blue, 255 }  ,fill opacity=1 ] (195.42,210.4) .. controls (194.58,210.4) and (193.91,209.69) .. (193.91,208.82) .. controls (193.91,207.95) and (194.58,207.25) .. (195.42,207.25) .. controls (196.26,207.25) and (196.94,207.95) .. (196.94,208.82) .. controls (196.94,209.69) and (196.26,210.4) .. (195.42,210.4) -- cycle ;
\draw  [draw opacity=0][fill={rgb, 255:red, 0; green, 0; blue, 255 }  ,fill opacity=1 ] (208.1,210.31) .. controls (207.26,210.31) and (206.58,209.61) .. (206.58,208.74) .. controls (206.58,207.87) and (207.26,207.16) .. (208.1,207.16) .. controls (208.94,207.16) and (209.62,207.87) .. (209.62,208.74) .. controls (209.62,209.61) and (208.94,210.31) .. (208.1,210.31) -- cycle ;
\draw  [draw opacity=0][fill={rgb, 255:red, 255; green, 0; blue, 0 }  ,fill opacity=1 ] (201.71,210.54) .. controls (200.87,210.54) and (200.19,209.84) .. (200.19,208.97) .. controls (200.19,208.1) and (200.87,207.39) .. (201.71,207.39) .. controls (202.54,207.39) and (203.22,208.1) .. (203.22,208.97) .. controls (203.22,209.84) and (202.54,210.54) .. (201.71,210.54) -- cycle ;

\draw  [draw opacity=0][fill={rgb, 255:red, 0; green, 0; blue, 255 }  ,fill opacity=1 ] (208.1,78.88) .. controls (208.94,78.88) and (209.62,79.58) .. (209.62,80.45) .. controls (209.62,81.32) and (208.94,82.02) .. (208.1,82.02) .. controls (207.26,82.02) and (206.58,81.32) .. (206.58,80.45) .. controls (206.58,79.58) and (207.26,78.88) .. (208.1,78.88) -- cycle ;
\draw  [draw opacity=0][fill={rgb, 255:red, 0; green, 0; blue, 255 }  ,fill opacity=1 ] (195.42,78.96) .. controls (196.26,78.96) and (196.94,79.67) .. (196.94,80.54) .. controls (196.94,81.41) and (196.26,82.11) .. (195.42,82.11) .. controls (194.58,82.11) and (193.91,81.41) .. (193.91,80.54) .. controls (193.91,79.67) and (194.58,78.96) .. (195.42,78.96) -- cycle ;
\draw  [draw opacity=0][fill={rgb, 255:red, 255; green, 0; blue, 0 }  ,fill opacity=1 ] (201.82,78.73) .. controls (202.65,78.73) and (203.33,79.44) .. (203.33,80.31) .. controls (203.33,81.18) and (202.65,81.88) .. (201.82,81.88) .. controls (200.98,81.88) and (200.3,81.18) .. (200.3,80.31) .. controls (200.3,79.44) and (200.98,78.73) .. (201.82,78.73) -- cycle ;

\draw    (135,111.43) .. controls (139.61,102.37) and (146.1,92.56) .. (157.98,84.39) ;
\draw [shift={(160.32,82.84)}, rotate = 147.7] [fill={rgb, 255:red, 0; green, 0; blue, 0 }  ][line width=0.08]  [draw opacity=0] (5.36,-2.57) -- (0,0) -- (5.36,2.57) -- (3.56,0) -- cycle    ;

\draw [line width=1.5]    (299.33,144.78) -- (405.32,144.73) ;
\draw [shift={(409.32,144.73)}, rotate = 179.98] [fill={rgb, 255:red, 0; green, 0; blue, 0 }  ][line width=0.08]  [draw opacity=0] (8.75,-4.2) -- (0,0) -- (8.75,4.2) -- (5.81,0) -- cycle    ;

\draw (129.7,85.25) node [anchor=north west][inner sep=0.75pt]  [font=\scriptsize,xscale=1,yscale=1]  {$4\beta $};
\draw (437.16,85.25) node [anchor=north west][inner sep=0.75pt]  [font=\scriptsize,xscale=1,yscale=1]  {$4\beta $};
\draw (158.2,182) node [anchor=north west][inner sep=0.75pt]  [font=\scriptsize,xscale=1,yscale=1]  {$x_{1}$};
\draw (141,198) node [anchor=north west][inner sep=0.75pt]  [font=\scriptsize,xscale=1,yscale=1]  {$x_{2}$};
\draw (243,139) node [anchor=north west][inner sep=0.75pt]  [font=\scriptsize,color={rgb, 255:red, 255; green, 0; blue, 0 }  ,opacity=1 ,xscale=0.9,yscale=0.9]  {$K_{1}$};
\draw (251,148.5) node [anchor=north west][inner sep=0.75pt]  [font=\scriptsize,color={rgb, 255:red, 0; green, 0; blue, 255 }  ,opacity=1 ,rotate=-10,xscale=0.9,yscale=0.9]  {$J_{1}$};
\draw (246,130) node [anchor=north west][inner sep=0.75pt]  [font=\scriptsize,color={rgb, 255:red, 0; green, 0; blue, 255 }  ,opacity=1 ,rotate=-350,xscale=0.9,yscale=0.9]  {$\text{-}J_{0}$};
\draw (195,193) node [anchor=north west][inner sep=0.75pt]  [font=\scriptsize,color={rgb, 255:red, 255; green, 0; blue, 0 }  ,opacity=1 ,xscale=0.9,yscale=0.9]  {$K_{2}$};
\draw (185,195) node [anchor=north west][inner sep=0.75pt]  [font=\scriptsize,color={rgb, 255:red, 0; green, 0; blue, 255 }  ,opacity=1 ,rotate=-10,xscale=0.9,yscale=0.9]  {$J_{2}$};
\draw (206,198) node [anchor=north west][inner sep=0.75pt]  [font=\scriptsize,color={rgb, 255:red, 0; green, 0; blue, 255 }  ,opacity=1 ,rotate=-350,xscale=0.9,yscale=0.9]  {$\text{-}J_{1}$};
\draw (145,139) node [anchor=north west][inner sep=0.75pt]  [font=\scriptsize,color={rgb, 255:red, 255; green, 0; blue, 0 }  ,opacity=1 ,xscale=0.9,yscale=0.9]  {$K_{3}$};
\draw (141,127) node [anchor=north west][inner sep=0.75pt]  [font=\scriptsize,color={rgb, 255:red, 0; green, 0; blue, 255 }  ,opacity=1 ,rotate=-10,xscale=0.9,yscale=0.9]  {$J_{3}$};
\draw (138,151.5) node [anchor=north west][inner sep=0.75pt]  [font=\scriptsize,color={rgb, 255:red, 0; green, 0; blue, 255 }  ,opacity=1 ,rotate=-350,xscale=0.9,yscale=0.9]  {$\text{-}J_{2}$};
\draw (195,86) node [anchor=north west][inner sep=0.75pt]  [font=\scriptsize,color={rgb, 255:red, 255; green, 0; blue, 0 }  ,opacity=1 ,xscale=0.9,yscale=0.9]  {$K_{0}$};
\draw (208,81.8) node [anchor=north west][inner sep=0.75pt]  [font=\scriptsize,color={rgb, 255:red, 0; green, 0; blue, 255 }  ,opacity=1 ,rotate=-10,xscale=0.9,yscale=0.9]  {$J_{0}$};
\draw (180,84.5) node [anchor=north west][inner sep=0.75pt]  [font=\scriptsize,color={rgb, 255:red, 0; green, 0; blue, 255 }  ,opacity=1 ,rotate=-350,xscale=0.9,yscale=0.9]  {$\text{-}J_{3}$};
\draw (465,182) node [anchor=north west][inner sep=0.75pt]  [font=\scriptsize,xscale=1,yscale=1]  {$x_{1}$};
\draw (449,198) node [anchor=north west][inner sep=0.75pt]  [font=\scriptsize,xscale=1,yscale=1]  {$x_{2}$};
\draw (309,97) node [anchor=north west][inner sep=0.75pt]  [xscale=0.9,yscale=0.9]  {$\displaystyle \prod _{j=0}^{3}\int \frac{dK_{j} dJ_{j}}{2\pi }$};

\end{tikzpicture}
\caption{The path integral from fig. \ref{repfigsph} can be realized by integrating over sources, coupled to just one of the oscillators, inside the path integral with Euclidean time periodicity $4\beta$ for both oscillators. Once again, the $\epsilon \rightarrow0$ limit is being suppressed for illustration clarity.}
\label{sourceaverageresult1}
\end{figure}
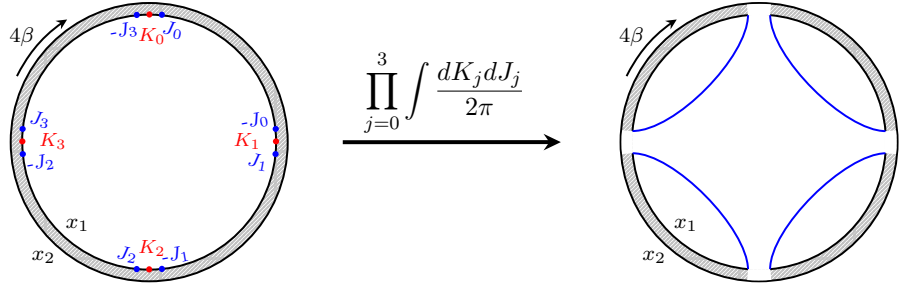

\ben
S_n=\frac{1}{1-n} \ln \left(\frac{Z_1(n \beta)Z_2(n \beta)}{Z_1(\beta)^n Z_2(\beta)^n}\right).
\een
 But since we want to calculate the entropy after tracing out $x_1$, we need to evaluate the path integral in fig. \ref{repfigsph} where $x_1$ is forced to live on Euclidean time circles of periodicity $\beta$ while $x_2$ lives on a Euclidean time circle of size $n\beta$. We can now perform this path integral by introducing sources that only couple to the position and momentum operators of $x_1$ and upon integrating over these sources we effectively produce the desired path integral as illustrated in fig. \ref{sourceaverageresult1}. The sources should be coupled to $x_1$ which is another way of saying that they are coupled to the linear combination $X_1 \cos{\theta} +X_2 \sin{\theta}$. Once we rewrite the position and momentum operators of $x_1$ as linear combinations of the position and momentum operators of the decoupled oscillators $X_1$ and $X_2$, the calculation of the effect of these insertions will follow an almost identical line to the one leading up to \eqref{deter}. The one crucial difference to our earlier result is that now one gets a contribution from each of the decoupled oscillators, so we get 
\begin{align*}
    \mathbf{K}_k &= \frac{m \omega_1}{2} \cdot \frac{\cos^2(\theta)\sinh(\omega_1 \beta)}{\cosh(\omega_1 \beta) - \cos\left(\frac{2 \pi k}{n}\right)}+\frac{m \omega_2}{2} \cdot \frac{\sin^2(\theta)\sinh(\omega_2 \beta)}{\cosh(\omega_2 \beta) - \cos\left(\frac{2 \pi k}{n}\right)}, \\[1em]
    \mathbf{J}_k &= \frac{2}{m \omega_1} \cdot \frac{\cos^2(\theta)\sinh(\omega_1 \beta) \sin^2\left(\frac{\pi k}{n}\right)}{1 + \cosh(\omega_1 \beta) - 2 \cos^2\left(\frac{\pi k}{n}\right)}+\frac{2}{m \omega_2} \cdot \frac{\sin^2(\theta)\sinh(\omega_2 \beta) \sin^2\left(\frac{\pi k}{n}\right)}{1 + \cosh(\omega_2 \beta) - 2 \cos^2\left(\frac{\pi k}{n}\right)}, \\[1em]
    \mathbf{B}_k &= \frac{i\cos^2(\theta)}{2} \cdot \frac{(\cosh(\omega_1 \beta) - 1)\left(1 + e^{-i \frac{2 \pi k}{n}}\right)}{\cosh(\omega_1 \beta) - \cos\left(\frac{2 \pi k}{n}\right)}+\frac{i\sin^2(\theta)}{2} \cdot \frac{(\cosh(\omega_2 \beta) - 1)\left(1 + e^{-i \frac{2 \pi k}{n}}\right)}{\cosh(\omega_2 \beta) - \cos\left(\frac{2 \pi k}{n}\right)}, \\[1em]
    \mathbf{B}^\top_k &= \frac{i\cos^2(\theta)}{2} \cdot \frac{(\cosh(\omega_1 \beta) - 1)\left(1 + e^{i \frac{2 \pi k}{n}}\right)}{\cosh(\omega_1 \beta) - \cos\left(\frac{2 \pi k}{n}\right)}+\frac{i\sin^2(\theta)}{2} \cdot \frac{(\cosh(\omega_2\beta) - 1)\left(1 + e^{i \frac{2 \pi k}{n}}\right)}{\cosh(\omega_2 \beta) - \cos\left(\frac{2 \pi k}{n}\right)}.
\end{align*}
And again we just need to calculate 
\ben
\det\left(\mathbf{\Omega}\right)=\det\left(\mathbf{K}\mathbf{J}-\mathbf{B}\mathbf{B}^\top\right),
\een
which is straightforward but a bit more tedious than before since the matrix elements look a bit more complicated. The Rényi entropy we are looking for is then given by 

\begin{align}
    S_n&=\frac{1}{1-n} \ln \left(\frac{Z_1(n \beta)Z_2(n \beta)}{\sqrt{\det\left(\mathbf{\Omega}\right)}\left(Z_1(\beta)Z_2(\beta)\right)^n}\right).
\end{align}

The exact expression looks a bit messy, but after spending all this effort to get here, we might as well just write it down, so here it is

\begin{align}
    S_n&=\frac{1}{2(1-n)}\left(n \ln \left(\frac{\eta}{(1-\eta)^2}\right)+\ln (2 (\cosh (n \ln (\eta))-1))\right),
\end{align}
and $\eta$ is given by 
\begin{equation*}
\eta=\frac{N}{D}-\sqrt{\frac{N^2}{D^2}-1},
\end{equation*}
with
\begin{align*}
N =\  \omega_1 \omega_2\cosh(\omega_2\beta) \biggl[-8\cos^2(\theta)
     +\Bigl(7&+\cos(4\theta)\Bigr)\cosh(\omega_1\beta)
     \biggr]- 8\omega_1\omega_2\,\sin^2(\theta)\cosh(\omega_1\beta)\\[1mm]
  &\quad + \sin^2(2\theta) \biggl[
    \sinh(\omega_2\beta)(\omega_1^2+\omega_2^2)\sinh(\omega_1\beta)
    + 2\omega_1\omega_2
\biggr],
\end{align*}
and
\begin{align*}
D =\  \omega_1\omega_2 \biggl[
    -7 - \cos(4\theta)
    + 8 \sin^2(\theta)\cosh(\omega_2\beta) - 8&\cos^2(\theta)\cosh(\omega_1\beta)
       \Bigl(\sin^2(\theta)\cosh(\omega_2\beta)-1\Bigr)
\biggr]\\[1mm]
  &\quad + \bigl(\omega_1^2+\omega_2^2\bigr)
       \sin^2(2\theta)\sinh(\omega_1\beta)\sinh(\omega_2\beta).
\end{align*}
Taking the limit $n\rightarrow 1$ gives the standard expression for von Neumann entropy. 

\section{Surgery via averaging } \label{sec:MIS}

In section \ref{sec:simple}, we saw how using some source insertions placed along the Euclidean time circle we were able to effectively cut and re-glue the path integral, transforming it from being performed over a Euclidean time circle of size $n\beta$ to the product of $n$ path integrals each performed over a Euclidean time circle of size $\beta$. We would like to further develop the idea of how one can perform quantum mechanical surgery operations in local theories without having to explicitly cut and glue the manifold. In particular, we will aim to provide a sketch of how these ideas can be applied to local field theory beyond the one dimensional setting of section \ref{sec:simple}.

In discussing field theories, the formulas we will write down will typically involve functional integrals over auxiliary sources, or fields living on codimension-one surfaces, instead of the standard integrals we had in section \ref{sec:simple}. These formulas should be understood at some finite cutoff, with the same cutoff or regularization applied to the auxiliary fields and the field theory itself. In other words, the auxiliary fields are to be treated like any other field in the theory, only living on codimension-one surfaces.\footnote{In some sense, this is similar to how one would think of the auxiliary fields used to enforce constraints in something like the non-linear Sigma model.} In any case, one relatively safe way to approach this discussion is to think of it as applying to a field theory that has been put on a lattice.

\subsection{Scalar fields as a first example}

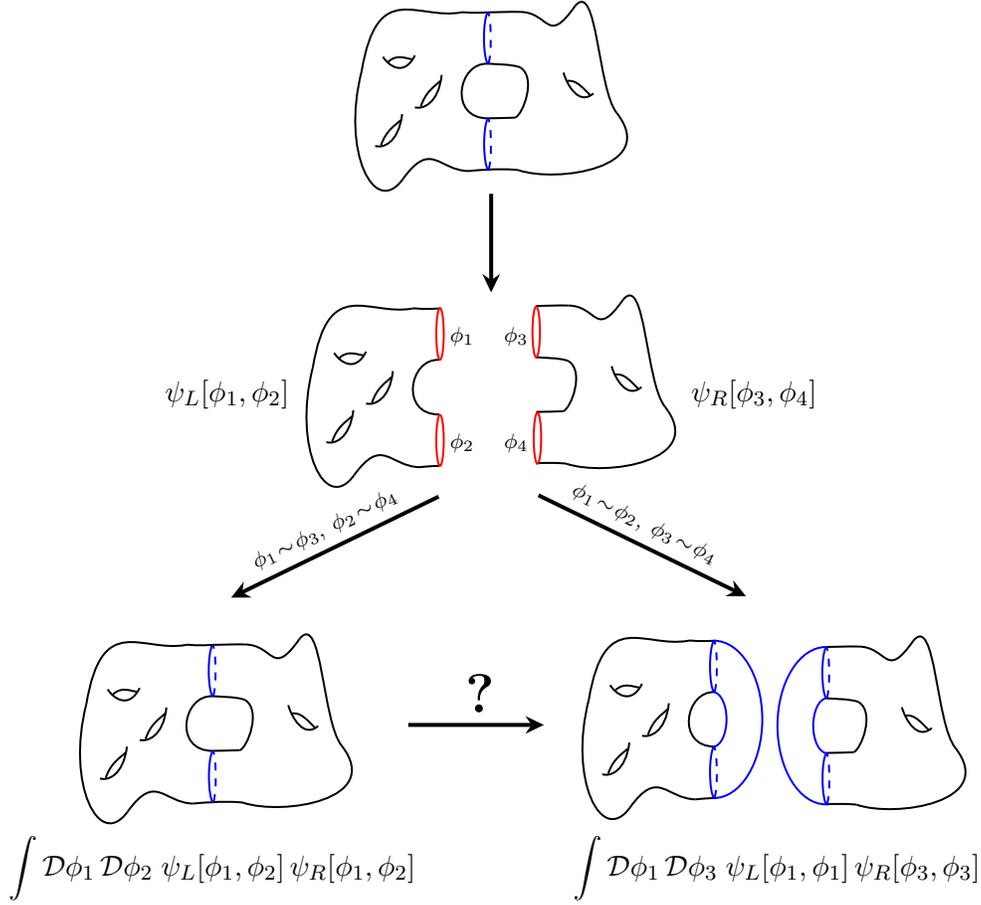
\begin{figure}[!t]
\centering
\tikzset{
pattern size/.store in=\mcSize, 
pattern size = 5pt,
pattern thickness/.store in=\mcThickness, 
pattern thickness = 0.3pt,
pattern radius/.store in=\mcRadius, 
pattern radius = 1pt}
\makeatletter
\pgfutil@ifundefined{pgf@pattern@name@_l1xqy0hsy}{
\pgfdeclarepatternformonly[\mcThickness,\mcSize]{_l1xqy0hsy}
{\pgfqpoint{0pt}{0pt}}
{\pgfpoint{\mcSize+\mcThickness}{\mcSize+\mcThickness}}
{\pgfpoint{\mcSize}{\mcSize}}
{
\pgfsetcolor{\tikz@pattern@color}
\pgfsetlinewidth{\mcThickness}
\pgfpathmoveto{\pgfqpoint{0pt}{0pt}}
\pgfpathlineto{\pgfpoint{\mcSize+\mcThickness}{\mcSize+\mcThickness}}
\pgfusepath{stroke}
}}
\makeatother

 
\tikzset{
pattern size/.store in=\mcSize, 
pattern size = 5pt,
pattern thickness/.store in=\mcThickness, 
pattern thickness = 0.3pt,
pattern radius/.store in=\mcRadius, 
pattern radius = 1pt}
\makeatletter
\pgfutil@ifundefined{pgf@pattern@name@_b8n76fz9o}{
\pgfdeclarepatternformonly[\mcThickness,\mcSize]{_b8n76fz9o}
{\pgfqpoint{0pt}{0pt}}
{\pgfpoint{\mcSize+\mcThickness}{\mcSize+\mcThickness}}
{\pgfpoint{\mcSize}{\mcSize}}
{
\pgfsetcolor{\tikz@pattern@color}
\pgfsetlinewidth{\mcThickness}
\pgfpathmoveto{\pgfqpoint{0pt}{0pt}}
\pgfpathlineto{\pgfpoint{\mcSize+\mcThickness}{\mcSize+\mcThickness}}
\pgfusepath{stroke}
}}
\makeatother

 
\tikzset{
pattern size/.store in=\mcSize, 
pattern size = 5pt,
pattern thickness/.store in=\mcThickness, 
pattern thickness = 0.3pt,
pattern radius/.store in=\mcRadius, 
pattern radius = 1pt}
\makeatletter
\pgfutil@ifundefined{pgf@pattern@name@_vp9iv54mp}{
\pgfdeclarepatternformonly[\mcThickness,\mcSize]{_vp9iv54mp}
{\pgfqpoint{0pt}{0pt}}
{\pgfpoint{\mcSize+\mcThickness}{\mcSize+\mcThickness}}
{\pgfpoint{\mcSize}{\mcSize}}
{
\pgfsetcolor{\tikz@pattern@color}
\pgfsetlinewidth{\mcThickness}
\pgfpathmoveto{\pgfqpoint{0pt}{0pt}}
\pgfpathlineto{\pgfpoint{\mcSize+\mcThickness}{\mcSize+\mcThickness}}
\pgfusepath{stroke}
}}
\makeatother

 
\tikzset{
pattern size/.store in=\mcSize, 
pattern size = 5pt,
pattern thickness/.store in=\mcThickness, 
pattern thickness = 0.3pt,
pattern radius/.store in=\mcRadius, 
pattern radius = 1pt}
\makeatletter
\pgfutil@ifundefined{pgf@pattern@name@_15z7agijo}{
\pgfdeclarepatternformonly[\mcThickness,\mcSize]{_15z7agijo}
{\pgfqpoint{0pt}{0pt}}
{\pgfpoint{\mcSize+\mcThickness}{\mcSize+\mcThickness}}
{\pgfpoint{\mcSize}{\mcSize}}
{
\pgfsetcolor{\tikz@pattern@color}
\pgfsetlinewidth{\mcThickness}
\pgfpathmoveto{\pgfqpoint{0pt}{0pt}}
\pgfpathlineto{\pgfpoint{\mcSize+\mcThickness}{\mcSize+\mcThickness}}
\pgfusepath{stroke}
}}
\makeatother
\tikzset{every picture/.style={line width=0.75pt}} 

\begin{tikzpicture}[x=0.75pt,y=0.75pt,yscale=-0.85,xscale=0.85]

\draw    (221.87,244.13) .. controls (231.14,188.7) and (254.65,213.84) .. (280.77,209.16) ;
\draw    (221.87,244.13) .. controls (209.65,291.76) and (229.45,331.57) .. (246.31,308.15) ;
\draw    (246.31,308.15) .. controls (263.16,284.73) and (267.46,302.84) .. (285.09,303.34) ;
\draw    (237.63,238.51) .. controls (242.51,233.83) and (247.74,236.33) .. (251.78,236.95) ;
\draw    (234.51,235.47) .. controls (237.88,240.93) and (247.06,247.57) .. (253.47,234.69) ;
\draw    (233.36,288.08) .. controls (236.32,279.28) and (241.44,276.46) .. (245.07,273.13) ;
\draw    (229.9,288.4) .. controls (234.2,289.96) and (243.79,286.97) .. (245.92,269.53) ;
\draw    (354.32,238.13) .. controls (353.81,238.13) and (370.19,238.3) .. (373.67,240.45) ;
\draw    (354.29,207.61) .. controls (367.92,207.4) and (369.45,207.17) .. (373.33,207.35) ;
\draw    (370.47,270.11) .. controls (371.45,270.23) and (354.49,270.5) .. (354.41,270.53) ;
\draw    (372.49,301.34) .. controls (369.1,300.58) and (356.86,300.7) .. (354.43,301.06) ;
\draw    (256.99,265.03) .. controls (259.8,256.26) and (264.95,253.27) .. (268.56,249.83) ;
\draw    (253.47,265.5) .. controls (257.9,266.83) and (267.59,263.47) .. (269.35,246.26) ;
\draw    (430.77,267.4) .. controls (456.26,300.09) and (395.47,308.96) .. (372.49,301.34) ;
\draw    (430.77,267.4) .. controls (419.18,252.94) and (419.35,187.67) .. (405.91,203.95) ;
\draw    (405.91,203.95) .. controls (389.21,227.76) and (390.96,207.44) .. (373.33,207.35) ;
\draw    (401.44,248.35) .. controls (406.91,247.37) and (411.26,253.29) .. (414.89,256.65) ;
\draw    (399.15,243.38) .. controls (401.18,250.77) and (408.38,263.26) .. (416.91,255.71) ;
\draw    (296.82,239.63) .. controls (278.99,238.5) and (273.19,272.3) .. (296.91,272.03) ;
\draw    (370.47,270.11) .. controls (375.19,268.2) and (383.19,246.6) .. (373.67,240.45) ;
\draw    (280.77,209.16) .. controls (283.72,208.67) and (283.24,209.85) .. (296.79,209.11) ;
\draw    (285.09,303.34) .. controls (292.54,302.67) and (290.54,302.79) .. (296.93,302.56) ;
\draw  [draw opacity=0][pattern=_l1xqy0hsy,pattern size=3pt,pattern thickness=0.75pt,pattern radius=0pt, pattern color={rgb, 255:red, 255; green, 0; blue, 0}] (299.53,224.19) .. controls (299.53,224.19) and (299.53,224.19) .. (299.53,224.19) .. controls (299.53,224.19) and (299.53,224.19) .. (299.53,224.19) .. controls (299.53,232.66) and (298.54,239.53) .. (297.31,239.53) .. controls (296.09,239.53) and (295.1,232.66) .. (295.1,224.19) .. controls (295.1,215.71) and (296.09,208.85) .. (297.31,208.85) .. controls (298.53,208.85) and (299.53,215.69) .. (299.53,224.15) -- (297.31,224.19) -- cycle ; \draw  [color={rgb, 255:red, 255; green, 0; blue, 0 }  ,draw opacity=1 ] (299.53,224.19) .. controls (299.53,224.19) and (299.53,224.19) .. (299.53,224.19) .. controls (299.53,224.19) and (299.53,224.19) .. (299.53,224.19) .. controls (299.53,232.66) and (298.54,239.53) .. (297.31,239.53) .. controls (296.09,239.53) and (295.1,232.66) .. (295.1,224.19) .. controls (295.1,215.71) and (296.09,208.85) .. (297.31,208.85) .. controls (298.53,208.85) and (299.53,215.69) .. (299.53,224.15) ;  
\draw  [draw opacity=0][pattern=_b8n76fz9o,pattern size=3pt,pattern thickness=0.75pt,pattern radius=0pt, pattern color={rgb, 255:red, 255; green, 0; blue, 0}] (299.53,287.43) .. controls (299.53,287.43) and (299.53,287.43) .. (299.53,287.43) .. controls (299.53,295.78) and (298.54,302.56) .. (297.31,302.56) .. controls (296.09,302.56) and (295.1,295.78) .. (295.1,287.43) .. controls (295.1,279.07) and (296.09,272.3) .. (297.31,272.3) .. controls (298.53,272.3) and (299.53,279.05) .. (299.53,287.39) -- (297.31,287.43) -- cycle ; \draw  [color={rgb, 255:red, 255; green, 0; blue, 0 }  ,draw opacity=1 ] (299.53,287.43) .. controls (299.53,287.43) and (299.53,287.43) .. (299.53,287.43) .. controls (299.53,295.78) and (298.54,302.56) .. (297.31,302.56) .. controls (296.09,302.56) and (295.1,295.78) .. (295.1,287.43) .. controls (295.1,279.07) and (296.09,272.3) .. (297.31,272.3) .. controls (298.53,272.3) and (299.53,279.05) .. (299.53,287.39) ;  
\draw  [draw opacity=0][pattern=_vp9iv54mp,pattern size=3pt,pattern thickness=0.75pt,pattern radius=0pt, pattern color={rgb, 255:red, 255; green, 0; blue, 0}] (356.62,223.04) .. controls (356.62,223.04) and (356.62,223.04) .. (356.62,223.04) .. controls (356.62,231.37) and (355.63,238.13) .. (354.4,238.13) .. controls (353.18,238.13) and (352.19,231.37) .. (352.19,223.04) .. controls (352.19,214.7) and (353.18,207.94) .. (354.4,207.94) .. controls (355.63,207.94) and (356.62,214.68) .. (356.62,223) -- (354.4,223.04) -- cycle ; \draw  [color={rgb, 255:red, 255; green, 0; blue, 0 }  ,draw opacity=1 ] (356.62,223.04) .. controls (356.62,223.04) and (356.62,223.04) .. (356.62,223.04) .. controls (356.62,231.37) and (355.63,238.13) .. (354.4,238.13) .. controls (353.18,238.13) and (352.19,231.37) .. (352.19,223.04) .. controls (352.19,214.7) and (353.18,207.94) .. (354.4,207.94) .. controls (355.63,207.94) and (356.62,214.68) .. (356.62,223) ;  
\draw  [draw opacity=0][pattern=_15z7agijo,pattern size=3pt,pattern thickness=0.75pt,pattern radius=0pt, pattern color={rgb, 255:red, 255; green, 0; blue, 0}] (357.35,286.13) .. controls (357.35,286.13) and (357.35,286.13) .. (357.35,286.13) .. controls (357.35,294.46) and (356.35,301.22) .. (355.13,301.22) .. controls (353.91,301.22) and (352.92,294.46) .. (352.92,286.13) .. controls (352.92,277.79) and (353.91,271.03) .. (355.13,271.03) .. controls (356.35,271.03) and (357.34,277.77) .. (357.35,286.09) -- (355.13,286.13) -- cycle ; \draw  [color={rgb, 255:red, 255; green, 0; blue, 0 }  ,draw opacity=1 ] (357.35,286.13) .. controls (357.35,286.13) and (357.35,286.13) .. (357.35,286.13) .. controls (357.35,294.46) and (356.35,301.22) .. (355.13,301.22) .. controls (353.91,301.22) and (352.92,294.46) .. (352.92,286.13) .. controls (352.92,277.79) and (353.91,271.03) .. (355.13,271.03) .. controls (356.35,271.03) and (357.34,277.77) .. (357.35,286.09) ;  
\draw    (250.82,68.29) .. controls (260.09,12.86) and (283.6,38) .. (309.72,33.31) ;
\draw    (250.82,68.29) .. controls (238.6,115.91) and (258.4,155.72) .. (275.26,132.3) ;
\draw    (275.26,132.3) .. controls (292.11,108.88) and (296.41,127) .. (314.04,127.5) ;
\draw    (266.58,62.67) .. controls (271.46,57.98) and (276.69,60.48) .. (280.73,61.1) ;
\draw    (263.46,59.62) .. controls (266.83,65.09) and (276.01,71.72) .. (282.42,58.84) ;
\draw    (262.31,112.23) .. controls (265.27,103.44) and (270.39,100.62) .. (274.02,97.28) ;
\draw    (258.85,112.55) .. controls (263.15,114.12) and (272.74,111.13) .. (274.87,93.68) ;
\draw    (325.77,63.79) .. controls (325.26,63.79) and (341.64,63.95) .. (345.12,66.1) ;
\draw    (325.74,33.26) .. controls (339.37,33.06) and (340.9,32.82) .. (344.78,33) ;
\draw    (341.92,95.77) .. controls (342.9,95.88) and (325.94,96.15) .. (325.86,96.18) ;
\draw    (343.94,127) .. controls (340.55,126.23) and (328.31,126.35) .. (325.88,126.71) ;
\draw    (285.94,89.18) .. controls (288.75,80.42) and (293.9,77.42) .. (297.51,73.99) ;
\draw    (282.42,89.65) .. controls (286.85,90.99) and (296.54,87.62) .. (298.3,70.41) ;
\draw    (402.22,93.05) .. controls (427.71,125.75) and (366.92,134.61) .. (343.94,127) ;
\draw    (402.22,93.05) .. controls (390.63,78.59) and (390.8,13.33) .. (377.36,29.6) ;
\draw    (377.36,29.6) .. controls (360.66,53.41) and (362.41,33.09) .. (344.78,33) ;
\draw    (372.89,74.01) .. controls (378.36,73.02) and (382.71,78.94) .. (386.34,82.31) ;
\draw    (370.6,69.03) .. controls (372.63,76.43) and (379.83,88.91) .. (388.36,81.36) ;
\draw    (325.77,63.79) .. controls (307.94,62.65) and (302.14,96.45) .. (325.86,96.18) ;
\draw    (341.92,95.77) .. controls (346.64,93.85) and (354.64,72.25) .. (345.12,66.1) ;
\draw  [draw opacity=0][dash pattern={on 3pt off 3pt}][line width=0.75]  (326.07,63.62) .. controls (326.86,61.76) and (327.44,55.78) .. (327.44,48.7) .. controls (327.44,40.68) and (326.7,34.08) .. (325.74,33.26) -- (325.54,48.7) -- cycle ; \draw  [color={rgb, 255:red, 0; green, 0; blue, 255}  ,draw opacity=1 ][dash pattern={on 3pt off 3pt}][line width=0.75]  (326.07,63.62) .. controls (326.86,61.76) and (327.44,55.78) .. (327.44,48.7) .. controls (327.44,40.68) and (326.7,34.08) .. (325.74,33.26) ;  
\draw  [draw opacity=0][dash pattern={on 3pt off 3pt}][line width=0.75]  (326.18,126.55) .. controls (326.98,124.68) and (327.56,118.71) .. (327.56,111.62) .. controls (327.56,103.61) and (326.81,97.01) .. (325.86,96.18) -- (325.65,111.62) -- cycle ; \draw  [color={rgb, 255:red, 0; green, 0; blue, 255}  ,draw opacity=1 ][dash pattern={on 3pt off 3pt}][line width=0.75]  (326.18,126.55) .. controls (326.98,124.68) and (327.56,118.71) .. (327.56,111.62) .. controls (327.56,103.61) and (326.81,97.01) .. (325.86,96.18) ;  
\draw    (309.72,33.31) .. controls (312.67,32.82) and (312.19,34) .. (325.74,33.26) ;
\draw    (314.04,127.5) .. controls (321.49,126.82) and (319.49,126.94) .. (325.88,126.71) ;
\draw  [draw opacity=0] (325.88,126.71) .. controls (324.53,125.36) and (323.52,119.04) .. (323.52,111.45) .. controls (323.52,104.62) and (324.34,98.82) .. (325.49,96.74) -- (326.42,111.45) -- cycle ; \draw  [color={rgb, 255:red, 0; green, 0; blue, 255}  ,draw opacity=1 ] (325.88,126.71) .. controls (324.53,125.36) and (323.52,119.04) .. (323.52,111.45) .. controls (323.52,104.62) and (324.34,98.82) .. (325.49,96.74) ;  
\draw  [draw opacity=0] (325.77,63.79) .. controls (324.42,62.44) and (323.4,56.12) .. (323.4,48.53) .. controls (323.4,41.69) and (324.23,35.89) .. (325.37,33.82) -- (326.31,48.53) -- cycle ; \draw  [color={rgb, 255:red, 0; green, 0; blue, 255}  ,draw opacity=1 ] (325.77,63.79) .. controls (324.42,62.44) and (323.4,56.12) .. (323.4,48.53) .. controls (323.4,41.69) and (324.23,35.89) .. (325.37,33.82) ;  

\draw [line width=1.5]    (327.69,141.02) -- (327.69,195.61) ;
\draw [shift={(327.69,199.61)}, rotate = 269.99] [fill={rgb, 255:red, 0; green, 0; blue, 0 }  ][line width=0.08]  [draw opacity=0] (11.07,-5.32) -- (0,0) -- (11.07,5.32) -- (7.35,0) -- cycle    ;

\draw    (87.51,443.89) .. controls (96.78,388.46) and (120.29,413.6) .. (146.41,408.91) ;
\draw    (87.51,443.89) .. controls (75.29,491.51) and (95.09,531.32) .. (111.95,507.9) ;
\draw    (111.95,507.9) .. controls (128.8,484.48) and (133.1,502.6) .. (150.73,503.1) ;
\draw    (103.27,438.27) .. controls (108.15,433.58) and (113.38,436.08) .. (117.42,436.7) ;
\draw    (100.15,435.22) .. controls (103.52,440.69) and (112.71,447.32) .. (119.11,434.44) ;
\draw    (99,487.83) .. controls (101.96,479.04) and (107.08,476.22) .. (110.71,472.88) ;
\draw    (95.54,488.15) .. controls (99.85,489.72) and (109.43,486.73) .. (111.56,469.28) ;
\draw    (162.46,439.39) .. controls (161.95,439.39) and (178.33,439.55) .. (181.81,441.7) ;
\draw    (162.44,408.86) .. controls (176.06,408.66) and (177.59,408.42) .. (181.47,408.6) ;
\draw    (178.61,471.37) .. controls (179.59,471.48) and (162.63,471.75) .. (162.55,471.78) ;
\draw    (180.63,502.6) .. controls (177.24,501.83) and (165,501.95) .. (162.57,502.31) ;
\draw    (122.63,464.78) .. controls (125.44,456.02) and (130.59,453.02) .. (134.2,449.59) ;
\draw    (119.12,465.25) .. controls (123.54,466.59) and (133.23,463.22) .. (134.99,446.01) ;
\draw    (238.92,468.65) .. controls (264.4,501.35) and (203.61,510.21) .. (180.63,502.6) ;
\draw    (238.92,468.65) .. controls (227.32,454.19) and (227.49,388.93) .. (214.05,405.2) ;
\draw    (214.05,405.2) .. controls (197.35,429.01) and (199.1,408.69) .. (181.47,408.6) ;
\draw    (209.58,449.61) .. controls (215.05,448.62) and (219.4,454.54) .. (223.03,457.91) ;
\draw    (207.29,444.63) .. controls (209.33,452.03) and (216.52,464.51) .. (225.05,456.96) ;
\draw    (162.46,439.39) .. controls (144.63,438.25) and (138.83,472.05) .. (162.55,471.78) ;
\draw    (178.61,471.37) .. controls (183.33,469.45) and (191.33,447.85) .. (181.81,441.7) ;
\draw  [draw opacity=0][dash pattern={on 3pt off 3pt}][line width=0.75]  (162.76,439.22) .. controls (163.55,437.36) and (164.13,431.38) .. (164.13,424.3) .. controls (164.13,416.28) and (163.39,409.68) .. (162.44,408.86) -- (162.23,424.3) -- cycle ; \draw  [color={rgb, 255:red, 0; green, 0; blue, 255}  ,draw opacity=1 ][dash pattern={on 3pt off 3pt}][line width=0.75]  (162.76,439.22) .. controls (163.55,437.36) and (164.13,431.38) .. (164.13,424.3) .. controls (164.13,416.28) and (163.39,409.68) .. (162.44,408.86) ;  
\draw  [draw opacity=0][dash pattern={on 3pt off 3pt}][line width=0.75]  (162.87,502.15) .. controls (163.67,500.28) and (164.25,494.31) .. (164.25,487.22) .. controls (164.25,479.21) and (163.5,472.61) .. (162.55,471.78) -- (162.35,487.22) -- cycle ; \draw  [color={rgb, 255:red, 0; green, 0; blue, 255} ,draw opacity=1 ][dash pattern={on 3pt off 3pt}][line width=0.75]  (162.87,502.15) .. controls (163.67,500.28) and (164.25,494.31) .. (164.25,487.22) .. controls (164.25,479.21) and (163.5,472.61) .. (162.55,471.78) ;  
\draw    (146.41,408.91) .. controls (149.36,408.42) and (148.89,409.6) .. (162.44,408.86) ;
\draw    (150.73,503.1) .. controls (158.18,502.42) and (156.18,502.54) .. (162.57,502.31) ;
\draw  [draw opacity=0] (162.57,502.31) .. controls (161.23,500.96) and (160.21,494.64) .. (160.21,487.05) .. controls (160.21,480.22) and (161.03,474.42) .. (162.18,472.34) -- (163.11,487.05) -- cycle ; \draw  [color={rgb, 255:red, 0; green, 0; blue, 255}  ,draw opacity=1 ] (162.57,502.31) .. controls (161.23,500.96) and (160.21,494.64) .. (160.21,487.05) .. controls (160.21,480.22) and (161.03,474.42) .. (162.18,472.34) ;  
\draw  [draw opacity=0] (162.46,439.39) .. controls (161.11,438.04) and (160.1,431.72) .. (160.1,424.13) .. controls (160.1,417.29) and (160.92,411.49) .. (162.07,409.42) -- (163,424.13) -- cycle ; \draw  [color={rgb, 255:red, 0; green, 0; blue, 255}  ,draw opacity=1 ] (162.46,439.39) .. controls (161.11,438.04) and (160.1,431.72) .. (160.1,424.13) .. controls (160.1,417.29) and (160.92,411.49) .. (162.07,409.42) ;  

\draw    (526.23,410.11) .. controls (539.85,409.91) and (541.68,409.67) .. (545.56,409.85) ;
\draw    (385.62,441.39) .. controls (394.89,385.96) and (418.4,411.1) .. (444.53,406.41) ;
\draw    (385.62,441.39) .. controls (373.4,489.01) and (393.2,528.82) .. (410.06,505.4) ;
\draw    (410.06,505.4) .. controls (426.91,481.98) and (431.21,500.1) .. (448.84,500.6) ;
\draw    (401.38,435.77) .. controls (406.27,431.08) and (411.49,433.58) .. (415.54,434.2) ;
\draw    (398.26,432.72) .. controls (401.63,438.19) and (410.82,444.82) .. (417.22,431.94) ;
\draw    (397.11,485.33) .. controls (400.08,476.54) and (405.2,473.72) .. (408.82,470.38) ;
\draw    (393.65,485.65) .. controls (397.96,487.22) and (407.54,484.23) .. (409.67,466.78) ;
\draw    (526.54,440.64) .. controls (526.04,440.64) and (542.42,440.8) .. (545.89,442.95) ;
\draw    (542.69,472.62) .. controls (543.68,472.73) and (526.72,473) .. (526.63,473.03) ;
\draw    (544.71,503.85) .. controls (541.32,503.08) and (529.09,503.2) .. (526.65,503.56) ;
\draw    (420.74,462.28) .. controls (423.56,453.52) and (428.7,450.52) .. (432.31,447.09) ;
\draw    (417.23,462.75) .. controls (421.65,464.09) and (431.34,460.72) .. (433.1,443.51) ;
\draw    (603,469.9) .. controls (628.48,502.6) and (567.7,511.46) .. (544.71,503.85) ;
\draw    (603,469.9) .. controls (591.4,455.44) and (591.57,390.18) .. (578.13,406.45) ;
\draw    (578.13,406.45) .. controls (561.44,430.26) and (563.19,409.94) .. (545.56,409.85) ;
\draw    (573.66,450.86) .. controls (579.14,449.87) and (583.49,455.79) .. (587.12,459.16) ;
\draw    (571.38,445.88) .. controls (573.41,453.28) and (580.61,465.76) .. (589.14,458.21) ;
\draw    (460.57,436.89) .. controls (442.74,435.75) and (436.94,469.55) .. (460.66,469.28) ;
\draw    (542.69,472.62) .. controls (547.42,470.7) and (555.42,449.1) .. (545.89,442.95) ;
\draw  [draw opacity=0][dash pattern={on 3pt off 3pt}][line width=0.75]  (526.84,440.47) .. controls (527.64,438.61) and (528.22,432.63) .. (528.22,425.55) .. controls (528.22,417.53) and (527.48,410.93) .. (526.52,410.11) -- (526.32,425.55) -- cycle ; \draw  [color={rgb, 255:red, 0; green, 0; blue, 255} ,draw opacity=1 ][dash pattern={on 3pt off 3pt}][line width=0.75]  (526.84,440.47) .. controls (527.64,438.61) and (528.22,432.63) .. (528.22,425.55) .. controls (528.22,417.53) and (527.48,410.93) .. (526.52,410.11) ;  
\draw  [draw opacity=0][dash pattern={on 3pt off 3pt}][line width=0.75]  (526.96,503.4) .. controls (527.75,501.53) and (528.33,495.56) .. (528.33,488.47) .. controls (528.33,480.46) and (527.59,473.86) .. (526.63,473.03) -- (526.43,488.47) -- cycle ; \draw  [color={rgb, 255:red, 0; green, 0; blue, 255} ,draw opacity=1 ][dash pattern={on 3pt off 3pt}][line width=0.75]  (526.96,503.4) .. controls (527.75,501.53) and (528.33,495.56) .. (528.33,488.47) .. controls (528.33,480.46) and (527.59,473.86) .. (526.63,473.03) ;  
\draw    (444.53,406.41) .. controls (447.47,405.92) and (447,407.1) .. (460.55,406.36) ;
\draw    (448.84,500.6) .. controls (456.29,499.92) and (454.29,500.04) .. (460.68,499.81) ;
\draw  [draw opacity=0] (526.65,503.56) .. controls (525.31,502.21) and (524.29,495.89) .. (524.29,488.3) .. controls (524.29,481.47) and (525.12,475.67) .. (526.26,473.59) -- (527.19,488.3) -- cycle ; \draw  [color={rgb, 255:red, 0; green, 0; blue, 255}  ,draw opacity=1 ] (526.65,503.56) .. controls (525.31,502.21) and (524.29,495.89) .. (524.29,488.3) .. controls (524.29,481.47) and (525.12,475.67) .. (526.26,473.59) ;  
\draw  [draw opacity=0] (526.54,440.64) .. controls (525.2,439.29) and (524.18,432.97) .. (524.18,425.38) .. controls (524.18,418.54) and (525,412.74) .. (526.15,410.67) -- (527.08,425.38) -- cycle ; \draw  [color={rgb, 255:red, 0; green, 0; blue, 255}  ,draw opacity=1 ] (526.54,440.64) .. controls (525.2,439.29) and (524.18,432.97) .. (524.18,425.38) .. controls (524.18,418.54) and (525,412.74) .. (526.15,410.67) ;  
\draw  [draw opacity=0][dash pattern={on 3pt off 3pt}][line width=0.75]  (460.21,436.72) .. controls (461,434.86) and (461.58,428.88) .. (461.58,421.8) .. controls (461.58,413.78) and (460.84,407.18) .. (459.88,406.36) -- (459.68,421.8) -- cycle ; \draw  [color={rgb, 255:red, 0; green, 0; blue, 255} ,draw opacity=1 ][dash pattern={on 3pt off 3pt}][line width=0.75]  (460.21,436.72) .. controls (461,434.86) and (461.58,428.88) .. (461.58,421.8) .. controls (461.58,413.78) and (460.84,407.18) .. (459.88,406.36) ;  
\draw  [draw opacity=0] (459.9,436.89) .. controls (458.56,435.54) and (457.54,429.22) .. (457.54,421.63) .. controls (457.54,414.79) and (458.37,408.99) .. (459.51,406.92) -- (460.44,421.63) -- cycle ; \draw  [color={rgb, 255:red, 0; green, 0; blue, 255}  ,draw opacity=1 ] (459.9,436.89) .. controls (458.56,435.54) and (457.54,429.22) .. (457.54,421.63) .. controls (457.54,414.79) and (458.37,408.99) .. (459.51,406.92) ;  
\draw  [draw opacity=0][dash pattern={on 3pt off 3pt}][line width=0.75]  (460.21,499.61) .. controls (461,497.75) and (461.58,491.77) .. (461.58,484.69) .. controls (461.58,476.67) and (460.84,470.07) .. (459.88,469.25) -- (459.68,484.69) -- cycle ; \draw  [color={rgb, 255:red, 0; green, 0; blue, 255}  ,draw opacity=1 ][dash pattern={on 3pt off 3pt}][line width=0.75]  (460.21,499.61) .. controls (461,497.75) and (461.58,491.77) .. (461.58,484.69) .. controls (461.58,476.67) and (460.84,470.07) .. (459.88,469.25) ;  
\draw  [draw opacity=0] (459.9,499.77) .. controls (458.56,498.42) and (457.54,492.11) .. (457.54,484.51) .. controls (457.54,477.68) and (458.37,471.88) .. (459.51,469.8) -- (460.44,484.51) -- cycle ; \draw  [color={rgb, 255:red, 0; green, 0; blue, 255} ,draw opacity=1 ] (459.9,499.77) .. controls (458.56,498.42) and (457.54,492.11) .. (457.54,484.51) .. controls (457.54,477.68) and (458.37,471.88) .. (459.51,469.8) ;  
\draw  [draw opacity=0] (459.22,499.8) .. controls (459.39,499.81) and (459.56,499.81) .. (459.73,499.81) .. controls (475.65,499.81) and (488.55,478.89) .. (488.55,453.08) .. controls (488.55,427.36) and (475.73,406.49) .. (459.88,406.36) -- (459.73,453.08) -- cycle ; \draw  [color={rgb, 255:red, 0; green, 0; blue, 255} ,draw opacity=1 ] (459.22,499.8) .. controls (459.39,499.81) and (459.56,499.81) .. (459.73,499.81) .. controls (475.65,499.81) and (488.55,478.89) .. (488.55,453.08) .. controls (488.55,427.36) and (475.73,406.49) .. (459.88,406.36) ;  
\draw  [draw opacity=0] (459.51,469.37) .. controls (463.86,468.97) and (467.33,461.84) .. (467.33,453.1) .. controls (467.33,444.63) and (464.07,437.66) .. (459.9,436.89) -- (459.12,453.1) -- cycle ; \draw  [color={rgb, 255:red, 0; green, 0; blue, 255}  ,draw opacity=1 ] (459.51,469.37) .. controls (463.86,468.97) and (467.33,461.84) .. (467.33,453.1) .. controls (467.33,444.63) and (464.07,437.66) .. (459.9,436.89) ;  
\draw  [draw opacity=0] (526.6,473.12) .. controls (522.25,472.72) and (518.78,465.59) .. (518.78,456.85) .. controls (518.78,448.38) and (522.04,441.41) .. (526.2,440.64) -- (526.98,456.85) -- cycle ; \draw  [color={rgb, 255:red, 0; green, 0; blue, 255}  ,draw opacity=1 ] (526.6,473.12) .. controls (522.25,472.72) and (518.78,465.59) .. (518.78,456.85) .. controls (518.78,448.38) and (522.04,441.41) .. (526.2,440.64) ;  
\draw  [draw opacity=0] (526.89,503.55) .. controls (526.72,503.56) and (526.55,503.56) .. (526.37,503.56) .. controls (510.46,503.56) and (497.56,482.64) .. (497.56,456.83) .. controls (497.56,431.11) and (510.38,410.24) .. (526.23,410.11) -- (526.37,456.83) -- cycle ; \draw  [color={rgb, 255:red, 0; green, 0; blue, 255} ,draw opacity=1 ] (526.89,503.55) .. controls (526.72,503.56) and (526.55,503.56) .. (526.37,503.56) .. controls (510.46,503.56) and (497.56,482.64) .. (497.56,456.83) .. controls (497.56,431.11) and (510.38,410.24) .. (526.23,410.11) ;  
\draw [line width=1.5]    (278.47,456.38) -- (354.97,456.38) ;
\draw [shift={(358.97,456.38)}, rotate = 180] [fill={rgb, 255:red, 0; green, 0; blue, 0 }  ][line width=0.08]  [draw opacity=0] (11.07,-5.32) -- (0,0) -- (11.07,5.32) -- (7.35,0) -- cycle    ;
\draw [line width=1.5]    (355.76,320.03) -- (475.29,378.25) ;
\draw [shift={(478.89,380.01)}, rotate = 205.97] [fill={rgb, 255:red, 0; green, 0; blue, 0 }  ][line width=0.08]  [draw opacity=0] (11.07,-5.32) -- (0,0) -- (11.07,5.32) -- (7.35,0) -- cycle    ;
\draw [line width=1.5]    (296.49,320.83) -- (176.95,379.05) ;
\draw [shift={(173.36,380.81)}, rotate = 334.03] [fill={rgb, 255:red, 0; green, 0; blue, 0 }  ][line width=0.08]  [draw opacity=0] (11.07,-5.32) -- (0,0) -- (11.07,5.32) -- (7.35,0) -- cycle    ;


\node[anchor=north west] at (35,516) {%
  \(\displaystyle 
    \int \mathcal{D}\phi_{1}\,\mathcal{D}\phi_{2}\;
          \psi_{L}[\phi_{1},\phi_{2}]\,
          \psi_{R}[\phi_{1},\phi_{2}]
  \)};

\node[anchor=north west] at (370,516) {%
  \(\displaystyle 
    \int \mathcal{D}\phi_{1}\,\mathcal{D}\phi_{3}\;
          \psi_{L}[\phi_{1},\phi_{1}]\,
          \psi_{R}[\phi_{3},\phi_{3}]
  \)};

\node[anchor=north west, font=\huge] at (306,420) {\textbf{?}};

\node[anchor=north west, font=\scriptsize, rotate=-26] at (376,303) {%
  \(\phi_{1}\!\sim\!\phi_{2},\;
    \phi_{3}\!\sim\!\phi_{4}\)};

\node[anchor=north west, font=\scriptsize, rotate=26] at (176,353) {%
  \(\phi_{1}\!\sim\!\phi_{3},\;
    \phi_{2}\!\sim\!\phi_{4}\)};

\node[anchor=north west, font=\scriptsize] at (297,214) {\(\phi_{1}\)};
\node[anchor=north west, font=\scriptsize] at (329,214) {\(\phi_{3}\)};
\node[anchor=north west, font=\scriptsize] at (297,277) {\(\phi_{2}\)};
\node[anchor=north west, font=\scriptsize] at (329,277) {\(\phi_{4}\)};

\node[anchor=north west] at (128,245) 
  { \(\displaystyle \psi_{L}[\phi_{1},\phi_{2}]\)};
\node[anchor=north west] at (440,245) 
  { \(\displaystyle \psi_{R}[\phi_{3},\phi_{4}]\)};

\end{tikzpicture}

\caption{Schematic representation of the kind of manifold surgery we are interested in. The manifold is cut in two across two identical surfaces with independent boundary conditions $\phi_i$ being placed on each of the boundaries resulting from the cut. The path integrals over the disconnected pieces represent functionals $\psi_{L}[\phi_{1},\phi_{2}]$ and $\psi_{R}[\phi_{3},\phi_{4}]$ of the boundary conditions we place on them. By picking which boundary conditions are identified, we can return to the original manifold or produce two disconnected pieces. The blue parts of the disconnected manifolds should not be interpreted as extra additions the manifolds; instead, they simply represent an identification of their ends. Our main task is to investigate ways of going directly from the original manifold to the disconnected pieces.}
\label{surgery1}
\end{figure}

An example of the general problem we are interested in can be described as follows: suppose we were given a path integral that is performed over a Euclidean manifold, with the theory being a $d+1$-dimensional local field theory of some real scalar fields that we collectively package into a ``vector'' that we denote by $\Phi$.\footnote{In that sense, each field corresponds to a different component of $\Phi$ with the goal here being merely to have a more compact notation. In particular, we make no assumption of any form of symmetry corresponding to rotating these components into each other.} We would like to split the manifold into two pieces such that each piece can be re-glued to itself to form two disconnected manifolds without introducing new boundaries that did not exist in the original manifold.  Typically, the way one would perform such an operation would be to explicitly disconnect the manifold at the surfaces of interest, with independent boundary conditions being placed on either side of the cuts. The path integrals over the disconnected pieces can be then thought of as functionals of the boundary conditions we place on them. The boundary conditions of the surfaces we wish to glue are then identified and summed over all possible choices. Depending on which boundary conditions are identified, we can glue the disconnected pieces in different ways with two possible results being reproducing the original manifold or producing two disconnected pieces multiplied together. A schematic representation of such an operation is shown in fig. \ref{surgery1}.\footnote{For a more detailed discussion of cutting and gluing in local quantum field theory see \cite{Dedushenko:2018aox, Dedushenko:2018tgx} and references therein.} The question now is: can we find a way to take us directly from the original manifold to the two disconnected pieces without having to explicitly cut and glue the manifold?

Following in the spirit of section \ref{sec:simple}, we will focus on operator insertions that we can place inside the path integral, in the hope that performing some form of averaging over them can implement the surgery operation for us.  Once again, we observe that gluing surfaces is the more straightforward task as we just need to introduce Dirac delta functionals in a convenient representation. For example, if we imagine that we were able to break the path integral over the original manifold $\mathcal{M}$ shown in fig. \ref{surgery1} into two pieces, $\mathcal{M}_L$ and $\mathcal{M}_R$, with all possible boundary conditions integrated over independently such that we have have a path integral evaluating the expression\footnote{Boundary conditions for $\Phi$ which live on codimension-one surfaces are denoted by the lower case $\phi$.}
\ben
\int \mathcal{D}\phi_{1}\,\mathcal{D}\phi_{2}\,\mathcal{D}\phi_{3}\,\mathcal{D}\phi_{4}\;
          \psi_{L}[\phi_{1},\phi_{2}]\,
          \psi_{R}[\phi_{3},\phi_{4}],
\een
with
\begin{equation*}
\psi_{L}[\phi_{1},\phi_{2}] = \int_{\substack{\Phi|_{\Sigma_{1}}=\phi_{1}\\\Phi|_{\Sigma_{2}}=\phi_{2}}}
\,\mathcal{D}\Phi\;e^{-S[\Phi;\mathcal{M}_L]},
\end{equation*}
\begin{equation*}
\psi_{R}[\phi_{3},\phi_{4}] = \int_{\substack{\Phi|_{\Sigma_{3}}=\phi_{3}\\\Phi|_{\Sigma_{4}}=\phi_{4}}}
\,\mathcal{D}\Phi\;e^{-S[\Phi;\mathcal{M}_R]},
\end{equation*}
then we can get the path integral corresponding to gluing each piece to itself by introducing Dirac delta functionals in the form of some auxiliary sources integrals as follows 
\begin{align*}
    \int \mathcal{D}J_1 \, \mathcal{D}J_2  \int \mathcal{D}\phi_{1}\,\mathcal{D}\phi_{2}\,\mathcal{D}\phi_{3}\,\mathcal{D}\phi_{4} \,& \psi_{L}[\phi_{1},\phi_{2}] \,
          \psi_{R}[\phi_{3},\phi_{4}] \exp\Bigg\{2\pi i\int d^{d}x\sqrt{\gamma} \;  \\
        &~~~\times \bigg[J_1(x) \cdot \Big( \phi_1(x)-\phi_2(x)\Big) + J_2(x) \cdot \Big(\phi_3(x)-\phi_4(x)\Big)\bigg]\Bigg\},
\end{align*}
where $x$ here represents some coordinates that parametrize the codimension-one surfaces at the cuts, while $\gamma$ is the determinant of the induced metric on this surface. Just like with any functional integration, the functional integral measure will require some regularization and some choice of normalization. We should make sure that this choice satisfies expected Dirac delta functional properties like 
\begin{equation*}
\int \mathcal{D}J \int\mathcal{D}\phi_1\,\mathcal{D}\phi_2 \;\psi[\phi_1,\phi_2]\, e^{2\pi i\int d^{d}x\sqrt{\gamma} \, J(x) \cdot \big( \phi_1(x)-\phi_2(x)\big)} = \int \mathcal{D}\phi_1 \;\psi[\phi_1,\phi_1], 
\end{equation*}
for any functional $\psi[\phi_1,\phi_2]$.

Now, similar to section \ref{sec:simple}, we can think of achieving this by turning on some sources, linearly coupled to the fields $\Phi$, and localized on a codimension-one surface at the boundaries. The sources themselves are identified according to the way we wish to glue the manifold and by performing functional integrals over the remaining freedom in choosing the source profile, we effectively introduce Dirac delta functionals that implement the desired gluing. These would be the direct analogue of the Euclidean time ``kick" sources from section \ref{sec:simple}.

We can also take a more abstract approach and simply think of this as introducing an operator-valued functional $Q_L[J]$ in the path integral over $\mathcal{M}_L$, that for any choice of a function $J(x)$, satisfies
\ben \label{insertionL}
\int_{\substack{\Phi|_{\Sigma_{1}}=\phi_{1}\\\Phi|_{\Sigma_{2}}=\phi_{2}}}
\,\mathcal{D}\Phi\;e^{-S[\Phi;\mathcal{M}_L]}\,Q_L[J]=\psi_{L}[\phi_{1},\phi_{2}]\,e^{2\pi i\int d^{d}x\sqrt{\gamma} \, J(x) \cdot \big( \phi_1(x)-\phi_2(x)\big)},
\een
and similarly for $\mathcal{M}_R$, we introduce $Q_R[J]$, which satisfies
\ben \label{insertionR}
\int_{\substack{\Phi|_{\Sigma_{3}}=\phi_{3}\\\Phi|_{\Sigma_{4}}=\phi_{4}}}
\,\mathcal{D}\Phi\;e^{-S[\Phi;\mathcal{M}_R]}\,Q_R[J]=\psi_{R}[\phi_{3},\phi_{4}]\,e^{2\pi i\int d^{d}x\sqrt{\gamma}\, J(x) \cdot \big( \phi_3(x)-\phi_4(x)\big)}.
\een
 The most straightforward way to construct $Q_L[J]$ and $Q_R[J]$ explicitly is to think of inserting codimension-one operators of the form $e^{2\pi i\int_{\Sigma} \, J\Phi}$ right before the boundary surfaces in $\mathcal{M}_L$ and $\mathcal{M}_R$. This is identical to the surface sources picture we just discussed earlier and appears to suggest that \eqref{insertionL} and \eqref{insertionR} are just a trivial rewriting of these sources. In reality, the operator insertion picture is a lot less restrictive in how we need to construct $Q_L[J]$ and $Q_R[J]$ since we are only requiring that they satisfy \eqref{insertionL} and \eqref{insertionR} when acting inside the path integrals over $\mathcal{M}_L$ and $\mathcal{M}_R$, respectively. By contrast, the surface sources picture will satisfy the relations in \eqref{insertionL} and \eqref{insertionR} for any choice of manifold, and in any local theory with the same field content. In other words, the surface sources construction offers a generic way to glue the manifolds, while focusing on operator insertions satisfying \eqref{insertionL} and \eqref{insertionR} may offer more freedom by focusing on operators that are good enough to glue the path integrals at hand. 

Now we need to focus on how to break off the original path integral so that instead of having an expression evaluating 
\ben
\int \mathcal{D}\phi_{1}\,\mathcal{D}\phi_{2}\;
          \psi_{L}[\phi_{1},\phi_{2}]\,
          \psi_{R}[\phi_{1},\phi_{2}],
\een
we need something along the lines of 
\ben
\int \mathcal{D}\phi_{1}\,\mathcal{D}\phi_{2}\,\mathcal{D}\phi_{3}\,\mathcal{D}\phi_{4}\;
          \psi_{L}[\phi_{1},\phi_{2}]\,
          \psi_{R}[\phi_{3},\phi_{4}].
\een
In section \ref{sec:simple}, we achieved this by carefully thinking of the phase space path integral and inserting sources, to be later integrated over, that are linearly coupled to the momentum operator at the location where we wanted to disconnect the manifold. Here, we just look for an operator-valued functional $P[K_1,K_2]$ such that for any two functions $K_1(x)$ and $K_2(x)$, inserting $P[K_1,K_2]$ implements a boundary fields shift such that we have
\ben
\int_{\substack{\Phi|_{\Sigma_{1}}=\phi_{1}\\\Phi|_{\Sigma_{2}}=\phi_{2}}}
\,\mathcal{D}\Phi\;e^{-S[\Phi;\mathcal{M}_L]}\,Q_L[J] \, P[K_1,K_2]=\psi'_{L}[\phi_{1},\phi_{2}]\,e^{2\pi i\int d^{d}x\sqrt{\gamma} \, J(x) \cdot \big( \phi_1(x)-\phi_2(x)\big)},
\een
with 
\ben
\psi'_{L}[\phi_{1},\phi_{2}]=\psi_{L}[\phi_{1}+K_1,\phi_{2}+K_2].
\een
Another option is to instead demand that
\ben
\int_{\substack{\Phi|_{\Sigma_{1}}=\phi_{1}\\\Phi|_{\Sigma_{2}}=\phi_{2}}}
\,\mathcal{D}\Phi\;e^{-S[\Phi;\mathcal{M}_R]}\,Q_R[J] \, P[K_1,K_2]=\psi'_{R}[\phi_{3},\phi_{4}]\,e^{2\pi i\int d^{d}x\sqrt{\gamma} \, J(x) \cdot \big( \phi_3(x)-\phi_4(x)\big)},
\een
with 
\ben
\psi'_{R}[\phi_{3},\phi_{4}]=\psi_{R}[\phi_{3}+K_1,\phi_{4}+K_2].
\een
We can also think of looking for two different insertions instead of $P[K_1,K_2]$ with one boundary fields shift happening for one of the cuts through acting on $\psi_{L}[\phi_{1},\phi_{2}]$, while the necessary shift for the other cut is implemented through the second insertion acting on $\psi_{R}[\phi_{3},\phi_{4}]$, such that when inserted in $\mathcal{M}$ we get some combination like $\psi_{L}[\phi_{1}+K_1,\phi_{2}] \, \psi_{R}[\phi_{1},\phi_{2}+K_2]$ or $\psi_{L}[\phi_{1},\phi_{2}+K_1] \, \psi_{R}[\phi_{1}+K_2,\phi_{2}]$. Regardless of which route we choose to take, it is clear that we can now easily relate the Euclidean path integral over $\mathcal{M}$ to a product of the path integrals over the boundary-less versions of $\mathcal{M}_L$ and $\mathcal{M}_R$ where we have each manifold's two boundaries identified together
\begin{align}
   \int \mathcal{D}J_1\,\mathcal{D}J_2\,\mathcal{D}K_1\,\mathcal{D}K_2\int \,\mathcal{D}\Phi\,\;e^{-S[\Phi;\mathcal{M}]}\,Q_L[J_1] \, P[K_1,K_2]\, Q_R[J_2]=\int_{\substack{\Phi|_{\Sigma_{1}}=\Phi|_{\Sigma_{2}}}}
\,\mathcal{D}\Phi\,\;e^{-S[\Phi;\mathcal{M}_L]}\; \nonumber\\ \times  \int_{\substack{\Phi|_{\Sigma_{3}}=\Phi|_{\Sigma_{4}}}}
\,\mathcal{D}\Phi\,\;e^{-S[\Phi;\mathcal{M}_R]}.
\end{align}
At its heart, this is a rather simple identity that simply says that averaging over the right combination of operator insertions can effectively cut and glue the Euclidean path integral for us. Averaging over the phase multiplying operators $Q_L[J_1]$ and $Q_R[J_2]$ implements gluing by introducing Dirac delta functionals into the path integral, while averaging over the shift operators $P[K_1,K_2]$ implements the desired cutting. 

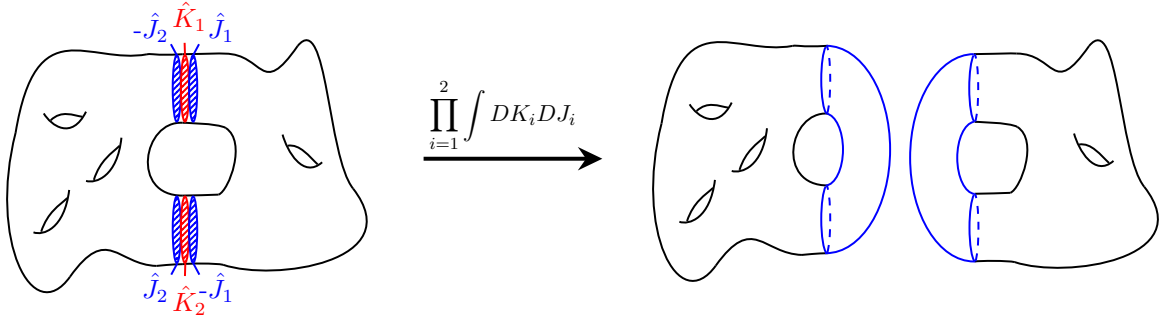
\begin{figure}[!t]
\centering
 
\tikzset{
pattern size/.store in=\mcSize, 
pattern size = 5pt,
pattern thickness/.store in=\mcThickness, 
pattern thickness = 0.3pt,
pattern radius/.store in=\mcRadius, 
pattern radius = 1pt}
\makeatletter
\pgfutil@ifundefined{pgf@pattern@name@_5ue7f3j79}{
\pgfdeclarepatternformonly[\mcThickness,\mcSize]{_5ue7f3j79}
{\pgfqpoint{0pt}{0pt}}
{\pgfpoint{\mcSize+\mcThickness}{\mcSize+\mcThickness}}
{\pgfpoint{\mcSize}{\mcSize}}
{
\pgfsetcolor{\tikz@pattern@color}
\pgfsetlinewidth{\mcThickness}
\pgfpathmoveto{\pgfqpoint{0pt}{0pt}}
\pgfpathlineto{\pgfpoint{\mcSize+\mcThickness}{\mcSize+\mcThickness}}
\pgfusepath{stroke}
}}
\makeatother

 
\tikzset{
pattern size/.store in=\mcSize, 
pattern size = 5pt,
pattern thickness/.store in=\mcThickness, 
pattern thickness = 0.3pt,
pattern radius/.store in=\mcRadius, 
pattern radius = 1pt}
\makeatletter
\pgfutil@ifundefined{pgf@pattern@name@_2fcvzd34z}{
\pgfdeclarepatternformonly[\mcThickness,\mcSize]{_2fcvzd34z}
{\pgfqpoint{0pt}{0pt}}
{\pgfpoint{\mcSize+\mcThickness}{\mcSize+\mcThickness}}
{\pgfpoint{\mcSize}{\mcSize}}
{
\pgfsetcolor{\tikz@pattern@color}
\pgfsetlinewidth{\mcThickness}
\pgfpathmoveto{\pgfqpoint{0pt}{0pt}}
\pgfpathlineto{\pgfpoint{\mcSize+\mcThickness}{\mcSize+\mcThickness}}
\pgfusepath{stroke}
}}
\makeatother

 
\tikzset{
pattern size/.store in=\mcSize, 
pattern size = 5pt,
pattern thickness/.store in=\mcThickness, 
pattern thickness = 0.3pt,
pattern radius/.store in=\mcRadius, 
pattern radius = 1pt}
\makeatletter
\pgfutil@ifundefined{pgf@pattern@name@_fux8la0xm}{
\pgfdeclarepatternformonly[\mcThickness,\mcSize]{_fux8la0xm}
{\pgfqpoint{0pt}{0pt}}
{\pgfpoint{\mcSize+\mcThickness}{\mcSize+\mcThickness}}
{\pgfpoint{\mcSize}{\mcSize}}
{
\pgfsetcolor{\tikz@pattern@color}
\pgfsetlinewidth{\mcThickness}
\pgfpathmoveto{\pgfqpoint{0pt}{0pt}}
\pgfpathlineto{\pgfpoint{\mcSize+\mcThickness}{\mcSize+\mcThickness}}
\pgfusepath{stroke}
}}
\makeatother

 
\tikzset{
pattern size/.store in=\mcSize, 
pattern size = 5pt,
pattern thickness/.store in=\mcThickness, 
pattern thickness = 0.3pt,
pattern radius/.store in=\mcRadius, 
pattern radius = 1pt}
\makeatletter
\pgfutil@ifundefined{pgf@pattern@name@_a968nvxni}{
\pgfdeclarepatternformonly[\mcThickness,\mcSize]{_a968nvxni}
{\pgfqpoint{0pt}{0pt}}
{\pgfpoint{\mcSize+\mcThickness}{\mcSize+\mcThickness}}
{\pgfpoint{\mcSize}{\mcSize}}
{
\pgfsetcolor{\tikz@pattern@color}
\pgfsetlinewidth{\mcThickness}
\pgfpathmoveto{\pgfqpoint{0pt}{0pt}}
\pgfpathlineto{\pgfpoint{\mcSize+\mcThickness}{\mcSize+\mcThickness}}
\pgfusepath{stroke}
}}
\makeatother

 
\tikzset{
pattern size/.store in=\mcSize, 
pattern size = 5pt,
pattern thickness/.store in=\mcThickness, 
pattern thickness = 0.3pt,
pattern radius/.store in=\mcRadius, 
pattern radius = 1pt}
\makeatletter
\pgfutil@ifundefined{pgf@pattern@name@_t98sjdqts}{
\pgfdeclarepatternformonly[\mcThickness,\mcSize]{_t98sjdqts}
{\pgfqpoint{0pt}{0pt}}
{\pgfpoint{\mcSize+\mcThickness}{\mcSize+\mcThickness}}
{\pgfpoint{\mcSize}{\mcSize}}
{
\pgfsetcolor{\tikz@pattern@color}
\pgfsetlinewidth{\mcThickness}
\pgfpathmoveto{\pgfqpoint{0pt}{0pt}}
\pgfpathlineto{\pgfpoint{\mcSize+\mcThickness}{\mcSize+\mcThickness}}
\pgfusepath{stroke}
}}
\makeatother

 
\tikzset{
pattern size/.store in=\mcSize, 
pattern size = 5pt,
pattern thickness/.store in=\mcThickness, 
pattern thickness = 0.3pt,
pattern radius/.store in=\mcRadius, 
pattern radius = 1pt}
\makeatletter
\pgfutil@ifundefined{pgf@pattern@name@_rp3sywncm}{
\pgfdeclarepatternformonly[\mcThickness,\mcSize]{_rp3sywncm}
{\pgfqpoint{0pt}{0pt}}
{\pgfpoint{\mcSize+\mcThickness}{\mcSize+\mcThickness}}
{\pgfpoint{\mcSize}{\mcSize}}
{
\pgfsetcolor{\tikz@pattern@color}
\pgfsetlinewidth{\mcThickness}
\pgfpathmoveto{\pgfqpoint{0pt}{0pt}}
\pgfpathlineto{\pgfpoint{\mcSize+\mcThickness}{\mcSize+\mcThickness}}
\pgfusepath{stroke}
}}
\makeatother
\tikzset{every picture/.style={line width=0.75pt}} 

\begin{tikzpicture}[x=0.75pt,y=0.75pt,yscale=-1.12,xscale=1.12]

\draw    (536.23,82.84) .. controls (549.85,82.64) and (551.68,82.41) .. (555.56,82.58) ;
\draw    (395.62,114.12) .. controls (404.89,58.69) and (428.4,83.83) .. (454.53,79.15) ;
\draw    (395.62,114.12) .. controls (383.4,161.74) and (403.2,201.56) .. (420.06,178.14) ;
\draw    (420.06,178.14) .. controls (436.91,154.72) and (441.21,172.83) .. (458.84,173.33) ;
\draw    (411.38,108.5) .. controls (416.27,103.82) and (421.49,106.31) .. (425.54,106.94) ;
\draw    (408.26,105.46) .. controls (411.63,110.92) and (420.82,117.56) .. (427.22,104.67) ;
\draw    (407.11,158.07) .. controls (410.08,149.27) and (415.2,146.45) .. (418.82,143.12) ;
\draw    (403.65,158.39) .. controls (407.96,159.95) and (417.54,156.96) .. (419.67,139.52) ;
\draw    (536.54,113.37) .. controls (536.04,113.37) and (552.42,113.54) .. (555.89,115.68) ;
\draw    (552.69,145.35) .. controls (553.68,145.46) and (536.72,145.74) .. (536.63,145.77) ;
\draw    (554.71,176.58) .. controls (551.32,175.82) and (539.09,175.93) .. (536.65,176.29) ;
\draw    (430.74,135.01) .. controls (433.56,126.25) and (438.7,123.25) .. (442.31,119.82) ;
\draw    (427.23,135.49) .. controls (431.65,136.82) and (441.34,133.46) .. (443.1,116.25) ;
\draw    (613,142.64) .. controls (638.48,175.33) and (577.7,184.2) .. (554.71,176.58) ;
\draw    (613,142.64) .. controls (601.4,128.18) and (601.57,62.91) .. (588.13,79.19) ;
\draw    (588.13,79.19) .. controls (571.44,102.99) and (573.19,82.68) .. (555.56,82.58) ;
\draw    (583.66,123.59) .. controls (589.14,122.61) and (593.49,128.52) .. (597.12,131.89) ;
\draw    (581.38,118.62) .. controls (583.41,126.01) and (590.61,138.5) .. (599.14,130.95) ;
\draw    (470.57,109.62) .. controls (452.74,108.49) and (446.94,142.29) .. (470.66,142.02) ;
\draw    (552.69,145.35) .. controls (557.42,143.44) and (565.42,121.84) .. (555.89,115.68) ;
\draw  [draw opacity=0][dash pattern={on 3pt off 3pt}][line width=0.75]  (536.84,113.21) .. controls (537.64,111.34) and (538.22,105.37) .. (538.22,98.28) .. controls (538.22,90.27) and (537.48,83.67) .. (536.52,82.84) -- (536.32,98.28) -- cycle ; \draw  [color={rgb, 255:red, 0; green, 0; blue, 255}  ,draw opacity=1 ][dash pattern={on 3pt off 3pt}][line width=0.75]  (536.84,113.21) .. controls (537.64,111.34) and (538.22,105.37) .. (538.22,98.28) .. controls (538.22,90.27) and (537.48,83.67) .. (536.52,82.84) ;  
\draw  [draw opacity=0][dash pattern={on 3pt off 3pt}][line width=0.75]  (536.96,176.13) .. controls (537.75,174.27) and (538.33,168.29) .. (538.33,161.21) .. controls (538.33,153.19) and (537.59,146.59) .. (536.63,145.77) -- (536.43,161.21) -- cycle ; \draw  [color={rgb, 255:red, 0; green, 0; blue, 255}  ,draw opacity=1 ][dash pattern={on 3pt off 3pt}][line width=0.75]  (536.96,176.13) .. controls (537.75,174.27) and (538.33,168.29) .. (538.33,161.21) .. controls (538.33,153.19) and (537.59,146.59) .. (536.63,145.77) ;  
\draw    (454.53,79.15) .. controls (457.47,78.66) and (457,79.83) .. (470.55,79.09) ;
\draw    (458.84,173.33) .. controls (466.29,172.66) and (464.29,172.77) .. (470.68,172.54) ;
\draw  [draw opacity=0] (536.65,176.29) .. controls (535.31,174.94) and (534.29,168.63) .. (534.29,161.03) .. controls (534.29,154.2) and (535.12,148.4) .. (536.26,146.32) -- (537.19,161.03) -- cycle ; \draw  [color={rgb, 255:red, 0; green, 0; blue, 255}  ,draw opacity=1 ] (536.65,176.29) .. controls (535.31,174.94) and (534.29,168.63) .. (534.29,161.03) .. controls (534.29,154.2) and (535.12,148.4) .. (536.26,146.32) ;  
\draw  [draw opacity=0] (536.54,113.37) .. controls (535.2,112.02) and (534.18,105.7) .. (534.18,98.11) .. controls (534.18,91.28) and (535,85.48) .. (536.15,83.4) -- (537.08,98.11) -- cycle ; \draw  [color={rgb, 255:red, 0; green, 0; blue, 255}  ,draw opacity=1 ] (536.54,113.37) .. controls (535.2,112.02) and (534.18,105.7) .. (534.18,98.11) .. controls (534.18,91.28) and (535,85.48) .. (536.15,83.4) ;  
\draw  [draw opacity=0][dash pattern={on 3pt off 3pt}][line width=0.75]  (470.21,109.46) .. controls (471,107.59) and (471.58,101.62) .. (471.58,94.53) .. controls (471.58,86.52) and (470.84,79.92) .. (469.88,79.09) -- (469.68,94.53) -- cycle ; \draw  [color={rgb, 255:red, 0; green, 0; blue, 255}  ,draw opacity=1 ][dash pattern={on 3pt off 3pt}][line width=0.75]  (470.21,109.46) .. controls (471,107.59) and (471.58,101.62) .. (471.58,94.53) .. controls (471.58,86.52) and (470.84,79.92) .. (469.88,79.09) ;  
\draw  [draw opacity=0] (469.9,109.62) .. controls (468.56,108.27) and (467.54,101.95) .. (467.54,94.36) .. controls (467.54,87.53) and (468.37,81.73) .. (469.51,79.65) -- (470.44,94.36) -- cycle ; \draw  [color={rgb, 255:red, 0; green, 0; blue, 255}  ,draw opacity=1 ] (469.9,109.62) .. controls (468.56,108.27) and (467.54,101.95) .. (467.54,94.36) .. controls (467.54,87.53) and (468.37,81.73) .. (469.51,79.65) ;  
\draw  [draw opacity=0][dash pattern={on 3pt off 3pt}][line width=0.75]  (470.21,172.35) .. controls (471,170.48) and (471.58,164.51) .. (471.58,157.42) .. controls (471.58,149.41) and (470.84,142.81) .. (469.88,141.98) -- (469.68,157.42) -- cycle ; \draw  [color={rgb, 255:red, 0; green, 0; blue, 255}  ,draw opacity=1 ][dash pattern={on 3pt off 3pt}][line width=0.75]  (470.21,172.35) .. controls (471,170.48) and (471.58,164.51) .. (471.58,157.42) .. controls (471.58,149.41) and (470.84,142.81) .. (469.88,141.98) ;  
\draw  [draw opacity=0] (469.9,172.51) .. controls (468.56,171.16) and (467.54,164.84) .. (467.54,157.25) .. controls (467.54,150.42) and (468.37,144.61) .. (469.51,142.54) -- (470.44,157.25) -- cycle ; \draw  [color={rgb, 255:red, 0; green, 0; blue, 255}  ,draw opacity=1 ] (469.9,172.51) .. controls (468.56,171.16) and (467.54,164.84) .. (467.54,157.25) .. controls (467.54,150.42) and (468.37,144.61) .. (469.51,142.54) ;  
\draw  [draw opacity=0] (469.22,172.54) .. controls (469.39,172.54) and (469.56,172.54) .. (469.73,172.54) .. controls (485.65,172.54) and (498.55,151.62) .. (498.55,125.82) .. controls (498.55,100.09) and (485.73,79.22) .. (469.88,79.09) -- (469.73,125.82) -- cycle ; \draw  [color={rgb, 255:red, 0; green, 0; blue, 255}  ,draw opacity=1 ] (469.22,172.54) .. controls (469.39,172.54) and (469.56,172.54) .. (469.73,172.54) .. controls (485.65,172.54) and (498.55,151.62) .. (498.55,125.82) .. controls (498.55,100.09) and (485.73,79.22) .. (469.88,79.09) ;  
\draw  [draw opacity=0] (469.51,142.1) .. controls (473.86,141.71) and (477.33,134.57) .. (477.33,125.83) .. controls (477.33,117.36) and (474.07,110.4) .. (469.9,109.62) -- (469.12,125.83) -- cycle ; \draw  [color={rgb, 255:red, 0; green, 0; blue, 255}  ,draw opacity=1 ] (469.51,142.1) .. controls (473.86,141.71) and (477.33,134.57) .. (477.33,125.83) .. controls (477.33,117.36) and (474.07,110.4) .. (469.9,109.62) ;  
\draw  [draw opacity=0] (536.6,145.85) .. controls (532.25,145.46) and (528.78,138.32) .. (528.78,129.58) .. controls (528.78,121.11) and (532.04,114.15) .. (536.2,113.37) -- (536.98,129.58) -- cycle ; \draw  [color={rgb, 255:red, 0; green, 0; blue, 255}  ,draw opacity=1 ] (536.6,145.85) .. controls (532.25,145.46) and (528.78,138.32) .. (528.78,129.58) .. controls (528.78,121.11) and (532.04,114.15) .. (536.2,113.37) ;  
\draw  [draw opacity=0] (536.89,176.29) .. controls (536.72,176.29) and (536.55,176.29) .. (536.37,176.29) .. controls (520.46,176.29) and (507.56,155.37) .. (507.56,129.57) .. controls (507.56,103.84) and (520.38,82.97) .. (536.23,82.84) -- (536.37,129.57) -- cycle ; \draw  [color={rgb, 255:red, 0; green, 0; blue, 255} ,draw opacity=1 ] (536.89,176.29) .. controls (536.72,176.29) and (536.55,176.29) .. (536.37,176.29) .. controls (520.46,176.29) and (507.56,155.37) .. (507.56,129.57) .. controls (507.56,103.84) and (520.38,82.97) .. (536.23,82.84) ;  

\draw [line width=1.5]    (288.47,129.98) -- (364.97,129.98) ;
\draw [shift={(368.97,129.98)}, rotate = 180] [fill={rgb, 255:red, 0; green, 0; blue, 0 }  ][line width=0.08]  [draw opacity=0] (11.07,-5.32) -- (0,0) -- (11.07,5.32) -- (7.35,0) -- cycle    ;
\draw    (104.74,118.34) .. controls (114.04,62.23) and (137.64,87.68) .. (163.87,82.93) ;
\draw    (104.74,118.34) .. controls (92.48,166.56) and (112.35,206.86) .. (129.27,183.15) ;
\draw    (129.27,183.15) .. controls (146.19,159.44) and (150.5,177.78) .. (168.2,178.29) ;
\draw    (120.56,112.65) .. controls (125.46,107.91) and (130.71,110.44) .. (134.77,111.07) ;
\draw    (117.43,109.57) .. controls (120.81,115.1) and (130.03,121.82) .. (136.46,108.78) ;
\draw    (116.27,162.84) .. controls (119.25,153.93) and (124.39,151.07) .. (128.03,147.7) ;
\draw    (112.8,163.16) .. controls (117.12,164.74) and (126.74,161.71) .. (128.88,144.05) ;
\draw    (179.97,113.79) .. controls (179.46,113.79) and (195.9,113.95) .. (199.39,116.13) ;
\draw    (179.95,82.88) .. controls (193.62,82.68) and (195.16,82.44) .. (199.05,82.62) ;
\draw    (196.18,146.16) .. controls (197.17,146.28) and (180.14,146.55) .. (180.06,146.59) ;
\draw    (198.21,177.78) .. controls (194.8,177.01) and (182.52,177.13) .. (180.08,177.49) ;
\draw    (139.99,139.5) .. controls (142.82,130.62) and (147.98,127.59) .. (151.61,124.11) ;
\draw    (136.47,139.97) .. controls (140.9,141.32) and (150.63,137.92) .. (152.4,120.5) ;
\draw    (256.71,143.42) .. controls (282.29,176.51) and (221.28,185.49) .. (198.21,177.78) ;
\draw    (256.71,143.42) .. controls (245.07,128.78) and (245.24,62.7) .. (231.75,79.18) ;
\draw    (231.75,79.18) .. controls (214.99,103.28) and (216.75,82.72) .. (199.05,82.62) ;
\draw    (227.26,124.13) .. controls (232.76,123.14) and (237.12,129.13) .. (240.77,132.54) ;
\draw    (224.97,119.1) .. controls (227.01,126.58) and (234.23,139.22) .. (242.79,131.58) ;
\draw    (179.97,113.79) .. controls (162.08,112.64) and (156.25,146.86) .. (180.06,146.59) ;
\draw    (196.18,146.16) .. controls (200.92,144.22) and (208.95,122.36) .. (199.39,116.13) ;
\draw    (163.87,82.93) .. controls (166.82,82.44) and (166.35,83.63) .. (179.95,82.88) ;
\draw    (168.2,178.29) .. controls (175.67,177.6) and (173.67,177.72) .. (180.08,177.49) ;
\draw  [color={rgb, 255:red, 0; green, 0; blue, 255}  ,draw opacity=1 ][pattern=_5ue7f3j79,pattern size=2.25pt,pattern thickness=0.75pt,pattern radius=0pt, pattern color={rgb, 255:red, 0; green, 0; blue, 255}] (175.5,98.3) .. controls (175.5,90.03) and (176.32,83.32) .. (177.34,83.32) .. controls (178.35,83.32) and (179.18,90.03) .. (179.18,98.3) .. controls (179.18,106.57) and (178.35,113.27) .. (177.34,113.27) .. controls (176.32,113.27) and (175.5,106.57) .. (175.5,98.3) -- cycle ;
\draw  [color={rgb, 255:red, 255; green, 0; blue, 0 }  ,draw opacity=1 ][pattern=_2fcvzd34z,pattern size=2.25pt,pattern thickness=0.75pt,pattern radius=0pt, pattern color={rgb, 255:red, 255; green, 0; blue, 0}] (179.18,98.3) .. controls (179.18,90.03) and (180,83.32) .. (181.02,83.32) .. controls (182.03,83.32) and (182.86,90.03) .. (182.86,98.3) .. controls (182.86,106.57) and (182.03,113.27) .. (181.02,113.27) .. controls (180,113.27) and (179.18,106.57) .. (179.18,98.3) -- cycle ;
\draw  [color={rgb, 255:red, 0; green, 0; blue, 255}  ,draw opacity=1 ][pattern=_fux8la0xm,pattern size=2.25pt,pattern thickness=0.75pt,pattern radius=0pt, pattern color={rgb, 255:red, 0; green, 0; blue, 255}] (182.86,98.3) .. controls (182.86,90.03) and (183.68,83.32) .. (184.7,83.32) .. controls (185.71,83.32) and (186.54,90.03) .. (186.54,98.3) .. controls (186.54,106.57) and (185.71,113.27) .. (184.7,113.27) .. controls (183.68,113.27) and (182.86,106.57) .. (182.86,98.3) -- cycle ;

\draw  [color={rgb, 255:red, 0; green, 0; blue, 255}  ,draw opacity=1 ][pattern=_a968nvxni,pattern size=2.25pt,pattern thickness=0.75pt,pattern radius=0pt, pattern color={rgb, 255:red, 0; green, 0; blue, 255}] (175.5,161.96) .. controls (175.5,153.69) and (176.32,146.99) .. (177.34,146.99) .. controls (178.35,146.99) and (179.18,153.69) .. (179.18,161.96) .. controls (179.18,170.23) and (178.35,176.94) .. (177.34,176.94) .. controls (176.32,176.94) and (175.5,170.23) .. (175.5,161.96) -- cycle ;
\draw  [color={rgb, 255:red, 255; green, 0; blue, 0 }  ,draw opacity=1 ][pattern=_t98sjdqts,pattern size=2.25pt,pattern thickness=0.75pt,pattern radius=0pt, pattern color={rgb, 255:red, 255; green, 0; blue, 0}] (179.18,161.96) .. controls (179.18,153.69) and (180,146.99) .. (181.02,146.99) .. controls (182.03,146.99) and (182.86,153.69) .. (182.86,161.96) .. controls (182.86,170.23) and (182.03,176.94) .. (181.02,176.94) .. controls (180,176.94) and (179.18,170.23) .. (179.18,161.96) -- cycle ;
\draw  [color={rgb, 255:red, 0; green, 0; blue, 255}  ,draw opacity=1 ][pattern=_rp3sywncm,pattern size=2.25pt,pattern thickness=0.75pt,pattern radius=0pt, pattern color={rgb, 255:red, 0; green, 0; blue, 255}] (182.86,161.96) .. controls (182.86,153.69) and (183.68,146.99) .. (184.7,146.99) .. controls (185.71,146.99) and (186.54,153.69) .. (186.54,161.96) .. controls (186.54,170.23) and (185.71,176.94) .. (184.7,176.94) .. controls (183.68,176.94) and (182.86,170.23) .. (182.86,161.96) -- cycle ;

\draw [color={rgb, 255:red, 0; green, 0; blue, 255}  ,draw opacity=1 ]   (174.68,78.37) -- (177.34,83.32) ;
\draw [color={rgb, 255:red, 0; green, 0; blue, 255}  ,draw opacity=1 ]   (187.35,78.37) -- (184.7,83.32) ;
\draw [color={rgb, 255:red, 255; green, 0; blue, 0 }  ,draw opacity=1 ]   (180.86,77.75) -- (181.02,83.32) ;
\draw [color={rgb, 255:red, 0; green, 0; blue, 255}  ,draw opacity=1 ]   (174.84,182.22) -- (177.49,177.28) ;
\draw [color={rgb, 255:red, 0; green, 0; blue, 255}  ,draw opacity=1 ]   (187.51,182.22) -- (184.85,177.28) ;
\draw [color={rgb, 255:red, 255; green, 0; blue, 0 }  ,draw opacity=1 ]   (181.01,182.85) -- (181.17,177.28) ;

\draw (290,93) node [anchor=north west][inner sep=0.75pt]  [ color={rgb, 255:red, 0; green, 0; blue, 0 }  ,opacity=1 ,xscale=0.8,yscale=0.8]  {$\displaystyle\prod_{i=1}^{2}\int  DK_{i} DJ_{i}$};
\draw (186,63) node [anchor=north west][inner sep=0.75pt]  [font=\small,color={rgb, 255:red, 0; green, 0; blue, 255 }  ,opacity=1 ,xscale=1,yscale=1]  {$\displaystyle \text{ }\hat{J}_{1}$};
\draw (156.5,63) node [anchor=north west][inner sep=0.75pt]  [font=\small,color={rgb, 255:red, 0; green, 0; blue, 255 }  ,opacity=1 ,xscale=1,yscale=1]  {$\displaystyle \text{-}\hat{J}_{2}$};
\draw (156.5,180) node [anchor=north west][inner sep=0.75pt]  [font=\small,color={rgb, 255:red, 0; green, 0; blue, 255 }  ,opacity=1 ,xscale=1,yscale=1]  {$\displaystyle \text{ }\hat{J}_{2}$};
\draw (186,180) node [anchor=north west][inner sep=0.75pt]  [font=\small,color={rgb, 255:red, 0; green, 0; blue, 255 }  ,opacity=1 ,xscale=1,yscale=1]  {$\displaystyle \text{-}\hat{J}_{1}$};
\draw (174,58) node [anchor=north west][inner sep=0.75pt]  [font=\small,color={rgb, 255:red, 255; green, 0; blue, 0 }  ,opacity=1 ,xscale=1,yscale=1]  {$\displaystyle \hat{K}_{1}$};
\draw (174,186) node [anchor=north west][inner sep=0.75pt]  [font=\small,color={rgb, 255:red, 255; green, 0; blue, 0 }  ,opacity=1 ,xscale=1,yscale=1]  {$\displaystyle \hat{K}_{2}$};

\end{tikzpicture}
\caption{Extended operators are inserted in the path integral over the single connected manifold. Extended operators are inserted in the path integral over the single connected manifold. By performing a specific functional average over these operators, we effectively break the path integral into two pieces, with each piece getting glued to itself to avoid introducing new boundaries.}
\label{sourcesurgery}
\end{figure}

Similar to how we can construct a generic version of $Q_L[J_1]$ and $Q_R[J_2]$ that would work for any manifold by thinking of codimension-one insertions and how they act locally, we can construct $P[K_1,K_2]$ from two codimension-one defects, one at each surface we desire to cut, that shift the field value as it crosses the surfaces. This would be the direct analogue to how we turned on sources linearly coupled to the momentum operator at the location we wanted to cut in section \ref{sec:simple}.  General definitions, that work independently of the choice of manifold, for both the insertions that multiply the fields by a phase profile and the ones that shift the field values are shown in fig. \ref{def}.
\\
\begin{figure}[!t]
\centering
 
\tikzset{
pattern size/.store in=\mcSize, 
pattern size = 5pt,
pattern thickness/.store in=\mcThickness, 
pattern thickness = 0.3pt,
pattern radius/.store in=\mcRadius, 
pattern radius = 1pt}
\makeatletter
\pgfutil@ifundefined{pgf@pattern@name@_bvkbfjv2b}{
\pgfdeclarepatternformonly[\mcThickness,\mcSize]{_bvkbfjv2b}
{\pgfqpoint{0pt}{0pt}}
{\pgfpoint{\mcSize+\mcThickness}{\mcSize+\mcThickness}}
{\pgfpoint{\mcSize}{\mcSize}}
{
\pgfsetcolor{\tikz@pattern@color}
\pgfsetlinewidth{\mcThickness}
\pgfpathmoveto{\pgfqpoint{0pt}{0pt}}
\pgfpathlineto{\pgfpoint{\mcSize+\mcThickness}{\mcSize+\mcThickness}}
\pgfusepath{stroke}
}}
\makeatother

 
\tikzset{
pattern size/.store in=\mcSize, 
pattern size = 5pt,
pattern thickness/.store in=\mcThickness, 
pattern thickness = 0.3pt,
pattern radius/.store in=\mcRadius, 
pattern radius = 1pt}
\makeatletter
\pgfutil@ifundefined{pgf@pattern@name@_3jsgkhgds}{
\pgfdeclarepatternformonly[\mcThickness,\mcSize]{_3jsgkhgds}
{\pgfqpoint{0pt}{0pt}}
{\pgfpoint{\mcSize+\mcThickness}{\mcSize+\mcThickness}}
{\pgfpoint{\mcSize}{\mcSize}}
{
\pgfsetcolor{\tikz@pattern@color}
\pgfsetlinewidth{\mcThickness}
\pgfpathmoveto{\pgfqpoint{0pt}{0pt}}
\pgfpathlineto{\pgfpoint{\mcSize+\mcThickness}{\mcSize+\mcThickness}}
\pgfusepath{stroke}
}}
\makeatother
\tikzset{every picture/.style={line width=0.75pt}} 

\begin{tikzpicture}[x=0.75pt,y=0.75pt,yscale=-1.4,xscale=1.4]

\draw    (304.66,126.03) -- (259.71,125.78) ;
\draw    (305.51,155.78) -- (259.71,155.53) ;
\draw   (300.07,140.9) .. controls (300.07,132.69) and (302.12,126.03) .. (304.66,126.03) .. controls (307.2,126.03) and (309.25,132.69) .. (309.25,140.9) .. controls (309.25,149.12) and (307.2,155.78) .. (304.66,155.78) .. controls (302.12,155.78) and (300.07,149.12) .. (300.07,140.9) -- cycle ;
\draw  [color={rgb, 255:red, 0; green, 0; blue, 255 }  ,draw opacity=1 ][pattern=_bvkbfjv2b,pattern size=2.25pt,pattern thickness=0.75pt,pattern radius=0pt, pattern color={rgb, 255:red, 0; green, 0; blue, 255}] (277.45,140.82) .. controls (277.45,132.65) and (279.51,126.03) .. (282.04,126.03) .. controls (284.58,126.03) and (286.64,132.65) .. (286.64,140.82) .. controls (286.64,148.99) and (284.58,155.61) .. (282.04,155.61) .. controls (279.51,155.61) and (277.45,148.99) .. (277.45,140.82) -- cycle ;
\draw  [dash pattern={on 3pt off 3pt}] (255.12,140.65) .. controls (255.12,132.44) and (257.17,125.78) .. (259.71,125.78) .. controls (262.25,125.78) and (264.3,132.44) .. (264.3,140.65) .. controls (264.3,148.87) and (262.25,155.53) .. (259.71,155.53) .. controls (257.17,155.53) and (255.12,148.87) .. (255.12,140.65) -- cycle ;
\draw  [draw opacity=0] (259.76,155.53) .. controls (257.19,155.53) and (255.12,148.87) .. (255.12,140.65) .. controls (255.12,132.44) and (257.19,125.78) .. (259.76,125.78) -- (259.76,140.65) -- cycle ; \draw   (259.76,155.53) .. controls (257.19,155.53) and (255.12,148.87) .. (255.12,140.65) .. controls (255.12,132.44) and (257.19,125.78) .. (259.76,125.78) ;  
\draw [color={rgb, 255:red, 0; green, 0; blue, 255 }  ,draw opacity=1 ]   (282.12,123.04) -- (282.04,126.03) ;
\draw    (259.82,158.15) .. controls (258.53,163.75) and (282.45,156.95) .. (282.45,165.55) ;
\draw    (305.08,158.15) .. controls (306.38,163.75) and (282.45,156.95) .. (282.45,165.55) ;

\draw [line width=0.75]    (336.92,144.83) -- (378.93,145.39) ;
\draw [shift={(381.93,145.43)}, rotate = 180.76] [fill={rgb, 255:red, 0; green, 0; blue, 0 }  ][line width=0.08]  [draw opacity=0] (5.36,-2.57) -- (0,0) -- (5.36,2.57) -- (3.56,0) -- cycle    ;

\draw    (304.66,50.87) -- (259.71,50.62) ;
\draw    (305.51,80.62) -- (259.71,80.37) ;
\draw   (300.07,65.75) .. controls (300.07,57.53) and (302.12,50.87) .. (304.66,50.87) .. controls (307.2,50.87) and (309.25,57.53) .. (309.25,65.75) .. controls (309.25,73.96) and (307.2,80.62) .. (304.66,80.62) .. controls (302.12,80.62) and (300.07,73.96) .. (300.07,65.75) -- cycle ;
\draw  [dash pattern={on 3pt off 3pt}] (255.12,65.5) .. controls (255.12,57.28) and (257.17,50.62) .. (259.71,50.62) .. controls (262.25,50.62) and (264.3,57.28) .. (264.3,65.5) .. controls (264.3,73.71) and (262.25,80.37) .. (259.71,80.37) .. controls (257.17,80.37) and (255.12,73.71) .. (255.12,65.5) -- cycle ;
\draw  [draw opacity=0] (259.76,80.37) .. controls (257.19,80.37) and (255.12,73.71) .. (255.12,65.5) .. controls (255.12,57.28) and (257.19,50.62) .. (259.76,50.62) -- (259.76,65.5) -- cycle ; \draw   (259.76,80.37) .. controls (257.19,80.37) and (255.12,73.71) .. (255.12,65.5) .. controls (255.12,57.28) and (257.19,50.62) .. (259.76,50.62) ;  
\draw    (259.82,83) .. controls (258.53,88.6) and (282.45,81.8) .. (282.45,90.4) ;
\draw    (305.08,83) .. controls (306.38,88.6) and (282.45,81.8) .. (282.45,90.4) ;

\draw [line width=0.75]    (336.92,69.68) -- (378.93,70.24) ;
\draw [shift={(381.93,70.28)}, rotate = 180.76] [fill={rgb, 255:red, 0; green, 0; blue, 0 }  ][line width=0.08]  [draw opacity=0] (5.36,-2.57) -- (0,0) -- (5.36,2.57) -- (3.56,0) -- cycle    ;

\draw    (304.66,201.17) -- (259.71,200.92) ;
\draw    (305.51,230.92) -- (259.71,230.68) ;
\draw   (300.07,216.05) .. controls (300.07,207.83) and (302.12,201.18) .. (304.66,201.18) .. controls (307.2,201.18) and (309.25,207.83) .. (309.25,216.05) .. controls (309.25,224.27) and (307.2,230.92) .. (304.66,230.92) .. controls (302.12,230.92) and (300.07,224.27) .. (300.07,216.05) -- cycle ;
\draw  [color={rgb, 255:red, 255; green, 0; blue, 0 }  ,draw opacity=1 ][pattern=_3jsgkhgds,pattern size=2.25pt,pattern thickness=0.75pt,pattern radius=0pt, pattern color={rgb, 255:red, 255; green, 0; blue, 0}] (277.45,215.97) .. controls (277.45,207.8) and (279.51,201.18) .. (282.04,201.18) .. controls (284.58,201.18) and (286.64,207.8) .. (286.64,215.97) .. controls (286.64,224.14) and (284.58,230.76) .. (282.04,230.76) .. controls (279.51,230.76) and (277.45,224.14) .. (277.45,215.97) -- cycle ;
\draw  [dash pattern={on 3pt off 3pt}] (255.12,215.8) .. controls (255.12,207.58) and (257.17,200.93) .. (259.71,200.93) .. controls (262.25,200.93) and (264.3,207.58) .. (264.3,215.8) .. controls (264.3,224.02) and (262.25,230.68) .. (259.71,230.68) .. controls (257.17,230.68) and (255.12,224.02) .. (255.12,215.8) -- cycle ;
\draw  [draw opacity=0] (259.76,230.68) .. controls (257.19,230.68) and (255.12,224.02) .. (255.12,215.8) .. controls (255.12,207.58) and (257.19,200.93) .. (259.76,200.93) -- (259.76,215.8) -- cycle ; \draw   (259.76,230.68) .. controls (257.19,230.68) and (255.12,224.02) .. (255.12,215.8) .. controls (255.12,207.58) and (257.19,200.93) .. (259.76,200.93) ;  
\draw [color={rgb, 255:red, 255; green, 0; blue, 0 }  ,draw opacity=1 ]   (282.12,198.19) -- (282.04,201.18) ;

\draw    (259.82,233.3) .. controls (258.53,238.9) and (282.45,232.1) .. (282.45,240.7) ;
\draw    (305.08,233.3) .. controls (306.38,238.9) and (282.45,232.1) .. (282.45,240.7) ;

\draw [line width=0.75]    (336.92,219.98) -- (378.93,220.54) ;
\draw [shift={(381.93,220.58)}, rotate = 180.76] [fill={rgb, 255:red, 0; green, 0; blue, 0 }  ][line width=0.08]  [draw opacity=0] (5.36,-2.57) -- (0,0) -- (5.36,2.57) -- (3.56,0) -- cycle    ;

\draw (349,128) node [anchor=north west][inner sep=0.75pt]  [ xscale=1,yscale=1]  {$\displaystyle \lim_{\epsilon \rightarrow 0}$};
\draw (349,53) node [anchor=north west][inner sep=0.75pt]  [xscale=1,yscale=1]  {$\displaystyle \lim_{\epsilon \rightarrow 0}$};
\draw (395.86,63.21) node [anchor=north west][inner sep=0.75pt]  [xscale=1,yscale=1]  {$\displaystyle \delta \left[ \phi _{1} -\phi _{2}\right]$};
\draw (310,63.21) node [anchor=north west][inner sep=0.75pt]  [font=\small,xscale=1,yscale=1]  {$\displaystyle \phi _{1}$};
\draw (243,63.21) node [anchor=north west][inner sep=0.75pt]  [font=\small,xscale=1,yscale=1]  {$\displaystyle \phi _{2}$};
\draw (280,91.54) node [anchor=north west][inner sep=0.75pt]  [font=\small,xscale=1,yscale=1]  {$\displaystyle \epsilon $};
\draw (396.9,213.51) node [anchor=north west][inner sep=0.75pt]  [xscale=1,yscale=1]  {$\displaystyle \delta \left[ \phi _{1} -\phi _{2} + K\right]$};
\draw (349,203) node [anchor=north west][inner sep=0.75pt]  [xscale=1,yscale=1]  {$\displaystyle \lim _{\epsilon \rightarrow 0}$};
\draw (243,213.51) node [anchor=north west][inner sep=0.75pt]  [font=\small,xscale=1,yscale=1]  {$\displaystyle \phi _{2}$};
\draw (310,213.51) node [anchor=north west][inner sep=0.75pt]  [font=\small,xscale=1,yscale=1]  {$\displaystyle \phi _{1}$};
\draw (280,241.85) node [anchor=north west][inner sep=0.75pt]  [font=\small,xscale=1,yscale=1]  {$\displaystyle \epsilon $};
\draw (278,186.5) node [anchor=north west][inner sep=0.75pt]  [font=\small,color={rgb, 255:red, 255; green, 0; blue, 0 }  ,opacity=1 ,xscale=1,yscale=1]  {$\displaystyle \hat{K}$};
\draw (399,138.36) node [anchor=north west][inner sep=0.75pt]  [xscale=1,yscale=1]  {$\displaystyle \delta \left[ \phi _{1} -\phi _{2}\right] \; e^{2\pi i\int d^{d}x\sqrt{\gamma}\, J( x) \cdot \phi _{1}( x)}$};
\draw (310,138.36) node [anchor=north west][inner sep=0.75pt]  [font=\small,xscale=1,yscale=1]  {$\displaystyle \phi _{1}$};
\draw (243,138.36) node [anchor=north west][inner sep=0.75pt]  [font=\small,xscale=1,yscale=1]  {$\displaystyle \phi _{2}$};
\draw (280,166.7) node [anchor=north west][inner sep=0.75pt]  [font=\small,xscale=1,yscale=1]  {$\displaystyle \epsilon $};
\draw (278,111) node [anchor=north west][inner sep=0.75pt]  [ font=\small,color={rgb, 255:red, 0; green, 0; blue, 255 }  ,opacity=1 ,xscale=1,yscale=1]  {$\displaystyle \hat{ J}$};

\end{tikzpicture}
\caption{Definitions of the surface phase and shift insertions through their actions inside a path integral evaluated over a small region that includes the insertions. Their action is defined in the limit where the region shrinks to zero along the perpendicular direction to the surface. The case with no insertions, which typically results in a Dirac delta functional gluing the two boundaries, is shown for comparison. To streamline our notation, we now denote the insertions by the phase function or shift function they induce, but with a hat on top to differentiate the operators from the functions.}
\label{def}
\end{figure}
Using these definitions, we can write down one concrete way to replace the question mark in fig. \ref{surgery1}. The claim is that if we can calculate the path integral with these insertions at and around the surfaces where we wish to perform the surgery, we can then average over them in a certain way that implements the surgery operation for us as shown in fig. \ref{sourcesurgery}. This is essentially the same logic we used in section \ref{sec:simple} with the obvious difference being that the averaging here will involve functional integrals over the insertions instead of the standard integrals we had before.

\subsubsection{Rényi entropy from surgery}

Let us now try to make contact with the analysis done in section \ref{sec:simple}. Starting with a local scalar field theory in flat space, suppose we are interested in using these ideas to calculate the Rényi entropy of some density matrix $\rho$. Our procedure will then always involve cutting and gluing $n$ identical pieces, each representing a copy of $\rho$. Any gluing of these these pieces, whether to each other or to themselves, should be done cyclically. Furthermore, in this case, the cutting and gluing operations are taking place along surfaces that can be naturally understood as spatial slices in the Lorentzian theory. From this perspective, the extended insertions $\hat{J}$ and $\hat{K}$ can be seen as operators acting on the Hilbert space of states in the theory. These operators can then be written explicitly as straightforward generalizations of the source insertions we used in section \ref{sec:simple}  
\begin{equation} \label{defJ}
    \hat{J}=e^{2\pi i \int d^{d}x \,J(x) \cdot \hat{\Phi}(0,x)},
\end{equation}
\begin{equation}\label{defA}
\hat{K}=e^{ i \int d^{d}x \,K(x)\cdot\hat{\Pi}(0,x)},
\end{equation}
where $\hat{\Phi}(t,x)$ and $\hat{\Pi}(t,x)$ are just the scalar field operators and the conjugate momenta satisfying the standard commutation relations 
\begin{align}
\left[\hat{\Phi}_a(0,x), \hat{\Pi}^b(0,y)\right]  & =i \delta_a^b \, \delta^{d}(x-y),  \nonumber \\\left[\hat{\Phi}_a(0,x), \hat{\Phi}_b(0,y)\right]  & =\left[\hat{\Pi}^a(0,x), \hat{\Pi}^b(0,y)\right]=0.
\end{align}
From here, if we focus on thermal partition functions for example, we can easily write the field theory generalization of \eqref{betatonbeta} 
\begin{equation}\label{therm}
   \frac{Z(n\beta)}{Z(\beta)^n}=\left(\prod_{j=0}^{n-1} \int \mathcal{D}K_j\, \mathcal{D}J_j\right)
\prod_{j=0}^{n-1}
\left\langle \hat{J}_j\,\hat{K}_j\,\hat{J}^{\dagger}_{j+1}\right\rangle_{\beta},
\end{equation}
with most of our analysis from section \ref{sec:simple} generalizing to the field theory case in a similar way. Again, we can perform some tests to see whether the functional integration measure is normalized correctly. For example, we should demand that the normalization satisfies 
\begin{equation}
    \int \mathcal{D}K\, \mathcal{D}J \;
\left\langle \hat{J}\,\hat{K}\,\hat{J}^{\dagger}\right\rangle_{\hat{\rho}}=1,
\end{equation}
for any choice of a state $\hat{\rho}$, that we calculate the expectation values in.

What about entanglement entropy? Our setup allows us to pick which degrees of freedom get manipulated by the insertions, which gives us a lot of freedom in how we can construct replica manifolds. In section \ref{sec:simple}, starting from a thermal state, we used this freedom to construct the replica manifold associated with the Rényi entropy of the reduced density matrix one gets after tracing out one of the two coupled oscillators.  We can do something similar here, if we wish to calculate the entanglement between different fields, by restricting the definitions of $\hat{J}$ and $\hat{K}$ to involve only a subset of the field operators and their conjugate momenta in \eqref{defJ} and \eqref{defA}. More explicitly, if we wish to calculate $\Tr{\hat{\rho}_r^n}$ where $\hat{\rho}_r$ is the normalized density matrix one gets from performing a partial trace of the thermal density matrix over a subset of fields , then we can write  
\begin{align} \label{Rényient1}
\Tr{\hat{\rho}_M^n}&=\left(\prod_{j=0}^{n-1} \int_{a \in M} \prod_{x,a} dK^a_j(x)\, dJ^a_j(x)\right) \nonumber\\
&\times\prod_{j=0}^{n-1}
\left\langle e^{2 \pi i {\sum_{a\in M}\int d^{d}xJ^a_j(x) \cdot \hat{\Phi}(0,x)}}\, e^{ i \sum_{a\in M}\int d^{d}x K^a_j(x) \cdot \hat{\Pi}(0,x)}\, e^{-2 \pi i \sum_{a\in M}\int d^{d}x J^a_{j+1}(x) \cdot \hat{\Phi}(0,x)}\right\rangle_{\beta},
\end{align}
where we now explicitly display the field index $a$, labeling the components of  the functions $J$ and $K$. By restricting $a$ to belong to the set of components we denoted by $M$, we effectively calculate $\Tr{\hat{\rho}_M^n}$ where the reduction was done by tracing over all components with $a \notin M$.

Alternatively, if the partial tracing was due a restriction to some subregion of space $\mathcal{R}$, then we can restrict the insertions' definitions by restricting the integrals in the exponents in \eqref{defJ} and \eqref{defA} to be only over a spatial subregion instead of the entire spatial slice\footnote{This expression is essentially identical to the one derived in \cite{Shiba:2014uia}.} 
\begin{align} \label{Rényient2}
\Tr{\hat{\rho}_\mathcal{R}^n}&=\left(\prod_{j=0}^{n-1} \int_{x \in \mathcal{R}} \prod_{x,a} dK^a_j(x)\, dJ^a_j(x)\right) \nonumber\\
&\times\prod_{j=0}^{n-1}
\left\langle e^{2 \pi i {\sum_{a}\int_{\mathcal{R}} d^{d}xJ^a_j(x) \cdot \hat{\Phi}(0,x)}}\, e^{ i \sum_{a}\int_{\mathcal{R}} d^{d}x K^a_j(x) \cdot \hat{\Pi}(0,x)}\, e^{-2 \pi i \sum_{a}\int_{\mathcal{R}} d^{d}x J^a_{j+1}(x) \cdot \hat{\Phi}(0,x)}\right\rangle_{\beta}.
\end{align}
Clearly, we can think of much more general ways of partitioning degrees of freedom and performing partial traces. For example, we might want to trace out some momentum modes, or we can think of doing a mix of tracing out some regions for some fields but different regions for others. We will try to argue for a more unified picture of how we can view many of these choices in section \ref{sec:constraints}.

\subsubsection{A field theory of defects}

Let us recap the picture that we constructed so far. We first argued that in local field theories, path integral surgery operations can be carried out by introducing certain operator insertions and averaging over them in a specific way. We then constructed a more concrete realization of this procedure by introducing the operator-valued functionals $\hat{K}_j$ and $\hat{J}_j$ and performing a functional average over their argument. We can also reinterpret this as coupling our system in a specific way to auxiliary fields $K_j$ and $J_j$ which live on codimension-one surfaces. By coupling these auxiliary fields to only a subset of the degrees of freedom of the fundamental fields, we can interpret the operation as a ``partial'' surgery where only certain degrees of freedom were affected by the surgery operation.

We can actually push this picture a bit further by defining a $d$-dimensional theory for the $K_j$ and $J_j$ fields where the action is given by 
\begin{equation} \label{defectaction}
S_{\hat{\rho},n}[\{K_j\},\{J_j\}]\equiv -\ln\left(\prod_{j=0}^{n-1}
\left\langle \hat{J}_j\,\hat{K}_j\,\hat{J}^{\dagger}_{j+1}\right\rangle_{\hat{\rho}}\right).
\end{equation}
In this language, every choice of state $\Psi$ and Rényi moment $n$ defines some action for the $K_j$ and $J_j$ fields. The surgery operation that takes us say from $\Tr{\rho}^n$ to $\Tr{\rho^n}$ then involves performing functional integrals over $K_j$ and $J_j$
\begin{equation} \label{Rényisurgery}
\Tr{\hat{\rho}^n}=\int\left(\prod_{j=0}^{n-1}  \mathcal{D}K_j\, \mathcal{D}J_j\right)\; e^{-S_{\hat{\rho},n}[\{K_j\},\{J_j\}]},
\end{equation}
which suggests that we can think of the calculation of $\Tr{\hat{\rho}^n}$ as a calculation in a field theory where the fundamental fields are $K_j$ and $J_j$ with the action being given by \eqref{defectaction}. In some sense, we can think of this as ``quantizing'' the defects. In fact, the general surgery operation shown in fig. \ref{sourcesurgery} can be framed in the same way. 

The big advantage of framing things in this way is that the theory we defined for the auxiliary fields $K_j$ and $J_j$ can inherit symmetries from the original theory we are calculating the Rényi entropy for. We can easily see this by observe that the way the action for $K_j$ and $J_j$ is defined in \eqref{defectaction} allows us to reinterpret transformations of $K_j$ and $J_j$ as transformations of $\hat{\Phi}(0,x)$ and $\hat{\Pi}(0,x)$. For example, if we simply focus on the operator $\hat{J}$, we see that the transformation $J^a \rightarrow R^a_{\,b}J^b $ is the same as the transformation $\hat{\Phi}_a(0,x) \rightarrow R^b_{\,a}\hat{\Phi}_b(0,x)$ where $R^a_{\,b}$ is some internal rotation in the field indices. Now if this transformation turns out to be something that leaves the expectation value in \eqref{defectaction} invariant, then the resulting action for $K_j$ and $J_j$ will inherit this invariance.\footnote{Clearly, this will depend on the state $\hat{\rho}$ as well rather than just the symmetries of the theory.}

This picture is very much analogous to the concept of ``gluing theories'' discussed in \cite{Dedushenko:2018aox,Dedushenko:2018tgx} where the fact that these codimension-one theories can inherit symmetries from their ``parent theories'' was used to derive a number of gluing formulas. While the theories we defined for $K_j$ and $J_j$ are not exactly the same object as the ``gluing theories'' in \cite{Dedushenko:2018aox,Dedushenko:2018tgx}, we can hope that similar formulas can be derived here using similar strategies. In our case, this would result in Rényi entropy formulas or surgery formulas relating the path integral on different manifolds similar to fig. \ref{sourcesurgery}.

 \subsubsection{Surgery in a different basis and the Heisenberg group}
 
So far we have mostly stuck to a picture where the $J_j$ sources couple the field operators while $K_j$ couple to their conjugate momenta. In reality, all we were trying to do was to find practical representations of Dirac delta functionals in some basis (for gluing) and an operator that effectively has all matrix elements equal to $1$ in that same basis (for cutting). Now, in decomposing the path integral into two parts, one can think of inserting many different possible resolutions of the identity, each corresponding to some choice of basis. For each such choice of basis, we can try to introduce the auxiliary fields $J_j$ and $K_j$ in the same way to implement our surgery operations. From this perspective, we see that we can get representations of Dirac deltas in this basis by coupling $J_j$ to a complete set of commuting operators with respect to which this basis is an eigenbasis. 
 
On the other hand, the cutting operation is somewhat more subtle as we need to find a practical representation of a matrix with all elements equal to 1 when written in that same basis. This is where the Heisenberg group with its clock and shift structure becomes very useful. The point here is that in a basis where the clock acts diagonally, a flat average over shifts has the exact right form to produce a matrix with all elements equal to 1. 

From here we see that the most straightforward way to implement our surgery technique is to find representations of the Heisenberg group to average over. The way we have been doing this so far is to look for canonically conjugate operators and exponentiating them. If we wish for our surgery to cut and glue all of the degrees of freedom in the theory, then we need our averaging to act non-degenerately on the entire Hilbert space. Of course, we are still left with a lot of freedom in picking which operators to average over even with all of these conditions satisfied.\footnote{From the perspective of the path integral, perhaps a more convenient language here would be to think of choices of polarization which would bring us closer to the framework discussed in \cite{Dedushenko:2018aox, Dedushenko:2018tgx}} 

So what happens if our averaging over some representations of the Heisenberg group acts degenerately on the Hilbert space? Clearly, we will end up with something of a partial surgery. In fact, this is the exact point we exploited in section \ref{sec:simple} to perform the surgery on one oscillator while the other one was unaffected. It would be interesting to further explore this kind of averaging in cases with more exotic operators that satisfy a Heisenberg group structure.
 
Beyond the use of the Heisenberg group, while we are far from being out of options, practical ways of implementing these surgery operations become less straightforward. We saw a simple example of this towards the end of section \ref{sec:simple} when we discussed surgery in the energy basis. As mentioned in the introduction, a subsequent paper will be mostly dedicated to exploring this direction.

\subsection{Gauge theories}
Typically, having discussed the treatment of scalar fields, a natural extension would be to consider fermions, followed by gauge fields. However, the problem with fermionic fields is that they do not obey canonical commutation relations, forcing us to have to adjust our Heisenberg group averaging strategy if we wish to incorporate them in our discussion. This obstacle is far from insurmountable, with the most natural strategies to overcome it involving (as one might expect) the introduction of Grassmann sources along similar lines to the approach of \cite{Moitra:2020cty, Moitra:2023gjm, Haldar:2020ymg}. That being said, such a discussion fits more naturally within the context of extending our results beyond averaging over the Heisenberg group in general, and so we will postpone it to the subsequent paper where we focus on these extensions.      
 
And so we now turn our attention to gauge theories. As one might expect, applying the ideas we have discussed so far to gauge theories also comes with additional subtleties. Given that our main interest in using these cutting and gluing techniques is to apply them to Rényi entropy calculations, we will focus on understanding what the picture looks like when we try to use it to relate $\Tr{\rho^n}$ to $\Tr{\rho}^n$. Of the multiple approaches one can choose to take here, our choice will be rather simple, but nevertheless instructive. Using free Maxwell theory as a concrete example, we will find that it is useful to work in an extended Hilbert space that includes the physical Hilbert space as a sector. We will then argue that the defect auxiliary theory will be itself a gauge theory.

\subsubsection{Working in an extended Hilbert space}

 We start by briefly reviewing how the extended Hilbert space picture looks like in the familiar setting of free Maxwell theory in Minkowski space. The action is given by 
\begin{align}
    S[A]&= \int d^4 x\, \mathcal{L}\nonumber\\
    &= -\frac{1}{4g^2} \int d^4 x \, F_{\mu \nu} F^{\mu \nu},
\end{align}
with
\begin{equation}
    F_{\mu \nu}\equiv\partial_\mu A_\nu-\partial_\nu A_\mu.
\end{equation}
This action is invariant under gauge transformations of the form
\begin{equation}
    A_\mu(x) \rightarrow A_\mu(x)+\partial_\mu \alpha(x),
\end{equation}
for any scalar function $\alpha(x)$.
To help us canonically quantize the theory, we exploit the gauge freedom to go to the so called temporal gauge where we set
\begin{equation}
A_0=0.
\end{equation}
We can then quantize the remaining spacial components $A_i$ in the standard way by promoting them and their conjugate momenta to operators satisfying the standard commutation relations
\begin{equation*}
\left[A_i(x), \pi^j(y)\right]=i\,\delta_i^j \, \delta^3(x-y),~~~~~~~~~~~~\left[A_i(x), A_j(y)\right]=\left[\pi^i(x), \pi^j(y)\right]=0
\end{equation*}
 Now having quantized the theory in this way, we have to demand that physical states in our Hilbert space must satisfy the Gauss law constraint
\begin{equation} \label{gauss}
G(x)\ket{\text{phys}}=0,~~~~~~~~~~~~~~~~~~~G(x) = \partial_i \pi^i(x).
\end{equation}
The fact that this condition is not satisfied automatically for all of the states in the Hilbert space we constructed is an indication that we built an extended Hilbert space that contains non-physical states in addition to the physical ones. Now if we wish to calculate the thermal partition function $Z(\beta)$ for example, we should make sure to only sum over states satisfying \eqref{gauss}. This can be conveniently achieved by defining a projector onto physical states $P_{\text{phys}}$ satisfying
\begin{equation} \label{physproj}
P_{\text{phys}}\ket{\text{phys}}=\ket{\text{phys}},~~~~~~~~~~~~~~~~~~~P_{\text{phys}}\ket{\text{non-phys}}=0.
\end{equation}
Then the partition function can be given by a trace in the extended Hilbert space, provided that we insert $P_{\text{phys}}$ inside the trace 
\begin{equation} 
Z(\beta)=\Tr{P_{\text{phys}}\,e^{-\beta H}}.
\end{equation}
This allows us to just apply the methods we developed so far by thinking of our formulas as being applied in the extended Hilbert space but to a density matrix that has been projected onto the physical sector of the Hilbert space. For instance, in the case of a thermal density matrix, we simply define the un-normalized density matrix as
\begin{equation} \label{gaugedensity}
    \rho= P_{\text{phys}}\,e^{-\beta H} \,P_{\text{phys}}.
\end{equation}
with the only remaining effect of the fact that this is a gauge theory being that the density matrix is given by \eqref{gaugedensity} instead of the usual $e^{-\beta H}$. From this perspective, in the extended Hilbert space, we can just treat this density matrix just like any other density matrix and write
\begin{align} \label{gauge1}
   \Tr{\rho^n}= &\left( \prod_{j=0}^{n-1} \int  \mathcal{D}\mathbf{K}_j \mathcal{D}\mathbf{J}_j \!\right) \! \prod_{j=0}^{n-1}\Tr{\rho \,  e^{2 \pi i\int d^3 x  \mathbf{J}_j(x)\cdot \boldsymbol{\pi}(x)}e^{i\int d^3 x  \mathbf{K}_j(x) \cdot \mathbf{A}(x)}e^{-2 \pi i\int d^3 x  \mathbf{J}_{j+1}(x) \cdot \boldsymbol{\pi}(x)}},
\end{align} 
where we have used a conjugate momentum basis implementation instead of the usual field basis. As usual, this comes with all the subtleties of how to make sense of these functional integrals and how to regularize them. Here we assume that, under appropriate regularization, we can think of these formulas as linear algebra identities in the extended Hilbert space with $\rho$ merely being some density matrix we are applying the identities to. 

Now as we argued earlier, we are free to interpret this from the perspective of a field theory for the $K_j$ and $J_j$ fields where the action is given by 
\begin{equation} \label{gaugeauxaction}
S_{\rho,n}[\{\mathbf{K}_j\},\{\mathbf{J}_j\}] \equiv -\ln{\left(\prod_{j=0}^{n-1}\Tr{\rho \,  e^{2 \pi i\int d^3 x  \mathbf{J}_j(x)\cdot \boldsymbol{\pi}(x)}e^{i\int d^3 x  \mathbf{K}_j(x) \cdot \mathbf{A}(x)}e^{-2 \pi i\int d^3 x  \mathbf{J}_{j+1}(x) \cdot \boldsymbol{\pi}(x)}}\right)}
\end{equation}
and $\Tr{\rho^n}$ corresponding to a calculation of the partition function in this theory.

\subsubsection{Gauge invariance in the auxiliary theory}

We would like to better understand the implications of gauge invariance in expressions like \eqref{gauge1}. At first glance, it appears that gauge invariance is not quite manifest in the way we wrote this expression. Nevertheless, we are counting on the projection $P_{\text{phys}}$ inside the definition of $\rho$ to do all the heavy lifting in keeping the trace gauge-invariant. Let us try to better understand what this implies. We start by focusing on a single copy of the trace  
\begin{equation} \label{tracegauge}
e^{-S_0[\mathbf{J}_j,\mathbf{K}_j,\mathbf{J}_{j+1}]}=\Tr{\rho \,  e^{2 \pi i\int d^3 x  \mathbf{J}_j(x)\cdot \boldsymbol{\pi}(x)}e^{i\int d^3 x  \mathbf{K}_j(x) \cdot \mathbf{A}(x)}e^{-2 \pi i\int d^3 x  \mathbf{J}_{j+1}(x) \cdot \boldsymbol{\pi}(x)}},
\end{equation}
where $S_0[\mathbf{J}_j,\mathbf{K}_j,\mathbf{J}_{j+1}]$ is what we get if we break down the action in \eqref{gaugeauxaction} into $n$ pieces 
\begin{equation} 
S_{\rho,n}[\{\mathbf{K}_j\},\{\mathbf{J}_j\}] =\sum_{j=0}^{n-1}S_0[\mathbf{J}_j,\mathbf{K}_j,\mathbf{J}_{j+1}].
\end{equation}
Now suppose we plug in a function for $\mathbf{J}_j(x)$ that is a gradient of some scalar function $\alpha_j(x)$
\begin{equation}
J^i_j(x)=\partial^i\alpha_j(x),
\end{equation}
then we have 
\begin{align}
\int d^3 x \ J^i_j(x)  \pi_i(x)&=\int d^3 x \ \partial_i\alpha_j(x)\pi^i(x) \nonumber\\
&=-\int d^3 x \ \alpha_j(x)\partial_i\pi^i(x).
\end{align}
But since the Gauss law constraint enforces $\partial_i\pi^i(x)\ket{\text{phys}}=0$, this tells us that 
\begin{equation} \label{auxgauge}
S_0[\mathbf{J}_j,\mathbf{K}_j,\mathbf{J}_{j+1}]=S_0[\mathbf{J}_j+\nabla\alpha_j,\;\mathbf{K}_j,\;\mathbf{J}_{j+1}+\nabla\alpha_{j+1}],
\end{equation}
for any two scalar functions $\alpha_j$ and $\alpha_{j+1}$.
On the other hand, we expect $P_{\text{phys}}$ to keep the trace gauge-invariant no matter which operators we insert next to $\rho$. By inspecting the operators in \eqref{tracegauge}, we see that under gauge transformations we pick up an extra factor 
\begin{equation}
\exp \left(i \int d^3 x \,K^i(x) \, \partial_i \alpha(x)\right)=\exp \left(-i \int d^3 x \,\alpha(x) \, \partial_i K^i(x)\right),
\end{equation}
which suggests that the trace will vanish unless we plug in a choice of function $\mathbf{K}(x)$ that satisfies 
\begin{equation}
\partial^i K_i(x)=0.
\end{equation}
The combination of this result with \eqref{auxgauge} tells us that we should think of the theory we defined for the auxiliary fields as a gauge theory with $\mathbf{J}(x)$ playing the role of the gauge field and $\mathbf{K}(x)$ its conjugate momentum. Of course, we can expect similar conclusions to hold in more general gauge theories, beyond the simple example of free Maxwell theory.

\section{Partial surgery and constraints}\label{sec:constraints}

Now that we sketched a picture of how we can effectively cut and glue path integrals by averaging over codimension-one operator insertions, we would like to develop a better understanding of what this picture means for entanglement entropy calculations. We saw how we can use these surgery techniques to calculate the Rényi entropy of some state by using formulas like \eqref{Rényisurgery} for example. We also saw how by restricting the sources to couple to only a subset of the degrees of freedom, we can evaluate entanglement Rényi entropy through expressions like \eqref{Rényient1} and \eqref{Rényient2}. 

In this section, we will attempt to formulate a more unified picture of these calculations and what they mean. The main idea is to interpret restricting the sources to couple to certain degrees of freedom as imposing a constraint on the $K_j$ and $J_j$ fields. We will show that, in the simple examples we considered so far, evaluating the Rényi entropy after partial tracing can be understood in terms of these constraints. While we will not attempt to classify general choices of constraints and which ones make sense from a quantum information theoretic perspective, we will argue for an interpretation of these constraints that allows us to go beyond the ones that correspond to a simple partial tracing.  

\subsection{Partial trace as a choice of constraint}

 So far, we have only discussed how this surgery picture can be applied in constructing replica manifolds for entanglement entropy calculations corresponding to a partial tracing of a subregion or a subset of fields everywhere in space. Obviously, we can come up with much more general ways of partitioning degrees of freedom by choosing what to couple the $K_j$ and $J_j$ fields to. We now propose a simple picture to view these different options in a unified way. The idea is to think of what we were doing so far as performing the functional integrals as in \eqref{therm} but with Dirac delta functional constraints. From this perspective, both of the cases in \eqref{Rényient1} and \eqref{Rényient2} can be written as 
\begin{align} \label{master}
\Tr{\hat{\rho}_r^n}=&\left(\prod_{j=0}^{n-1} \int \mathcal{D}K_j\, \mathcal{D}J_j\;\delta\left[G_r\!\left(K_j\right)\right]\,\delta\left[G_r\!\left(J_j\right)\right]\,\det\left(\frac{\delta G_r}{ \delta K_j}\right)\det\left(\frac{\delta G_r}{ \delta J_j}\right)\!\right)\prod_{j=0}^{n-1}
\left\langle \hat{J}_{j}\,\hat{K}_{j}\,\hat{J}^{\dagger}_{j+1}\right\rangle_{\hat{\rho}},
\end{align}
where $G_r$ is a function chosen such that it satisfies   
\begin{equation}
G_r(f(x)) = 0\quad~~~\forall x,
\end{equation}
only for the set of functions $f(x)$ that are consistent with the desired restriction.
For example, the restriction to a subset of fields can now be implemented through the choice
\begin{equation}
G_M(f^a(x))=\begin{cases}
f^a(x),\;\; \;\; a\notin M,\\
0, \;\;\;\;\;\;\;\;\;~\;a\in M,
\end{cases}
\end{equation}
while the restriction to a subregion corresponds to the choice
\begin{equation} \label{regionrestriction}
G_\mathcal{R}(f^a(x))=\begin{cases}
f^a(x),\;\; \;\; x\notin\mathcal{R},\\
0, \;\;\;\;\;\;\;\;\;~\;x\in \mathcal{R}.
\end{cases}
\end{equation}
A mix of both where which region gets traced out is different for different subsets of fields corresponds to a restriction that can be written as 
\begin{equation} \label{regionrestriction}
G_{\mathcal{R}^a}(f^a(x))=\begin{cases}
f^a(x),\;\; \;\; x\notin\mathcal{R}^a,\\
0, \;\;\;\;\;\;\;\;\;~\;x\in \mathcal{R}^a,
\end{cases}
\end{equation}
where we are simply now allowing the choice of region to depend on the field index $a$. 

In general, some care should be taken in evaluating \eqref{master} given how we wrote it down. This point will be discussed in \ref{sec:zeromodes}. For now, we simply point out that when we pick a choice of $G_r$, we will have a set of functions ``parallel'' to the surface defined by the constraint, $f^{||}(x)$, and a set of functions that are orthogonal to this surface, $f^{\perp}(x)$. We should make sure that the functional integral measure reduces to 
\begin{equation}\label{collapse}
\prod_{j=0}^{n-1} \int \mathcal{D}K^{||}_j\, \mathcal{D}J^{||}_j,
\end{equation}
with $J^{||}_j$ coupling to a subset of operators and $K^{||}_j$ coupling to their corresponding conjugate momenta. We assume that in \eqref{master}, this kind of reduction always takes place or that we simply set any zero modes factors, coming from both the delta functional and functional determinant, to $1$ from the start. 

Notice that so far we are constructing  $\Tr{\hat{\rho}_r^n}$ by starting first with $n$ separate copies of $\Tr{\hat{\rho}}$ and then using the insertions to cut and glue the degrees of freedom we want to keep in $\hat{\rho}_r$. We can also go the other way, as we did in section \ref{sec:simple}, by starting with $\Tr{\hat{\rho}^n}$ and then cut and glue the ``complement'' degrees of freedom that we want to trace out in defining $\hat{\rho}_r$. In this case, we need to introduce ``opposite'' constraints with the definitions of $G_r$ getting replaced with ones where we make the switch $\in \leftrightarrow \notin$ everywhere.

\subsubsection{State $\rightarrow$ action, partial trace $\rightarrow$ constraint}

Suppose we now go back to the auxiliary field theory picture with the action we defined in \eqref{defectaction}. Our formula for $\Tr{\hat{\rho}_r^n}$ then becomes 
\begin{equation} \label{constrainedsurgery}
\Tr{\hat{\rho}_r^n}=\int\left(\prod_{j=0}^{n-1}  \mathcal{D}K_j\, \mathcal{D}J_j\;\delta\left[G_r\!\left(K_j\right)\right]\,\delta\left[G_r\!\left(J_j\right)\right]\,\det\left(\frac{\delta G_r}{ \delta K_j}\right)\det\left(\frac{\delta G_r}{ \delta J_j}\right)\!\right) \; e^{-S_{\hat{\rho},n}[\{K_j\},\{J_j\}]}.
\end{equation}
So we see that from this perspective, the choice of the state and Rényi moment defines the action, while the choice of which degrees of freedom we keep determines the constraint.

\subsection{Interpretation of more general constraints}
So far, we have only discussed choices of constraints that can be easily reinterpreted as implementing a standard partial trace on the density matrix. More precisely, these are constraints that can be represented in terms of Dirac delta constraints, which can then be removed by choosing a convenient basis of functions for the functional integrals over $K_j$ and $J_j$ in terms of which the functional integral just collapses to the form \eqref{collapse}. From this perspective, the constraints just reproduce the same result we would have gotten if we had picked only a subset of field operators to couple $J_j$ to and coupled $K_j$ to their conjugate momenta. 

Now suppose we just forget about the partial trace origins that led us to \eqref{constrainedsurgery} and instead take it as is. One might then be tempted to consider more general constraints to impose on the $K_j$ and $J_j$ auxiliary fields, leading to an extended notion of what one might mean by performing a constrained or ``partial'' surgery. From the perspective of the Rényi entropy calculations, a potential interpretation is that we should think of these constraints as generally encoding the limitations of an observer trying to probe the information contained in the density matrix. As it stands, this is merely a suggestion of how one might think of these constraints. It would be interesting to investigate if any more general choice of these constraints can be concretely linked to some quantum information-theoretic quantity.

\section{An application: color entanglement} \label{sec:gaugetheory}

One interesting application of the framework we developed so far is how it can allow us to define notions of color entanglement in non-Abelian gauge theories. In many examples, the number of degrees of freedom in, say, a $U(N)$ gauge theory can scale like $N^2$. This fact is at the heart of many of the fascinating features one can get when considering the large-$N$ limit of these theories, including, as the gauge/gravity duality teaches us, the potential emergence of a gravitational theory out of a theory that did not have gravity to begin with. We would like to better understand the nature of entanglement between these degrees of freedom.

The task of defining color entanglement, and entanglement entropy in gauge theory in general, is faced with some extra challenges resulting from the fact that the physical Hilbert space of a gauge theory typically does not admit a factorization corresponding to the entanglement entropy we are interested in. The reason for that is simple: in a gauge theory, physical states are forced to obey a Gauss law constraint that spoils our would-be factorization by linking the degrees of freedom between the two tensor factors. This makes the extended Hilbert space picture we discussed in section \ref{sec:MIS} even more appealing when discussing entanglement entropy as the extended Hilbert space, as opposed to the physical Hilbert space, can admit the desired factorization.

While this approach allows us to define the partial trace as a well-defined operation in the extended Hilbert space that just happens to be performed on a gauge-invariant state embedded in it, defining a manifestly gauge-invariant notion of color entanglement remains challenging. The issue is that, in addition to the state being gauge-invariant, we want to have a gauge-invariant definition of subsystem partitioning within the extended Hilbert space. Naturally, this is challenging when the very thing we are trying to probe is entanglement between color degrees of freedom. Using the framework we developed so far, we will now propose some possible ways to make progress.

\subsection{Setting up the problem}

Since for this discussion we only care about color degrees of freedom rather than spatial degrees of freedom, we will avoid unnecessary complications by considering matrix quantum mechanics. More concretely, we consider a theory of a single $N \times N$ Hermitian matrix $\Phi$ and a non-dynamical Hermitian matrix $A_0$.\footnote{Technically, the number of physical degrees of freedom in this model is determined by the single matrix eigenvalues, which scale like $N$ and not $N^2$. Nevertheless, the main lessons we learn here can be easily extended to other models.} The real-time action is given by
\begin{equation}
S\left[\Phi, A_0\right]=\int d t \operatorname{Tr}\left(\frac{1}{2}\left(D_t \Phi\right)^2-V(\Phi)\right),
\end{equation}
with
\begin{equation}
    D_t \Phi \equiv \partial_t{\Phi}-i\left[A_0, \Phi\right].
\end{equation}
The action is invariant under any time dependent unitary transformation acting as
\begin{align*}
\Phi(t)& \rightarrow U(t) \,\Phi(t)\, U^{\dagger}(t),\\
\quad A_0(t)& \rightarrow U(t) \, A_0(t) \, U^{\dagger}(t)+i U(t)\, \partial_t{U}^{\dagger}(t),
\end{align*}
which tells us that we are dealing with a gauge theory with the gauge group being $U(N)$. 

Once again, we exploit the gauge freedom to set 
\begin{equation}
A_0=0,
\end{equation}
and proceed with standard canonical quantization where the matrix elements of $\Phi$ are promoted to operators along with their conjugate momenta satisfying
\begin{equation}
\left[\Phi_{i j}, \Pi_{k l}\right]=i \delta_{i l} \delta_{j k},~~~~~~~~~~~\left[\Phi_{i j}, \Phi_{k l}\right]=\left[\Pi_{i j}, \Pi_{k l}\right]=0.
\end{equation}
Now having set $A_0=0$, we need to impose the Gauss law constraint that would have resulted from its equation of motion on physical states
\begin{equation}
G^a\ket{\text{phys}}=0  ~~~~~~\forall a,~~~~~~~~~~~~~~~~~~ G^a= \Tr\left( i[\Phi, \Pi]\,T^a\right),
\end{equation}
where $T^a$ are the $U(N)$ Hermitian generators.\footnote{The index $a$ here takes integer values from $0$ to $N^2-1$ with $T^0$ being proportional to the identity and generating the $U(1)$ Abelian subgroup of $U(N)$. For $a \neq 0$, $T^a$ is Hermitian and traceless.}

We now define the usual projector onto physical states $P_{\text{phys}}$ according to the same definition in \eqref{physproj}, with the unnormalized thermal density matrix $\rho$ defined just like in \eqref{gaugedensity}. We can now proceed in the same way we did before and, for some normalized density matrix $\hat{\rho}$, we can write\footnote{To avoid notational confusion between the trace in the Hilbert space and the trace in the space of $N \times N$ matrices, we use the wiggly brackets trace ``$\Tr\{\dots\}$'' to denote the Hilbert space trace while the curved brackets trace ``$\Tr(\dots)$'' denotes the color index trace.} 
\begin{equation} \label{matrixsurgery}
   \Tr{\hat{\rho}^n}= \left( \prod_{j=0}^{n-1} \int_{\text{Herm}}  dK_j \, dJ_j\right)\prod_{j=0}^{n-1}\left\langle e^{2 \pi i  \Tr\left(J_j \Phi\right)}e^{i\Tr\left(K_j \Pi \right)}e^{-2 \pi i\Tr\left(J_{j+1} \Phi\right)}\right\rangle_{\hat{\rho}},
\end{equation} 
where the auxiliary sources $K_j$ and $J_j$ that we are integrating over are now $N \times N$ Hermitian matrices in $0$ dimensions with the integration measure we are using being defined as
\begin{equation}
    \int_{\text{Herm}}  d\Omega\equiv {2^{\frac{1}{2}N(N-1)}} \int \prod_{i=1}^N d\Omega_{ii} \prod_{1\leq i<j\leq N}\operatorname{dRe} \Omega_{ij} \operatorname{dIm} \Omega_{ij},
\end{equation} 
for a Hermitian matrix $\Omega$ we are integrating over.

The sources in \eqref{matrixsurgery} show up inside traces and it is easy to see that everything in the above expression will be manifestly gauge-invariant if we think of $K_j$ and $J_j$ themselves as transforming under the same $U(N)$ group that $\Phi$ and $\Pi$ transform under.\footnote{Technically, our approach so far only guarantees invariance under $U(N)$ transformations that are performed in the same way simultaneously in all $n$ copies of the system. Independent $U(N)$ invariance per replica will be restored shortly.}

Now let us turn our attention to how we can define color entanglement. The proper way to proceed when defining any entanglement entropy is to think in terms of operator algebras and how a restriction to some sub-algebra of observables can help us define a reduced density matrix $\hat{\rho}_r$. From this perspective, our task becomes to find choices of sub-algebras that somehow probe the structure of entanglement between color degrees of freedom. This leads to some tension, as observables are built out of gauge-invariant operators with the color degrees of freedom packaged into invariant combinations that do not make it particularly easy to split them into subsystems. Some of the attempts to circumvent this difficulty have ranged from focusing on matrix eigenvalues and utilizing notions of target-space entanglement \cite{Das:2020xoa,Das:2020jhy, Frenkel:2023aft, Hampapura:2020hfg, Frenkel:2021yql}, arguing for a natural gauge fixing condition to base the subsystem definition upon \cite{Gautam:2022akq}, or using the concept of relational observables to define the relevant sub-algebra \cite{Fliss:2025kzi}.

Our approach here will be much simpler. Instead of trying to face the problem head-on, we will completely sidestep the task of defining $\hat{\rho}_r$ properly and focus on figuring out if there is a way to come up with a replica manifold construction that is both completely analogous to standard entanglement entropy calculations and probes something about the structure of entanglement between color degrees of freedom. More precisely, using our surgery technique, we will try to define a quantity that is manifestly gauge-invariant and can be reasonably called $\Tr{\hat{\rho}_r^n}$, even if we do not have a proper definition of $\hat{\rho}_r$ yet. If we manage to achieve this goal, then we can go back and try to make sense of the corresponding $\hat{\rho}_r$ in the standard way.

The simplest way one might think of defining an entanglement entropy between color degrees of freedom is to say that we are going to define a reduced density matrix by tracing out all degrees of freedom corresponding to matrix elements outside some block in the top left corner of the matrix $\Phi$ in some basis. We can achieve this by introducing a constraint in \eqref{matrixsurgery} that restricts $K_j$ and $J_j$ to only have non-zero matrix elements inside this block. This is mathematically valid within the extended Hilbert space, and in fact, if we calculate $\Tr{\hat{\rho}_r^n}$ in this way, we expect it to be gauge-invariant in the sense that it would not depend on the choice of the basis we use to pick this block if the full density matrix we start out with is gauge-invariant. That being said, there is nothing manifestly gauge-invariant about the definition itself, since matrix elements that are in the top left corner of the matrix $\Phi$ will no longer be in that corner if we apply a gauge transformation.

If we want a manifestly gauge-invariant definition of $\Tr{\hat{\rho}_r^n}$, then perhaps we should introduce gauge-invariant constraints on $K_j$ and $J_j$. One option is to try to introduce constraints that enforce trace relations on $K_j$ and $J_j$ that can only be satisfied by matrices of lower rank. For example, if we had some $2 \times 2$ matrix $\Omega$, then introducing a Dirac delta function constraint of the form
\begin{equation} \label{trace1}
    \delta\left(\Tr\left(\Omega^2\right)-\Tr\left(\Omega\right)^2\right)
\end{equation}
would force $\Omega$ to be rank $\leq1 $. If $\Omega$ were a $3 \times 3$ matrix, then we can force it to have rank $\leq2 $ by using
\begin{equation} \label{trace2}
    \delta\left(\Tr\left(\Omega\right)^3-3 \Tr\left(\Omega\right)\Tr\left(\Omega^2\right)+2\Tr\left(\Omega^3\right)\right)
\end{equation}
If we want to go all the way from $3 \times 3$ to rank one, then we can use both \eqref{trace1} and \eqref{trace2} simultaneously. Clearly, our claim here is that by enforcing the standard set of trace relations that a $M \times M$ matrix would have to satisfy but a bigger $N \times N$ can escape, we effectively force the matrix to lower its rank without ever explicitly picking a particular block. While this approach has the advantage of being manifestly gauge-invariant all the way, it faces multiple complications in practical applications. One example of such a complication is that we need to enforce $N-M$ trace relations, which tends to become increasingly complicated as the matrices get larger. 

\subsection{Introducing auxiliary bifundamental matter}

The approach we will actually take instead is based on a simple trick. We start by introducing $N$ linearly independent vectors which we label by $Q^a$ with the superscript label going from $1$ to $N$ to differentiate the different vectors. We can write the components as
\begin{equation} \label{Q}
Q^a \equiv [q_1^a, q^a_2, \dots, q^a_N]^\top.
\end{equation}
Now for any integral over Hermitian matrices, we have the following identity\footnote{This identity is further discussed in \ref{sec:matrixid}.}   
\begin{align} \label{birep}
\int_{\text{Herm}} {d\Omega} \;F\left(\Omega\right) & \nonumber = \int_{\text{Herm}} d\Lambda\int\left(\prod_{a=1}^N   dX^a  \right) \left[\det_{ab}\left(\Tr\!\left(Q^a {Q^b}^{\dagger}\right)\right)\right]^N \;F\left(\sum_{a=1}^N X^a {Q^a}^{\dagger}\right)   \nonumber \\&~~~~~~~~~~~~~~~~~~~~~~~~~~~~~~~~~~~~~~\times \exp\left[-\pi\sum_{a=1}^N \Tr\left(X^a  Q^{a^\dagger}\Lambda  -  \Lambda Q^a X^{a^\dagger} \right)\right], 
\end{align}
for any function of the matrices $F(\Omega)$. The determinant here, as its subscript indicates, is not defined in the space where the matrices live, but rather in the superscript ($a$ and $b$) space. As for the vectors $X^a$, they are another set of $N$ vectors we are integrating over with the following definitions
\begin{equation}
    \int  dX^a\equiv \int\prod_{i=1}^N  \operatorname{dRe} x_i^a \operatorname{dIm} x_i^a, ~~~~~~~~~~~~~~~~~~X^a \equiv [x_1^a, x^a_2, \dots, x^a_N]^\top.
\end{equation}
Now we go back to \eqref{matrixsurgery} and use \eqref{birep} to make the replacements 
\begin{equation*}
K_j \rightarrow X^a_j, ~~~~~~~~~~~~~~~~~~~~~~~~~~~~~~ J_j \rightarrow Y_j^a,
\end{equation*}
which gives us
\begin{align}
\Tr{\hat{\rho}^n}
&=
\left(\prod_{j=0}^{n-1}\int_{\text{Herm}} d\Omega_j\, d\Lambda_j \int \prod_{a=1}^{N} dX_j^a\, dY_j^a \, \right)\left[\det_{ab}\!\left(\Tr\!\left(Q^a {Q^b}^{\dagger}\right)\right)\right]^{2nN}\nonumber\\
&\quad \times \prod_{j=0}^{n-1}\Bigg\{
\left\langle \  e^{2 \pi i \sum_{a=1}^N \Tr\left(Y^a_j{Q^a}^{\dagger} \Phi \right)}e^{i\sum_{a=1}^N\Tr\left(X^a_j{Q^a}^{\dagger} \Pi \right)}e^{-2 \pi i \sum_{a=1}^N\Tr\left(Y^a_{j+1}{Q^{a\dagger}} \Phi \right)}\right\rangle_{\hat{\rho}}\nonumber \\
&\qquad\qquad\qquad\ ~~~~~~~~~~~~~~~~~ \times \exp\!\left[-\pi\sum_{a=1}^{N}\Tr\!\left(X_j^a  Q^{a^\dagger}\Omega_j  + Y_j^a  Q^{a^\dagger}\Lambda_j - \text{H.c.}\right)\right]
\Bigg\}.
\end{align}
We can try to improve this expression in multiple ways. First, we can streamline our notation a bit by realizing that the objects we are dealing with now can carry two different kinds of indices. We originally had indices labeling matrix elements and vector components in the space of $N \times N$ matrices we started out with. We now have these additional superscript indices which we used to label the different vectors we introduced to replace the matrix integrals. We can think of these indices as living in a different space where we can define standard operations for them like the trace and the determinant. Indeed, this is exactly how we defined $\det_{ab}\left(\Tr\!\left(Q^a {Q^b}^{\dagger}\right)\right)$. To differentiate between the different trace operations we can now define, we will give the original trace we have been using so far a subscript $N$ while the new trace is given a subscript $M$. Now, objects like $Q$, $X_j$ and $Y_j$ should all be thought of as $N \times M$ matrices and acting on them with left multiplication corresponds to contraction of the $N$ index while right multiplication corresponds to contraction of the $M$ index. Their Hermitian conjugates then behave in the opposite way.

With these conventions, the expression now becomes
\begin{align}
\Tr{\hat{\rho}^n}
&=
\left(\prod_{j=0}^{n-1}\int_{\text{Herm}_N}   d\Omega_j\, d\Lambda_j \int   dX_j\, dY_j     \right)\left[\det_{M}\!\left(Q^{\dagger} Q\right)\right]^{2nN}\nonumber\\
&\times\Bigg\{\! \prod_{j=0}^{n-1}
\left\langle \!  e^{2 \pi i  \Tr_N\left(Y_j Q^{\dagger}\Phi \right)}e^{i\Tr_N\left(X_j Q^{\dagger}\Pi \right)}e^{-2 \pi i \Tr_N\left(Y_{j+1} {Q^{\dagger}}\Phi \right)}\!\right\rangle_{\!\hat{\rho}} \,e^{\! -\pi \!\Tr_N\!\left(\!X_j Q^{\dagger}\Omega_j  + Y_j Q^{\dagger}\Lambda_j  - \text{H.c.}\!\right)
}\!\Bigg\}, 
\end{align}
where we now have suppressed the superscript indices, with their role understood implicitly, on equal footing with the original matrix indices. For example, the integral measure over $X$ should be now understood according to
\begin{equation}
\int dX \equiv  \int \prod_{a=1}^M \prod_{i=1}^N  \operatorname{dRe} x_i^a \operatorname{dIm} x_i^a.
\end{equation}
Notice that although we have started to write everything in a way that would allow us to have $M$ not equal to $N$, so far we are choosing them to be equal in value. 

Before we proceed further, now that we freed up the location of the superscript index, it will be helpful to change our notation a bit to keep things more compact. By choosing the following notational replacements 
\begin{align}
Y\rightarrow X^1,~~~~~~~~~~X\rightarrow X^2,~~~~~~~~~~\Lambda\rightarrow \Omega^1,~~~~~~~~~~\Omega\rightarrow \Omega^2,
\end{align}
we can rewrite our expression as 
\begin{align}
\Tr{\hat{\rho}^n}
&=
\left(\prod_{j=0}^{n-1}\prod_{\alpha=1}^{2}\int_{\text{Herm}_N}   d\Omega^\alpha_j\int   dX^\alpha_j     \right)\left[\det_{M}\!\left(Q^{\dagger} Q\right)\right]^{2nN}\nonumber\\
&\times\Bigg\{\! \prod_{j=0}^{n-1}
\left\langle \!  e^{2 \pi i  \Tr_N\left(X^1_j Q^{\dagger}\Phi \right)}e^{i\Tr_N\left(X^2_j Q^{\dagger}\Pi \right)}e^{-2 \pi i \Tr_N\left(X^1_{j+1} {Q^{\dagger}}\Phi \right)}\!\right\rangle_{\!\hat{\rho}} \,e^{\! -\pi \sum_{\alpha=1}^2\!\Tr_N\!\left(\!X^\alpha_j Q^{\dagger}\Omega^\alpha_j    - \text{H.c.}\!\right)}\!\Bigg\}.
\end{align}

Now since nothing here should depend on the particular choice of $Q$, we might as well integrate over it  
\begin{align}\label{nohat}
\Tr{\hat{\rho}^n}
=C^{-1}&
\int_{\text{Herm}_M}   dV \int dQ\left(\prod_{j=0}^{n-1}\prod_{\alpha=1}^{2}\int_{\text{Herm}_N}   d\Omega^\alpha_j\int   dX^\alpha_j     \right)\nonumber\\
&\times\Bigg\{\! \prod_{j=0}^{n-1}
\left\langle \!  e^{2 \pi i  \Tr_N\left(X^1_j Q^{\dagger}\Phi \right)}e^{i\Tr_N\left(X^2_j Q^{\dagger}\Pi \right)}e^{-2 \pi i \Tr_N\left(X^1_{j+1} {Q^{\dagger}}\Phi \right)}\!\right\rangle_{\!\hat{\rho}}\Bigg\}  \nonumber\\
&\times\exp\bigg[ i\Tr_M\!\left(V\left[Q^{\dagger}Q-I_M\right]\right)- \pi \sum_{j=0}^{n-1} \sum_{\alpha=1}^2\!\Tr_N\!\left(\!X^\alpha_j Q^{\dagger}\Omega^\alpha_j    - \text{H.c.}\!\right) \!
\bigg],
\end{align}
where $I_M$ is the identity matrix in the $M$ space, and the constant $C$ given by
\begin{align}
C=\int_{\text{Herm}_M}   dV \int dQ\;\exp\bigg[ i\Tr_M\!\left(V\left[Q^{\dagger}Q-I_M\right]\right)\bigg].
\end{align}
Notice how we introduced a constraint on $Q$ through the auxiliary $M\times M$ Hermitian matrix $V$, allowing us to drop the determinant factor.

Clearly, the expression we now have for $\Tr{\hat{\rho}^n}$ is invariant under the $U(N)$ transformations that correspond to the gauge group of the matrix model, as it should be. However, it appears that the expression is also invariant under unitary transformations, $U(M)$, that act on the superscript indices, making the entire expression invariant under the group $U(N) \times U(M)$. If we take the $U(M)$ transformations seriously, then we can say that the $\Tr{\rho^n}$ calculation can be seen as a calculation in a theory with a $U(N)\times U(M)$ invariance. From this perspective, the fields $Q$ and $X^\alpha_j$ transform in the bifundamental representation of $U(N)\times U(M)$
\begin{equation*}
Q \mapsto U_N Q U^{\dagger}_M, \qquad ~~~~~~~~~~X^\alpha_j \mapsto U_N X^\alpha_j  U^{\dagger}_M.
\end{equation*}
On the other hand, the fields $\Omega^\alpha_j$ transform in the adjoint representation of $U(N)$ but are $U(M)$ singlets while $V$ is a $U(N)$ singlet transforming in the adjoint representation of $U(M)$.

\subsection{A notion of color entanglement?}

 So why go through all of this trouble to rewrite \eqref{matrixsurgery}? The answer is that now we are finally well positioned to propose a definition of a quantity that, at least in spirit, can be interpreted as $\Tr{\rho_M^n}$ with $\rho_M$ being the ``density matrix'' resulting from the reduction of $\hat{\rho}$ due to a restriction to a block of size $M$. All we need to do is choose $M < N$ in \eqref{nohat} instead of $M = N$.\footnote{In general, such a step should also involve keeping track of any changes to normalization/Jacobian factors.} What this does is that it restricts the surgery operation to be of lower rank than the full $N \times N$ matrix since $M$ can be traced back to the number of linearly independent vectors we used to replace the auxiliary matrices performing the surgery. 

One way to better see this is to first recall that we have a constraint enforcing the condition  
\begin{equation}
Q^{\dagger} Q=I_M
\end{equation}
If we think of $Q$ as a set of $M$ vectors with $N$ components $Q$ as we did in \eqref{Q}, then this condition just implies that they are orthogonal and normalized to unity. If $N=M$ this just means that they form a complete orthonormal basis. The integral over them then simply averages over all such choices of basis. On the other hand, if we choose $M<N$ then they can only form a basis within a $M$-dimensional subspace. The fact that we integrate over them freely otherwise means that we, in addition to averaging over all choices of basis within a fixed $M$-dimensional subspace, we are also averaging over all choices of the $M$-dimensional subspace out of the full $N$-dimensional space.  
 
Now, the auxiliary $N\times N$ matrices we are coupling to $\Phi$ and $\Pi$ are formed of multiplications of the form $X^\alpha Q^{\dagger}$. The Hermiticity condition we are imposing on these matrices, combined with $Q$ forming a basis within some $M$-dimensional subspace, means that in some choice of basis, we can think of these auxiliary matrices as being restricted to live in an $M\times M$ block. The integral over $Q$ averages over choices of any such block/subspace. In this way, the surgery picks out degrees of freedom corresponding to a block that is smaller in size than a full-rank matrix, but in a manifestly invariant way.

So what about the reduced density matrix here? In terms of the actual value, the entanglement entropy we calculate in this way is essentially the same as the one we would have gotten by explicitly performing a partial trace over degrees of freedom outside of an $M\times M$ block in some basis. Our calculation averages over all possible choices of such a block, but if the full density matrix we start with is gauge-invariant, then we expect to get the same value for $\Tr{\rho_M^n}$ for any such choice anyway, and the averaging does not add much to the final answer. So effectively, we are calculating the entropy of the reduced density matrix one gets by starting with an invariant density matrix and just naively performing a partial trace over a subset of color degrees of freedom.\footnote{For related discussions on how one can make more sense of such an object, see \cite{Fliss:2025kzi, Gautam:2022akq}.}

\subsection{Relation to quiver gauge theory}

 We now end this section with a discussion of what is perhaps the most interesting feature of this construction, which is how it can be related to quiver gauge theory. Let us go back to \eqref{nohat} and use the fact that inside this equation, we are allowed to make the replacement (up to a constant factor\footnote{An infinite factor actually for $M<N$.})
\begin{align} \label{replacement}
F\!\left(\!X^1 Q^{\dagger}\!\right) \rightarrow \!\int_{\text{Herm}_N} \! d\Omega^3 \! \int_{\text{Herm}_M}  \! dW \! \int \! dX^3    F\!\left(\!X^3{Q}^{\dagger}\!\right)     e^{ i \!\Tr_M\!\left( W\!\left[X^{3^{\dagger}} \! Q  -   X^{1^{\dagger}} \!Q \right]\right) -\pi \!\Tr_N\!\left(\Omega^3\!\left[X^3 Q^{\dagger}  -  Q X^{3^{\dagger}}\right] \right)\!} 
\end{align}
to rewrite our formula as\footnote{In the next few lines we will not keep track of the correct normalization of the expression. We removed the ``hat'' on $\rho$ in the left hand side here to differentiate it from the normalized case.} 
\begin{align} \label{comp}
\Tr{\rho^n}
=C^{-1}&
\int_{\text{Herm}_M}  \! dV  \int dQ\left(\prod_{j=0}^{n-1}\int_{\text{Herm}_M} dW_j\left[\prod_{\alpha=1}^{3}\int_{\text{Herm}_N}   d\Omega^\alpha_j\int   dX^\alpha_j   \right]\right)\nonumber\\
&\times\Bigg\{\! \prod_{j=0}^{n-1}
\left\langle \!  e^{2 \pi i  \Tr_N\left(X^1_j Q^{\dagger}\Phi \right)}e^{i\Tr_N\left(X^2_j Q^{\dagger}\Pi \right)}e^{-2 \pi i \Tr_N\left(X^3_j {Q^{\dagger}}\Phi \right)}\!\right\rangle_{\!\hat{\rho}}\Bigg\} e^{i\Tr_M\!\left(V\!\left[Q^{\dagger}Q-I_M\right]\right)}  \nonumber\\
&\times\exp\!\left[\sum_{j=0}^{n-1}\left\{i \Tr_M\left( W_j\!\left[X^{3{\dagger}}_j  Q  -   X^{1^{\dagger}}_{j+1} Q \right]\right)- \pi \sum_{\alpha=1}^3 \!\Tr_N\!\left(\!X^\alpha_j {Q}^{\dagger}\!\Omega^\alpha_j   - \text{H.c.}\!\right)\right\} \right] .
\end{align}
Now compare this to the following expression
\begin{align} 
C^{-n}&
  \left(\prod_{j=0}^{n-1}\int_{\text{Herm}_M}  \! dV_j \,dW_j\int dQ_j  \,\left[\prod_{\alpha=1}^3   \int_{\text{Herm}_N} \!  d\Omega^\alpha_j  \int dX^\alpha_j\,\right]    \right)\nonumber\\
&\times\Bigg\{\! \prod_{j=0}^{n-1}
\left\langle \!  e^{2 \pi i  \Tr_N\left(X^1_j Q_j^{\dagger}\Phi \right)}e^{i\Tr_N\left(X^2_j Q_j^{\dagger}\Pi \right)}e^{-2 \pi i \Tr_N\left(X^3_j {Q_j^{\dagger}}\Phi \right)}\!\right\rangle_{\!\hat{\rho}} e^{i\Tr_M\!\left(V_j\!\left[Q_j^{\dagger}Q_j-I_M\right]\right)}  \Bigg\}\nonumber\\
&\times\exp\!\left[\sum_{j=0}^{n-1}\left\{i \Tr_M\left( W_j\!\left[X^{3{\dagger}}_j  Q_j  -   X^{1^{\dagger}}_{j+1} Q_{j+1} \right]\right)- \pi \sum_{\alpha=1}^3 \!\Tr_N\!\left(\!X^\alpha_j Q_j^{\dagger}\Omega^\alpha_j   - \text{H.c.}\!\right)\right\}\right] ,
\end{align}
where we allow the $Q$ fields to be different for each replica. At a first glance, this is not really the same as \eqref{comp}, mainly because our claim about the replacement \eqref{replacement} does not work in the same way anymore. Instead, the replacement can only be done up to a $U(N)$ transformation.  Nevertheless, the fact that all the expectation values are gauge-invariant allows us to essentially cancel the effect of this unitary transformation, leading us to conclude that this form is equivalent to \eqref{comp}. 

We can make further progress by having a different $Q$ next to $X^1$, $X^2$, and $X^3$ inside every expectation value if we accept the price of introducing more constraints.\footnote{Overall, the constraints we introduced throughout this section are more than the minimum necessary to get the job done, but proceeding in this way makes the structure more transparent.} The expression we then get, up to more overall factors, is given by\footnote{For more details on how this form is essentially equivalent to \eqref{nohat}, see \ref{sec:quiver}.}
\begin{align} \label{masterquiver}
Z_n&=
   \left(\prod_{j=0}^{n-1} \int_{\text{Herm}_M}  \!  dL_j\, dW_j\left[\prod_{\alpha=1}^3 \int_{\text{Herm}_M}dV^\alpha_j  \int_{\text{Herm}_N} \!  d\Omega^\alpha_j\, d\Lambda_j^\alpha \int dQ^{\alpha}_j  \, dX^\alpha_j\,\right]    \right)\nonumber\\
&\times\Bigg\{\! \prod_{j=0}^{n-1}
\left\langle \!  e^{i  \Tr_N\left(X^1_j {Q^{1^{\dagger}}_j}\Phi \right)}e^{i\Tr_N\left(X^2_j {Q^{2^{\dagger}}_j}\Pi \right)}e^{- i \Tr_N\left(X^3_j {{Q^3_j}^{\dagger}}\Phi \right)}\!\right\rangle_{\!\hat{\rho}} \! e^{i\Tr_M\!\left(L_j\!\left[\left(\prod_{k=0}^{n-1}\!Q^{1^{\dagger}}_{j+k} Q^3_{j+k}\right)-I_M\!\right]+\text{H.c.}\right)}\!\Bigg\}  \nonumber\\
&\times\exp\!\Bigg[\sum_{j=0}^{n-1}\Bigg\{ i \Tr_M\!\left( W_j \!\left[X^{3^{\dagger}}_j  {Q^3_j}  -   {X^{1^{\dagger}}_{j+1}} {Q^{1}_{j+1}} \right]\right)+\sum_{\alpha=1}^{2}i\Tr_N\!\left( \Lambda_j^\alpha\!\left[ Q^\alpha_j{Q^\alpha_j}^{\dagger}-Q^{\alpha+1}_j{Q^{\alpha+1}_j}^{\dagger}\right]\right)\!\nonumber\\&~~~~~~~~~~~~~~~~~~~~~~~~~~~~~~~~~~~~~+\sum_{\alpha=1}^{3}i\Tr_M\left(V^\alpha_j\!\left[{Q^\alpha_j}^{\dagger}\!Q^\alpha_j-I_M\right]\right)-   \!\Tr_N\!\left(\!X^\alpha_j {Q^\alpha_j}^{\dagger}\!\Omega^\alpha_j   - \text{H.c.}\!\right) \bigg]\!\Bigg\}\Bigg],
\end{align}
which we have identified with some replica partition function $Z_n$ and simplified a bit further using the fact that we are not keeping track of the exact normalization. In terms of an effective action, this can then be rewritten as 
\begin{equation}
Z_n=
  \left(\prod_{j=0}^{n-1} \int_{\text{Herm}_M}  \!  dL_j\,dW_j\left[\prod_{\alpha=1}^3 \int_{\text{Herm}_M}dV^\alpha_j  \int_{\text{Herm}_N} \!  d\Omega^\alpha_j\, d\Lambda_j^\alpha \int dQ^{\alpha}_j  \, dX^\alpha_j\,\right]    \right) e^{-\left(S_{\hat{\rho}}+S_{\text{aux}}\right)},
 \end{equation}
with
\begin{equation*}
S_{\hat{\rho}}=-\ln\left(\prod_{j=0}^{n-1}\left\langle \!  e^{ i  \Tr_N\left(X^1_j {Q^{1^{\dagger}}_j}\Phi \right)}e^{i\Tr_N\left(X^2_j {Q^{2^{\dagger}}_j}\Pi \right)}e^{-i \Tr_N\left(X^3_j {{Q^3_j}^{\dagger}}\Phi \right)}\!\right\rangle_{\!\hat{\rho}} \right),
\end{equation*}
and
\begin{align*}
S_{\text{aux}}=&-\! \sum_{j=0}^{n-1}\Bigg\{ \!i \! \Tr_M\!\left( \!L_j\!\left[\left(\prod_{k=0}^{n-1}\!Q^{1^{\dagger}}_{j+k} Q^3_{j+k}\!\right)\!-\!I_M\!\right]\!+\text{H.c.} \!\right)+i \! \Tr_M\!\left(W_j \!\left[X^{3^{\dagger}}_j  {Q^3_j}  -   {X^{1^{\dagger}}_{j+1}} {Q^{1}_{j+1}} \right]\right)\!\nonumber\\&~~~~~~~~~~~~~~~~~~~~~~~~~~~~~~~~+\sum_{\alpha=1}^{3}\bigg[i\!\Tr_M\!\left(V^\alpha_j\!\left[{Q^\alpha_j}^{\dagger}\!Q^\alpha_j-I_M\right]\right)-   \!\Tr_N\!\left(\!X^\alpha_j {Q^\alpha_j}^{\dagger}\!\Omega^\alpha_j   - \text{H.c.}\!\right) \bigg]\\&~~~~~~~~~~~~~~~~~~~~~~~~~~~~~~~~~~~~~~~~~~~~~~~~~~~~~~~~+\sum_{\alpha=1}^{2}i\!\Tr_N\!\left( \Lambda_j^\alpha\!\left[ Q^\alpha_j{Q^\alpha_j}^{\dagger}-Q^{\alpha+1}_j{Q^{\alpha+1}_j}^{\dagger}\right]\right)\!\!\Bigg\}.
\end{align*}
\begin{figure}[t]
    \centering

\tikzset{every picture/.style={line width=0.75pt}} 

\begin{tikzpicture}[x=0.75pt,y=0.75pt,yscale=-1,xscale=1]

\draw  [color={rgb, 255:red, 0; green, 0; blue, 255 }  ,draw opacity=1 ] (280.63,129.7) .. controls (280.63,91.34) and (312.11,60.25) .. (350.94,60.25) .. controls (389.76,60.25) and (421.24,91.34) .. (421.24,129.7) .. controls (421.24,168.05) and (389.76,199.15) .. (350.94,199.15) .. controls (312.11,199.15) and (280.63,168.05) .. (280.63,129.7) -- cycle ;
\draw  [color={rgb, 255:red, 0; green, 0; blue, 0 }  ,draw opacity=1 ][fill={rgb, 255:red, 255; green, 255; blue, 255 }  ,fill opacity=1 ] (265.79,129.26) .. controls (265.79,120.59) and (272.86,113.56) .. (281.58,113.56) .. controls (290.3,113.56) and (297.37,120.59) .. (297.37,129.26) .. controls (297.37,137.93) and (290.3,144.96) .. (281.58,144.96) .. controls (272.86,144.96) and (265.79,137.93) .. (265.79,129.26) -- cycle ;
\draw  [color={rgb, 255:red, 0; green, 0; blue, 0 }  ,draw opacity=1 ][fill={rgb, 255:red, 255; green, 255; blue, 255 }  ,fill opacity=1 ] (335.08,60.37) .. controls (335.08,51.7) and (342.16,44.67) .. (350.88,44.67) .. controls (359.6,44.67) and (366.67,51.7) .. (366.67,60.37) .. controls (366.67,69.04) and (359.6,76.07) .. (350.88,76.07) .. controls (342.16,76.07) and (335.08,69.04) .. (335.08,60.37) -- cycle ;
\draw  [color={rgb, 255:red, 0; green, 0; blue, 0 }  ,draw opacity=1 ][fill={rgb, 255:red, 255; green, 255; blue, 255 }  ,fill opacity=1 ] (404.38,129.26) .. controls (404.38,120.59) and (411.46,113.56) .. (420.18,113.56) .. controls (428.9,113.56) and (435.97,120.59) .. (435.97,129.26) .. controls (435.97,137.93) and (428.9,144.96) .. (420.18,144.96) .. controls (411.46,144.96) and (404.38,137.93) .. (404.38,129.26) -- cycle ;
\draw  [color={rgb, 255:red, 0; green, 0; blue, 0 }  ,draw opacity=1 ][fill={rgb, 255:red, 255; green, 255; blue, 255 }  ,fill opacity=1 ] (335.08,198.15) .. controls (335.08,189.48) and (342.16,182.45) .. (350.88,182.45) .. controls (359.6,182.45) and (366.67,189.48) .. (366.67,198.15) .. controls (366.67,206.82) and (359.6,213.85) .. (350.88,213.85) .. controls (342.16,213.85) and (335.08,206.82) .. (335.08,198.15) -- cycle ;
\draw  [color={rgb, 255:red, 0; green, 0; blue, 255 }  ,draw opacity=1 ][fill={rgb, 255:red, 255; green, 255; blue, 255 }  ,fill opacity=1 ] (284.37,79.7) .. controls (284.37,71.05) and (291.44,64.03) .. (300.15,64.03) .. controls (308.86,64.03) and (315.92,71.05) .. (315.92,79.7) .. controls (315.92,88.36) and (308.86,95.37) .. (300.15,95.37) .. controls (291.44,95.37) and (284.37,88.36) .. (284.37,79.7) -- cycle ;
\draw  [color={rgb, 255:red, 0; green, 0; blue, 255 }  ,draw opacity=1 ][fill={rgb, 255:red, 255; green, 255; blue, 255 }  ,fill opacity=1 ] (385.74,79.8) .. controls (385.74,71.15) and (392.8,64.13) .. (401.51,64.13) .. controls (410.23,64.13) and (417.29,71.15) .. (417.29,79.8) .. controls (417.29,88.46) and (410.23,95.47) .. (401.51,95.47) .. controls (392.8,95.47) and (385.74,88.46) .. (385.74,79.8) -- cycle ;
\draw  [color={rgb, 255:red, 0; green, 0; blue, 255 }  ,draw opacity=1 ][fill={rgb, 255:red, 255; green, 255; blue, 255 }  ,fill opacity=1 ] (385.95,179.69) .. controls (385.95,171.04) and (393.01,164.02) .. (401.72,164.02) .. controls (410.44,164.02) and (417.5,171.04) .. (417.5,179.69) .. controls (417.5,188.35) and (410.44,195.36) .. (401.72,195.36) .. controls (393.01,195.36) and (385.95,188.35) .. (385.95,179.69) -- cycle ;
\draw  [color={rgb, 255:red, 0; green, 0; blue, 255 }  ,draw opacity=1 ][fill={rgb, 255:red, 255; green, 255; blue, 255 }  ,fill opacity=1 ] (284.64,179.69) .. controls (284.64,171.04) and (291.71,164.02) .. (300.42,164.02) .. controls (309.13,164.02) and (316.19,171.04) .. (316.19,179.69) .. controls (316.19,188.35) and (309.13,195.36) .. (300.42,195.36) .. controls (291.71,195.36) and (284.64,188.35) .. (284.64,179.69) -- cycle ;

\draw (393.22,172.09) node [anchor=north west][inner sep=0.75pt]  [font=\normalsize,color={rgb, 255:red, 0; green, 0; blue, 255 }  ,opacity=1 ]  {$M$};
\draw (291.92,172.09) node [anchor=north west][inner sep=0.75pt]  [font=\normalsize,color={rgb, 255:red, 0; green, 0; blue, 255 }  ,opacity=1 ]  {$M$};
\draw (291.65,72.1) node [anchor=north west][inner sep=0.75pt]  [font=\normalsize,color={rgb, 255:red, 0; green, 0; blue, 255 }  ,opacity=1 ]  {$M$};
\draw (393.01,72.2) node [anchor=north west][inner sep=0.75pt]  [font=\normalsize,color={rgb, 255:red, 0; green, 0; blue, 255 }  ,opacity=1 ]  {$M$};
\draw (343.88,52.77) node [anchor=north west][inner sep=0.75pt]  [font=\normalsize]  {$N$};
\draw (343.88,190.55) node [anchor=north west][inner sep=0.75pt]  [font=\normalsize]  {$N$};
\draw (413.18,121.66) node [anchor=north west][inner sep=0.75pt]  [font=\normalsize]  {$N$};
\draw (274.58,121.66) node [anchor=north west][inner sep=0.75pt]  [font=\normalsize]  {$N$};
\draw (337,117) node [anchor=north west][inner sep=0.75pt]  [font=\huge]  {$Z_{4}$};

\end{tikzpicture}

\caption{Schematic representation of the kind of quiver diagrams that may be relevant to our entropy calculations. Objects drawn in blue are $0$-dimensional. The diagram shows  four replicas connected cyclically by multiple intermediate auxiliary $U(M)$ nodes.}
\label{fig:quiver}
\end{figure}
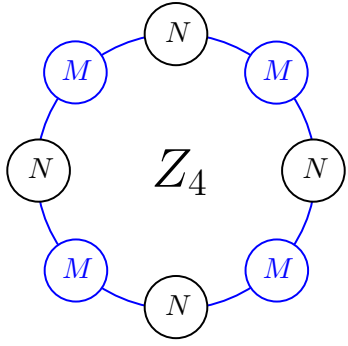
So what is the point of all of these manipulations, leading us to a more complicated-looking expression? It appears that we are now left with a picture that looks increasingly similar to constructions of quiver gauge theory \cite{Douglas:1996sw}. The aim of these manipulations was to get an expression that is manifestly invariant under independent $U(N)$ transformations per replica, and have these replicas connected together by intermediate terms that are manifestly gauge-invariant. It turns out that these terms can come along with some extra $U(M)$ group invariances. From this perspective, one would say that in trying to perform a surgery operation that builds $\Tr{\hat{\rho}_M^n}$ out of $n$ replica copies of the theory, we found that we can achieve this by connecting them through a combination of a 0-dimensional $U(M)$ node and bifundamental auxiliary fields. Of course, the $n$ copies of the theory themselves can be seen as 1-dimensional $U(N)$ nodes along with the auxiliary $\Omega_j$ and $\Lambda_j$ matrices. The quantity $\Tr{\hat{\rho}_M^n}$ then simply corresponds to
 \begin{equation}
     \Tr{\hat{\rho}_M^n}=\frac{Z_n}{Z_1^n}
 \end{equation}
effectively getting mapped to a calculation of a path integral in a quiver gauge theory. A schematic illustration of a quiver diagram relevant to this construction is shown in fig. \ref{fig:quiver}.

From the perspective of string theory, quiver gauge theories arise naturally in situations where D-branes are placed at orbifold singularities.\footnote{Replica manifolds can also be thought of in the language of orbifolds, which may offer an interesting point of contact.} It would be very interesting if one can come up with a concrete construction that, at least in some limit, can be related to a color entanglement calculation in the sense we outlined.

\section{A look ahead: large-N limit and holography} \label{sec:lookahead}
In this section, we would like to discuss a bird’s eye view of future directions and what all of this might be good for. As mentioned in the introduction, the author's interest in developing these ideas originally stemmed from an interest in understanding holographic duals of more general kinds of entanglement entropy that can be defined in the boundary theory. With this goal in mind, it may be instructive to try to understand the interplay between the techniques we developed so far and the large-$N$ expansion in simple examples. Our example of choice will be a simple quantum mechanical vector model. We will then use the lessons we learn from this simple example to argue for what this may mean when applied in holographic models in general.

\subsection{Another simple example: the $O(N)$ vector model} 

One of the simplest examples that can be used as an illustration of large-$N$ techniques is the linear sigma model.\footnote{Standard treatments of this model as well as large N expansions in general can be found in \cite{Coleman:1980nk}.} This is a model of $N$ scalar fields $\phi_a$ that can be organized into an $N$-component vector $\Phi$ with the system enjoying an internal symmetry corresponding to rotations of this vector. We will simplify our discussion by considering the quantum mechanics version of this model which has no spatial dimensions, just time.

We will begin with a brief review of some basic facts about how this model can be treated in the large-$N$ limit. Then we will use it as a simple illustration of how entanglement entropy calculations involving internal degrees of freedom can be represented in the large-$N$ limit. 

\subsubsection{Brief review of some basic facts}
Let us begin with a quick review of this model in Euclidean time. The Lagrangian which can be written as
\begin{equation} \label{largen}
L_E=\frac{1}{2}\left( \partial_\tau \Phi\cdot \partial_\tau\Phi+ m^2 \Phi \cdot \Phi+ \frac{g}{4N}\left(\Phi \cdot \Phi\right)^2\right),
\end{equation}
where the quartic interaction term was written with an explicit factor of $1/N$ since the precise way we will be taking the large-$N$ limit involves keeping $g$ in \eqref{largen} constant as $N \rightarrow \infty$ and not the entire quartic coupling. 

The standard treatment of the large-$N$ expansion in this model typically starts by decoupling the fields using a technique called the Hubbard–Stratonovich transformation. The idea is to introduce an auxiliary field $\sigma$ and rewrite $L_E$ as 
\begin{equation} 
L_E=\frac{1}{2}\left( \partial_\tau \Phi\cdot \partial_\tau\Phi+ \left(m^2+\sigma\right) \Phi \cdot \Phi- \frac{N}{g}\sigma^2\right).
\end{equation}
This Lagrangian is equivalent to the one given in \eqref{largen}, but is much more useful in setting up the large-$N$ expansion. Consider for example the thermal partition function of the system at a temperature $1/\beta$
\begin{align*}
    Z(\beta)&=\int \mathcal{D}\sigma \mathcal{D}\Phi~e^{-\int_\beta  L_E~d\tau}\\
    &=\int \mathcal{D}\sigma \mathcal{D}\Phi~e^{-\frac{1}{2}\int_\beta  ~d\tau\left[\Phi\cdot\left(-\partial_\tau^2+m^2+\sigma\right)\Phi-\frac{N}{g}\sigma^2\right]}.
\end{align*} 
The upshot here is that the path integral over $\Phi$ can be done right away and is simply given in terms of the functional determinant of the operator inside the brackets in the above equation. Notice that this operator does not mix the different components of $\Phi$ and as such we can evaluate the determinant for one component and raise the answer to the power $N$. The partition function can then be written as a path integral over $\sigma$ only but using an effective action $S_{\text{eff}}[\sigma]$ that includes the effects of integrating $\Phi$ out
\begin{align} \label{auxpartition}
    Z(\beta)=\int \mathcal{D}\sigma~e^{-NS_{\text{eff}}[\sigma]},
\end{align} 
with
\ben\label{effectiveaction}
S_{\text{eff}}[\sigma]=\frac{1}{2}\left(\Tr\Big(\ln\left(-\partial_\tau^2+m^2+\sigma\right)\Big)-\int_\beta  \frac{\sigma^2}{g}d\tau\right).
\een
Notice that with the explicit factor of $N$ placed outside of the action in \eqref{auxpartition}, $S_{\text{eff}}[\sigma]$ itself has no $N$ dependence. Using this form, we can start setting up calculations as an expansion in $1/N$. If we are only interested in the $N\rightarrow \infty$ limit of $Z(\beta)$, we just need to find the configuration of $\sigma$ that minimizes $S_{\text{eff}}[\sigma]$. We can find extrema of $S_{\text{eff}}[\sigma]$ by demanding that first order variations with respect to $\sigma$ vanish, which leads to a condition typically called the gap equation
\ben \label{gap}
G_{\sigma,\beta}(\tau, \tau)=\frac{2\sigma}{g},
\een
with 
\ben \label{greendef}
G_{\sigma,\beta}(\tau, \tau')\equiv\frac{1}{-\partial_\tau^2+m^2+\sigma}.
\een
One way to understand the explicit $\beta$ subscript on the Green's function here is that it is a result of the fact that the differential operator we inverted on the right side of \eqref{greendef} is only defined on periodic functions of $\tau$ with periodicity $\beta$. In other words, we must have
\begin{equation}
G_{\sigma, \beta}\left(\tau+\beta, \tau^{\prime}\right)=G_{\sigma, \beta}\left(\tau, \tau^{\prime}\right).
\end{equation}
Now, for a constant configuration $\sigma_0$, the condition in \eqref{gap} becomes
\ben\label{constantgap}
\frac{\sigma_0}{g}=\frac{1}{4 \sqrt{m^2+\sigma_0}} \operatorname{coth}\left(\frac{\beta \sqrt{m^2+\sigma_0}}{2}\right).
\een
This equation, at least implicitly, determines how the value of $\sigma_0$ depends on the temperature and other parameters in the original Lagrangian. Assuming we manage to determine the value of $\sigma_0$ satisfying \eqref{gap}, we can then use this value to obtain the leading large-$N$ approximation of the partition function
\begin{align} \label{Nauxpartition}
    Z(\beta)\approx~e^{-NS_{\text{eff}}(\sigma_0)}.
\end{align} 
The effective action can now also be written in a much simpler form  
\ben
S_{\text{eff}}(\sigma_0)=\ln \left(2 \sinh \left(\frac{\beta\sqrt{m^2+\sigma_0} }{2}\right)\right)-\frac{\beta}{2}  \frac{\sigma_0^2}{g}.
\een
From here, evaluating the thermal entropy or any other thermodynamic quantity is straightforward.  
\subsubsection{Tracing out components}
Now that we have covered the standard large-$N$ treatment of this elementary model, let us see what we get from a partial trace over the degrees of freedom corresponding to a subset of the components of $\Phi$.

How should we proceed? Our strategy will be very simple: we will just repeat what we did for the coupled harmonic oscillators in section \ref{sec:simple}. Namely, we will start with the partition function of the theory on a Euclidean time circle of length $n\beta$, then we will couple each of $M$ components of $\Phi$ to the exact same source given in \eqref{finalsource}. Integrating over these sources then will force these $M$ components to live on smaller Euclidean time circles of length $\beta$. This then gives us $Z_n=\Tr{\rho_{N-M}^n}$, with $\rho_{N-M}$ being the (unnormalized) reduced density matrix seen by an observer that only has access to the remaining $N-M$ components.

\begin{equation}\label{ONreduced}
Z_n=\lim_{\epsilon \rightarrow 0}\left( \prod_{j=0}^{n-1} \prod_{a=1}^{M} \int_{-\infty}^{\infty} \frac{d K^a_{j} dJ^a_{j}}{2\pi}\right)\int \mathcal{D}\sigma \mathcal{D}\Phi~e^{-\int_{n\beta} ~d\tau \big\{ L_{E}+Q-\mathcal{J}_M\cdot\Phi \big\}},
\end{equation}
where $\mathcal{J}_M$ is a vector with only $M$ non-zero components given by
\ben
\mathcal{J}_M=\sum_{a=1}^{M}\sum_{j=0}^{n-1}\Bigg\{K^a_{j}\delta'(\tau-\tau_j)+iJ^a_{j}\Big[\delta\big(\tau-(\tau_j+\epsilon)\big)-\delta\big(\tau-(\tau_{j+1}-\epsilon)\big)\Big]\Bigg\}.
\een
while $Q$ is given by
\ben
Q=\frac{1}{2}\sum_{a=1}^{M}\left(\sum_{j=0}^{n-1}K^a_{j}\delta(\tau-\tau_j)\right)^2.
\een
The good news here is that having reduced the action to a quadratic form in $\Phi$ by introducing the auxiliary field, we can repeat the same standard analysis from section \ref{sec:simple} on how to deal with the sources. Now if we define
\begin{align}\label{entaction}
\nonumber E_M[\sigma] \equiv \ln\Bigg[&\lim_{\epsilon \rightarrow 0}\Bigg(
\prod_{j=0}^{n-1} \prod_{a=1}^{M} \int_{-\infty}^{\infty} \frac{d K^a_{j}\,dJ^a_{j}}{2\pi}\Bigg) \\
&\times \exp\Bigg\{  \frac{1}{2} \int_{n\beta} d\tau \int_{n\beta}d\tau'\,
\mathcal{J}_M(\tau)\cdot G_{\sigma,n\beta}(\tau,\tau')\, \mathcal{J}_M(\tau')-\int_{n\beta}d\tau\, Q(\tau)\Bigg\}\Bigg],
\end{align}
then we can write $Z_n$ as
\ben \label{ONreduced2}
Z_n=\int \mathcal{D}\sigma e^{-NS_{n\beta}[\sigma]+E_M[\sigma]},
\een
where, for future convenience, we have removed the usual ``$\text{eff}$'' subscript on the effective action and replaced it with the length of the Euclidean time circle it is defined on. Now if we can evaluate $E_M[\sigma]$, the calculation reduces to a path integral over $\sigma$ with some modified effective action, and we can reasonably hope that in the large-$N$ limit we can get an answer via saddle approximation. In this way, leading large-$N$ information about the entanglement entropy between the components of $\Phi$ may be encoded in  classical configurations of $\sigma$ that minimize the modified effective action.

From here we can try to find a way to evaluate \eqref{entaction} directly, which, in principle, is similar to what we did in section \ref{sec:simple}. This time, however, the calculation is complicated by the fact that the Green's function has a functional dependence on $\sigma$. Instead of trying to brute force our way towards an answer, we will make a simple argument as to what the answer must be. 

We start by pointing out the simple observation that $ G_{\sigma,n\beta}(\tau,\tau')$, just like its inverse, does not mix the different components of $\mathcal{J}_M$. This again implies that we can evaluate \eqref{entaction} independently for each component of $\mathcal{J}_M$ and sum up the contributions, which implies that we have
\ben
E_M[\sigma]=ME_1[\sigma].
\een
Now if we believe that the effect we get from performing the integral over the sources in \eqref{ONreduced} is to force $M$ components of $\Phi$ to live on smaller Euclidean time circles of size $\beta$, then if we pick $M=N$, we should just get $Z(\beta)^n$. But since we know exactly what $Z(\beta)^n$ should look like, this implies that we have 
\ben
E_1[\sigma]=S_{n\beta}[\sigma]-nS_{\beta}[\sigma].
\een
We are actually being quite cavalier here with what we exactly mean by $nS_{\beta}[\sigma]$, and in the next few lines it should really be understood in a somewhat schematic sense, but we will make the definition precise shortly. Now we see that \eqref{ONreduced2} becomes
\ben \label{ONzn}
Z_n=\int \mathcal{D}\sigma e^{-N\big[(1-\lambda)S_{n\beta}[\sigma]+\lambda nS_{\beta}[\sigma]\big]},
\een
with $\lambda \equiv M/N$. If we take the limit $N\rightarrow \infty$ as before we can approximate $Z_n$ by its dominant saddle corresponding to some configuration $\sigma_n$. Now if we use \eqref{Rényi1} to evaluate the Rényi entropy $R_n$ we get\footnote{We changed the notation here in referring to the Rényi entropy to avoid confusion with the action. }
\begin{align} \label{ONRényi}
\nonumber R_n&=\frac{1}{1-n} \Big(\ln Z_n -n\ln Z_1\Big) \\
&=\frac{N}{n-1}\Big((1-\lambda)S_{n\beta}[\sigma_n]+\lambda nS_{\beta}[\sigma_n]-nS_{\beta}[\sigma_1]\Big).
\end{align}
This looks \textit{almost} identical to a purely thermal Rényi entropy calculation multiplied by the fraction of components that are left after reducing the density matrix. Indeed, if we are not careful, we might assume that this is just telling us that, if we focus on the leading contribution to the Rényi entropy at large-$N$, we will find that the result of tracing out components   behaves exactly the same as if the different components of $\Phi$ are completely decoupled, giving us just the entropy of the fraction of the system we are looking at. This conclusion would not be quite accurate. The key point here lies in the saddle configuration $\sigma_n$. This is a minimum of the entire exponent in \eqref{ONzn}, and not either of the individual actions $S_{n\beta}[\sigma]$ and $S_{\beta}[\sigma]$. In contrast with the thermal Rényi entropy calculation, where we would have 
\begin{align} \label{ONthermalRényi}
R_{\beta,n}=\frac{N}{n-1}\Big(S_{n\beta}[\sigma_{n\beta}]-nS_{\beta}[\sigma_1]\Big).
\end{align}
where $\sigma_{n\beta}$ is the configuration that minimizes $S_{n\beta}[\sigma]$.\footnote{In both \eqref{ONRényi} and \eqref{ONthermalRényi}, $\sigma_1$ minimizes $S_{\beta}[\sigma]$.}

Let us go back to \eqref{ONzn} to try to be more careful in writing the exponent. Let us split $\tau$ into $n$ intervals each of length $\beta$, with $\tau_j$ denoting $\tau$ within the interval from $j\beta$ to $(j+1)\beta$. We also define $\sigma_j$ as the part of the function sigma that is defined on $\tau_j$, that is $\sigma_j=\sigma$ for any $\tau$ inside the $\tau_j$ range but is undefined otherwise. Now we can rewrite \eqref{ONzn} in a more accurate way as
\begin{align}\label{ONPP}
\nonumber Z_n=\int \mathcal{D}\sigma \exp\Bigg\{-N\Bigg[&(1-\lambda)\left(\Tr\Big(\ln\left(-\partial_\tau^2+m^2+\sigma\right)_{n\beta}\Big)-\int_{n\beta}  \frac{\sigma^2}{g}d\tau\right)\\
&~~~~~~+\lambda \sum_{j=0}^{n-1}\left(\Tr\Big(\ln\left(-\partial_{\tau_j}^2+m^2+\sigma_j\right)_{\beta}\Big)-\int_{\beta}  \frac{\sigma_j^2}{g}d\tau_j\right)\Bigg]\Bigg\},
\end{align}
where the subscripts that we put on the operators in the brackets are there to remind us of the periodicity condition on the functions that the operator is defined on. To make sure everything is well-defined, we need the exponent to be a well-defined functional of $\sigma$. Looking at an object like the trace in the second line of \eqref{ONPP}, we see that it can only take periodic functions with periodicity $\beta$ as input. But $\sigma$ can be any function with periodicity $n\beta$, so this functional is instructed to take the values of $\sigma$ on the $j$-th interval ($\sigma_j$), make a periodic function of periodicity $\beta$ out of it, and then use it as input. We now see that the second line of \eqref{ONPP} is not really $\lambda nS_{\beta}[\sigma]$, but is something that is perhaps better written as $\lambda\sum_{j=0}^{n-1} S_{\beta}[\sigma_j]$.

We can now vary the exponent of \eqref{ONPP} with respect to $\sigma$ to get the gap equation analogue for our calculation
\begin{equation}
(1-\lambda)G_{\sigma,n \beta}(\tau, \tau)+\lambda G_{\sigma_j, \beta}(\tau, \tau)=\frac{2 \sigma(\tau)}{g},~~~~~~~~~~~~~~~~\tau \in [j\beta,(j+1)\beta].
\end{equation}
The first Green's function on the left depends on the choice of $\sigma$ on the entire $n\beta$ interval, while the second Green's function only depends on $\sigma_j$. Of course, what we ended up with here is not a single equation, but rather $n$ equations (one for each choice of $j$) that should be solved simultaneously. One may argue that we should look for configurations displaying a $\mathbb{Z}_n$ symmetry in $\tau$ since the exponent in \eqref{ONPP} is symmetric under $\tau \rightarrow \tau+j\beta$ for any integer $j$. As these are just periodic functions with periodicity $\beta$, we can write $\sum_{j=0}^{n-1} S_{\beta}[\sigma_j]$ as $nS_{\beta}[\sigma]$ for these configurations, offering some justification for how we wrote \eqref{ONRényi} to  make contact with the thermal Rényi entropy formula. Restricting to these configurations, the gap equation becomes
\begin{equation}
(1-\lambda)G_{\sigma,n \beta}(\tau, \tau)+\lambda G_{\sigma, \beta}(\tau, \tau)=\frac{2 \sigma(\tau)}{g}.
\end{equation}
Regardless of which gap equation we choose to use, the lesson here is clear: the large-$N$ saddle can be affected by the partial tracing over internal degrees of freedom in a nontrivial way. This is not at all surprising, but the nice thing about this picture is that we have this parameter $\lambda$ that can gradually take us from saddles corresponding to $n$ separate solutions, each with Euclidean time periodicity $\beta$, to a single saddle with time periodicity $n\beta$. More importantly, we know how to understand this in terms of integrals over sources that generate this saddle shift. 

While the result we got for the vector model here may not be particularly interesting, there is nothing about this picture that we cannot carry over to large-$N$ theories with adjoint fields. After all, large-$N$ saddles should tell us something about Rényi entropy calculations (if the result has the right scaling with $N$) regardless of which degrees of freedom we partially trace. As is typically the case, figuring out how to get an actual answer in large-$N$ theories with adjoint fields is nowhere near as simple as it is for the vector model. Nevertheless, we at least now know how to set up such a calculation in terms of sources to be integrated over.  

\subsection{Some holographic musings}

We would like to now end with a discussion of how all of this can be relevant in the context of holography and the AdS/CFT correspondence. In the introduction, we started by motivating our investigations by pointing out the GKPW dictionary tells us how to relate sources on the CFT side to boundary conditions on the string theory/gravity side. Let us reproduce the equation here for convenience 
\begin{equation} \label{GKPW2}
\left\langle e^{\int d^d x \phi_0(\vec{x}) \mathcal{O}(\vec{x})}\right\rangle_{CFT}=
Z_{\text{string/gravity}}\Big[\left.\phi\right|_{\partial \text {AdS }}=\phi_{0}\Big]
\end{equation}
Our framework now suggests that we may be able to express Rényi entropy calculations as particular choices of correlated averaging over sources on the CFT side. More precisely, we might seek to have a correlated average that manipulates multiple copies of the boundary theory, by cutting and then cyclically gluing some of their degrees of freedom. Alternatively, we can look for a correlated averaging of multiple sources inside one copy of the boundary. These two pictures correspond to the difference between trying to go from $\Tr{\rho}^n$ to $\Tr{\rho^n}$ or trying to do the opposite. Regardless of which way we choose to proceed, we now have some hope of being able to translate the calculation to some manipulations on the string theory/gravity side. 

As mentioned in the introduction, the operators we discussed averaging over in this paper are different from the kind of operators one typically discusses in the context of the AdS/CFT correspondence. Instead, one usually thinks in terms of sources and deformations corresponding to operators built out of already traced objects, with the source sitting outside of any trace. In contrast, if we consider the construction in section \ref{sec:gaugetheory} for example, we see that we had to place the sources inside traces of the matrix and its conjugate momentum.

 Discussing how to extend this framework to more general kinds of averaging will be the focus of a subsequent paper. For now, we will ignore all of this and just try to ask what kind of picture one expects to get if we were able to construct Rényi entropy calculations in terms of an averaging that can be understood from the perspective of the AdS/CFT dictionary. 

 Performing some form of averaging in the context of holography has become rather ubiquitous in the literature in recent years \cite{Saad:2019lba, Stanford:2019vob, Maloney:2020nni, Afkhami-Jeddi:2020ezh, Dong:2021wot, Chandra:2022bqq, Marolf:2020xie, Cotler:2020ugk, Jafferis:2025vyp, Barbar:2023ncl, Barbar:2025krh}. Here we are interested in a specific kind of averaging, designed to calculate some Rényi entropy as a mathematical identity from the field theory perspective. The interesting question then becomes: can we in general perform this kind of calculation in the bulk? The fact that integrations over sources that are similar to what we are discussing have already been discussed in the context of holography \cite{Aharony:2015aea, Gubser:2002vv,Hartman:2006dy,Betzios:2019rds, Betzios:2021fnm, VanRaamsdonk:2020tlr, Gao:2016bin}, may give us some encouragement that the answer to this question may be (at least in some cases) yes\footnote{Recently, a relevant kind of averaging over defects localized on a lower-dimensional submanifold in CFT has been discussed in \cite{Shimamori:2025afk}. Special thanks to Zohar Komargodski for pointing this out.} In any case, we will not attempt to answer this question concretely here, but instead we will discuss a related question. Namely, assuming that it is possible to capture this kind of calculation from the bulk's perspective, what would that calculation possibly look like? 

As a start, we should make sure that we are getting results that make sense from the boundary theory point of view. This is where the simple vector model we discussed can serve as a useful template. If we view the gravitational solutions in AdS/CFT as large-$N$ saddles of some boundary calculations, then we expect a similar situation, at least in essence, to our vector model example. More precisely, we expect that whatever source integration we come up with may feed back into the large-$N$ effective action leading to a different saddle/gravitational solution. In that sense, a new gravitational solution may be able to capture the leading term in a large-$N$ expansion of $\Tr{\rho_r^n}$ for some reduced density matrix $\rho_r$ in the boundary theory. We should stress that in principle, the source integration must be done at finite $N$ and only afterwards do we seek a large-$N$ expansion of the resulting quantity.  

Similar to our treatment of the vector model, as we define our choice of the partial trace, we can at times choose to parametrize the size of the subsystem we trace out by some parameter $\lambda$. This parametrization is done in such a way that we have $0\leq\lambda\leq1$ and at $\lambda=1$ corresponding to the subsystem becoming the full system (everything traced out) while $\lambda=0$ corresponds to nothing getting traced out. In this case, we know what the correct calculation for the $0$ and $1$ end points of the range of $\lambda$ should correspond to. Namely, $\lambda=1$ should result in $\Tr{\rho}^n$ with $\rho$ being the full unreduced density matrix while $\lambda=0$ must give us $\Tr{\rho^n}$. Of course, choosing $\lambda$ between $0$ and $1$ is what we are after. Assuming that we found a choice in this range that can be captured by a semiclassical bulk computation, then we may expect connected geometries like Euclidean wormholes to play a role in getting us the correct answer. Indeed, in many of the cases where similar averaging was discussed in the literature, wormhole solutions played an important role \cite{Betzios:2019rds, Betzios:2021fnm, VanRaamsdonk:2020tlr, Gao:2016bin}.

Alternatively, we may circumvent direct use of the AdS/CFT dictionary and instead look for holographic constructions that can be related to the quiver gauge theory structure discussed in section \ref{sec:gaugetheory}. Indeed, it was argued in \cite{VanRaamsdonk:2020tlr} that gauge theories connected by bifundamental matter through an auxiliary gauge theory, potentially arising from D-brane constructions, may be dual to wormhole geometries. This is almost exactly the kind of situation we found in section \ref{sec:gaugetheory}, suggesting that perhaps the kind of color entanglement we discussed can be related to some wormhole solutions.  

\section{Discussion}

We saw in this paper how a certain correlated averaging over codimension-one defects can effectively cut and glue path integrals. This can then be used to perform partial surgery operations to represent ``generalized replica manifolds'' corresponding to a wide range of entanglement entropy calculations. We then showed how this framework can be used to link a certain notion of entanglement entropy between color degrees of freedom to quiver gauge theories. Finally, we discussed how this framework may potentially find useful applications in large-$N$ calculations and holography. The natural next step now is to discuss how this framework can be extended to averaging over more general choices of defects/operators. This will be the subject addressed in a subsequent paper. 
\appendix

\section*{Acknowledgments}
I am especially grateful to my advisor, Sumit Das, for his many useful comments and guidance throughout this project. I also thank Ahmed Barbar, Hardik Bohra, Nico Cooper, Alexander Frenkel,
Zohar Komargodski, Jacob McNamara,
Ganpathy Murthy, Alfred Shapere and Joseph Straley for helpful discussions and comments. This manuscript was finalized during a visit to the Simons Center for Geometry and Physics (SCGP) at Stony Brook University, whose hospitality is gratefully acknowledged. This work is partially supported by National Science Foundation grants NSF-PHY/211673 and NSF-PHY/2410647.

\section{Constraints and zero modes} \label{sec:zeromodes}
In section \ref{sec:MIS}, we pointed out that one should be somewhat careful in evaluating the functional integrals in \eqref{master}. Here, we expand a little on this point.
Consider as an example the constraint given in \eqref{regionrestriction}. Now suppose that we take a very simple-minded approach where we think of the theory as being on a lattice with a finite number of lattice points. Then, what we really want to insert in the functional integral are factors of the form
\begin{equation*}
\prod_{x\notin \mathcal{R}} \delta(K_j(x))\, \delta(J_j(x)),
\end{equation*}
instead, the way we wrote \eqref{master}, we are inserting factors of the following form
\begin{equation*}
\prod_{x} \delta(\Theta_\mathcal{R}(x
)K_j(x))\, \delta(\Theta_\mathcal{R}(x
)J_j(x))\det\left(\frac{\delta \left(\Theta_\mathcal{R}(x
)K_j(x)\right)}{ \delta K_j}\right)\det\left(\frac{\delta \left(\Theta_\mathcal{R}(x
)J_j(x)\right)}{ \delta J_j}\right),
\end{equation*}
with
\begin{equation*}
\Theta_\mathcal{R}(x)=\begin{cases}
1,\; \;\;\;\;\;\;\;\; x\notin\mathcal{R},\\
0, \;\;\;\;\;\;\;\;\;x\in \mathcal{R}.
\end{cases}
\end{equation*}
The issue here is that the functional determinant has an infinite number of zero modes corresponding to functions that are strictly $0$ outside of $\mathcal{R}$. On the other hand, the delta functions contain infinite volume factors, corresponding to the same set of functions
\begin{equation*}
\prod_{x} \delta(\Theta_\mathcal{R}(x
)K_j(x))\, \delta(\Theta_\mathcal{R}(x
)J_j(x))=\left(\prod_{x \notin \mathcal{R}} \delta(K_j(x))\, \delta(J_j(x))\right)\left(\prod_{x \in \mathcal{R}} \delta(0)\, \delta(0)\right).
\end{equation*}
In the way we wrote \eqref{master}, we assume that both the zero modes of the functional determinant and the infinite volume factors can be regularized, say by introducing some parameter $\epsilon$, in such a way that they cancel exactly in the limit $\epsilon \rightarrow 0$. For example, if we introduce some regularized delta function $\delta_{\epsilon}(0)$, such that it is finite for $\epsilon \neq 0$ but satisfies
\begin{equation*}
\lim_{\epsilon \rightarrow 0}\delta_{\epsilon}(0)=\delta(0),
\end{equation*}
then we can adjust the functional determinant factors to be 
\begin{equation*}
\det\left(\frac{\delta \left(\Theta_\mathcal{R}(x
)K_j(x)\right)}{ \delta K_j}+\frac{1}{ \delta_\epsilon(0)}\right)\,\det\left(\frac{\delta \left(\Theta_\mathcal{R}(x
)J_j(x)\right)}{ \delta J_j}+\frac{1}{ \delta_\epsilon(0)}\right),
\end{equation*}
and then by taking the $\epsilon \rightarrow 0$ limit we get\footnote{The limit $\epsilon \rightarrow 0$ here is taken before the lattice spacing in the functional integration measure is taken to $0$.} 
\begin{equation*}
\lim_{\epsilon \rightarrow 0}\Bigg\{\left(\prod_{x \in \mathcal{R}} \delta_\epsilon(0)\, \delta_\epsilon(0)\right)\det\left(\frac{\delta \left(\Theta_\mathcal{R}(x
)K_j(x)\right)}{ \delta K_j}+\frac{1}{ \delta_\epsilon(0)}\right)\det\left(\frac{\delta \left(\Theta_\mathcal{R}(x
)J_j(x)\right)}{ \delta J_j}+\frac{1}{ \delta_\epsilon(0)}\right)\Bigg\}=1.
\end{equation*}
 Of course, another way to avoid all of this is to define the Dirac delta functional constraint from start in such a way that it does not include all of these Dirac deltas with $0$ argument when regularized on a lattice. This would also mean that we should use determinants with zero modes omitted, typically written as 
\begin{equation*}
\det\text{}^\prime\left(\frac{\delta \left(\Theta_\mathcal{R}(x
)K_j(x)\right)}{ \delta K_j}\right)\,\det\text{}^\prime\left(\frac{\delta \left(\Theta_\mathcal{R}(x
)J_j(x)\right)}{ \delta J_j}\right)
\end{equation*}
where the prime means with zero modes omitted. 

\section{Matrix integral identity}\label{sec:matrixid}
Our starting point here will be the simple identity
\begin{align}
1=\int \left(\prod_{i,a=1}^N  d^2x_i^a\right) \left[\det_{ab}\left(\sum_{i=1}^N q_i^a \left(q^b_i\right)^*\right)\right]^N\prod_{i,j=1}^N\delta^{(2)}\left(\Omega_{ij}-\sum_{a=1}^N x_i^a \left(q^a_j\right)^*\right)
\end{align}
valid for any complex matrix $\Omega$ as long as we choose the $N$ vectors given by 
\begin{equation}
Q^a \equiv [q_1^a, q^a_2, \dots, q^a_N]^\top
\end{equation}
to be linearly independent. As explained in the main text, the determinant here is defined in the superscript ($a$ and $b$) space. Using the fact that we can plug this anywhere without changing the expression, we can use this factor to replace integrals over $\Omega$ with ones over the vectors $\hat{x}_i$
\begin{equation}
\int  d\Omega \; F\left(\Omega_{ij}\right)=\int \left(\prod_{i,a=1}^N  d^2x_i^a\right) \left[\det\left(\sum_{i=1}^N q_i^a \left(q^b_i\right)^*\right)\right]^N F\left(\sum_{a=1}^N x_i^a \left(q^a_j\right)^*\right)
\end{equation}
Now if the integral was only over Hermitian matrices, we get 
\begin{align}
\int_{\text{Herm}} {d\Omega} \;F\left(\Omega_{ij}\right) & \nonumber = \int\left(\prod_{i,a=1}^N   d^2x_i^a  \right){2^{\frac{1}{2}N(N-1)}}\left[\det\left(\sum_{i=1}^N q_i^a \left(q^b_i\right)^*\right)\right]^N  \;F\left(\sum_{a=1}^N x_i^a \left(q^a_j\right)^*\right)  \nonumber \\&\times \left[\prod_{i=1}^N \delta \left(\operatorname{Im}\left(\sum_{a=1}^N x_i^a \left(q^a_i\right)^*\right)\right) \right]\left[\prod_{1\leq i<j \leq N} \delta^{(2)}\left(\sum_{a=1}^N \left[x_i^a \left(q^a_j\right)^*- q_i^a \left(x^a_j\right)^*\right]\right) \right]
\end{align}
The Dirac delta functions above can be written in a more compact way using the fact that we have
\begin{align}
\int_{\text{Herm}} d\Lambda \;e^{-\pi\sum_{a=1}^N \Tr\left(X^a  Q^{a^\dagger}\Lambda  -  \Lambda Q^a X^{a^\dagger} \right)}={2^{\frac{1}{2}N(N-1)}}&\left[\prod_{i=1}^N \delta \left(\operatorname{Im}\left(\sum_{a=1}^N x_i^a \left(q^a_i\right)^*\right)\right) \right] \nonumber \\
 \times & \left[\prod_{1\leq i<j \leq N} \delta^{(2)}\left(\sum_{a=1}^N \left[x_i^a \left(q^a_j\right)^*- q_i^a \left(x^a_j\right)^*\right]\right) \right]
\end{align}
Now putting everything together and using a more compact notation we get the desired identity
\begin{align}
\int_{\text{Herm}} {d\Omega} \;F\left(\Omega\right) & \nonumber = \int_{\text{Herm}} d\Lambda\int\left(\prod_{a=1}^N   dX^a  \right) \left[\det_{ab}\left(\Tr\!\left(Q^a {Q^b}^{\dagger}\right)\right)\right]^N \;F\left(\sum_{a=1}^N X^a {Q^a}^{\dagger}\right)   \nonumber \\&~~~~~~~~~~~~~~~~~~~~~~~~~~~~~~~~~~~~~~\times \exp\left[-\pi\sum_{a=1}^N \Tr\left(X^a  Q^{a^\dagger}\Lambda  -  \Lambda Q^a X^{a^\dagger} \right)\right] 
\end{align}

\section{Quiver to color entanglement} \label{sec:quiver}

We would like to go back and offer more justification for why \eqref{masterquiver} is actually equivalent to \eqref{nohat} up to some overall normalization. For convenience, let us reproduce the two formulas we are comparing here  
\begin{align} \label{quiver1}
    &\left(\prod_{j=0}^{n-1} \int_{\text{Herm}_M}  \!  dL_j\, dW_j\left[\prod_{\alpha=1}^3 \int_{\text{Herm}_M}dV^\alpha_j  \int_{\text{Herm}_N} \!  d\Omega^\alpha_j\, d\Lambda_j^\alpha \int dQ^{\alpha}_j  \, dX^\alpha_j\,\right]    \right)\nonumber\\
&\times\Bigg\{\! \prod_{j=0}^{n-1}
\left\langle \!  e^{i  \Tr_N\left(X^1_j {Q^{1^{\dagger}}_j}\Phi \right)}e^{i\Tr_N\left(X^2_j {Q^{2^{\dagger}}_j}\Pi \right)}e^{- i \Tr_N\left(X^3_j {{Q^3_j}^{\dagger}}\Phi \right)}\!\right\rangle_{\!\hat{\rho}} \! e^{i\Tr_M\!\left(L_j\!\left[\left(\prod_{k=0}^{n-1}\!Q^{1^{\dagger}}_{j+k} Q^3_{j+k}\right)-I_M\!\right]+\text{H.c.}\right)}\!\Bigg\}  \nonumber\\
&\times\exp\!\Bigg[\sum_{j=0}^{n-1}\Bigg\{ i \Tr_M\!\left( W_j \!\left[X^{3^{\dagger}}_j  {Q^3_j}  -   {X^{1^{\dagger}}_{j+1}} {Q^{1}_{j+1}} \right]\right)+\sum_{\alpha=1}^{2}i\Tr_N\!\left( \Lambda_j^\alpha\!\left[ Q^\alpha_j{Q^\alpha_j}^{\dagger}-Q^{\alpha+1}_j{Q^{\alpha+1}_j}^{\dagger}\right]\right)\!\nonumber\\&~~~~~~~~~~~~~~~~~~~~~~~~~~~~~~~~~~~~~+\sum_{\alpha=1}^{3}i\Tr_M\left(V^\alpha_j\!\left[{Q^\alpha_j}^{\dagger}\!Q^\alpha_j-I_M\right]\right)-   \!\Tr_N\!\left(\!X^\alpha_j {Q^\alpha_j}^{\dagger}\!\Omega^\alpha_j   - \text{H.c.}\!\right) \bigg]\!\Bigg\}\Bigg],
\end{align}
which we claim is equivalent to 
\begin{align} \label{quiver2}
&\int_{\text{Herm}_M}   dV \int dQ\left(\prod_{j=0}^{n-1}\prod_{\alpha=1}^{2}\int_{\text{Herm}_N}   d\Omega^\alpha_j\int   dX^\alpha_j     \right)\nonumber\\
&~~~~~~~~\times\Bigg\{\! \prod_{j=0}^{n-1}
\left\langle \!  e^{2 \pi i  \Tr_N\left(X^1_j Q^{\dagger}\Phi \right)}e^{i\Tr_N\left(X^2_j Q^{\dagger}\Pi \right)}e^{-2 \pi i \Tr_N\left(X^1_{j+1} {Q^{\dagger}}\Phi \right)}\!\right\rangle_{\!\hat{\rho}}\Bigg\}  \nonumber\\
&~~~~~~~~\times\exp\bigg[ i\Tr_M\!\left(V\left[Q^{\dagger}Q-I_M\right]\right)- \pi \sum_{j=0}^{n-1} \sum_{\alpha=1}^2\!\Tr_N\!\left(\!X^\alpha_j Q^{\dagger}\Omega^\alpha_j    - \text{H.c.}\!\right) \!
\bigg].
\end{align}
If we think of all formulas from the perspective of the extended Hilbert space, we can imagine embedding all the replica copies in the same Hilbert space where we can compare all the matrices to each other. From this perspective, the fundamental difference between these two equations is that in \eqref{quiver2} we are using the same bifundamental $Q$ everywhere while in \eqref{quiver1} every replica comes with three new $Q$ bifundamentals. It should be clear that choosing all of $Q$ fields to be the same in \eqref{quiver1} is both consistent with all the constraints and effectively reduces it to \eqref{quiver2} (again up to a normalization).  

To show equivalence, however, we need to show that any general valid choice of the $Q$ fields in \eqref{quiver1} can be made equivalent to choosing them all to be the same by exploiting gauge transformations and so allowing freedom in this choice only produces an overall gauge group volume.
Now, the condition 
\begin{equation} \label{cond1}
{Q^\alpha_j}^{\dagger}\!Q^\alpha_j=I_M
\end{equation}
implies that any ${Q^\alpha_j}$ can be related to any other by a (left-acting) $U(N)$ transformation. On the other hand, the expression \eqref{quiver1} is independent under $n$ independent $U(N)$ transformations (one inside each expectation value). We can exploit those transformations to set 
\begin{equation} \label{fix1}
{Q^3_j}={Q^1_{j+1}}
\end{equation}
for all the replicas except the last connection closing the ring between ${Q^3_{n-1}}$ and ${Q^1_{0}}$. Next we point out that the constraint 
\begin{equation} \label{cond2}
 Q^\alpha_j{Q^\alpha_j}^{\dagger}=Q^{\alpha+1}_j{Q^{\alpha+1}_j}^{\dagger}
\end{equation}
forces all of the $Q^\alpha_j{Q^\alpha_j}^{\dagger}$ matrices within the same replica to live in the same subspace in the $N \times N$ matrices space. Which means that any two choices of $Q$ within the same replica can be made equal by (right-acting) $U(M)$ transformation. We have many independent copies of those available but they link the different replicas since ${Q^3_j}$ and ${Q^1_{j+1}}$ have to transform together. We now leave $Q_0^1$ to be whatever it is and use these transformations to align all the other different $Q$ fields inside each replica to be the same
\begin{equation}
{Q^1_{j}}={Q^2_j}={Q^3_j}
\end{equation}
except $Q_{n-1}^3$, which we do not try to align with $Q_{n-1}^1$ and $Q_{n-1}^2$ since transforming it by a $U(M)$ transformation will also change $Q^1_{0}$. Since these transformations can only transform ${Q^3_j}$ and ${Q^1_{j+1}}$ together, they do not spoil \eqref{fix1}.
Now we have exhausted all the freedom we had to align all the $Q$ fields with spoiling what we already fixed. The problem now becomes finding a way to get $Q_{n-1}^3$ to be the same as all the others but realizing that we have one more constraint in our formula
\begin{equation}
\left(\prod_{k=0}^{n-1}\!Q^{1^{\dagger}}_{j+k} Q^3_{j+k}\right)=I_M
\end{equation}
We see that $Q_{n-1}^3$ has to be automatically equal to $Q^1_{0}$ since all the other $Q$ matrices have been made equal to $Q^1_{0}$ by gauge transformations and both \eqref{cond1} and \eqref{cond2} are enforced on all $Q$ matrices. We now see that we can gauge fix \eqref{quiver1} to essentially become equivalent to \eqref{quiver2}.

\end{document}